\documentclass[journal,10pt]{IEEEtran}

\usepackage{outline}
\usepackage{pmgraph}
\usepackage[normalem]{ulem}
\usepackage[utf8]{inputenc}
\usepackage{amsbsy,amsthm,amsmath,amssymb,stmaryrd,pict2e,picture}
\usepackage{hyperref}
\hypersetup{colorlinks,allcolors=black}
\usepackage{enumerate}

\usepackage{graphicx}
\usepackage{times}
\usepackage{xcolor}
\usepackage{xspace}
\usepackage{bm} % bold
\usepackage[justification=centering]{caption}
\usepackage{subcaption}
\usepackage{epstopdf}
\AppendGraphicsExtensions{.tiff}
\graphicspath{{fig/}} % Directory in which figures are stored

\usepackage{epsfig}
\usepackage{tikz}
\usepackage{algpseudocode}
\usepackage[ruled,vlined]{algorithm2e}
\usepackage{mathrsfs}
\usepackage{setspace}
\usepackage[backend=biber,maxbibnames=10,giveninits=true,doi=true,isbn=false,url=true,
eprint=false,
style=numeric-comp,sorting=none%,backref=true % probably should comment out for T-CI
]{biblatex}

\long\def\comment#1{}
\newcommand{\fref}[1] {Fig.~\ref{#1}\xspace}

\newcommand{\cdft}{\xmath{c_{\text{DFT}}}}
\DeclareMathOperator*{\argminop}{arg\,min}
\newcommand{\argmin}[1]{\argminop_{#1}} % better spacing
\newcommand{\diag}{\operatorname{diag}}

\newcommand{\etal}{\textit{et al. }}
\newcommand{\xmath}[1] {\ensuremath{#1}\xspace}
\newcommand{\blmath}[1] {\xmath{\bm{#1}}}
\newcommand{\A}{\blmath{A}}

\newcommand{\D}{\blmath{D}}

\newcommand{\I}{\blmath{I}}

\newcommand{\T}{\blmath{T}}

\newcommand{\W}{\blmath{W}}

\newcommand{\x}{\blmath{x}}

\newcommand{\y}{\blmath{y}}
\newcommand{\z}{\blmath{z}}

\newcommand{\ie}{\text{i.e.}}
\newcommand{\eg}{\text{e.g.}}

\newcommand{\subdot} {_{\bm{\cdot}}}

\newcommand{\bi} {\xmath{b_i}}

\newcommand{\ti} {\xmath{t_i}}

\newcommand{\vi} {\xmath{v_i}}
\newcommand{\yi} {\xmath{y_i}}

\newcommand{\defequ} {\triangleq}
\newcommand{\sumi} {\xmath{\sum_i \; }}
\newcommand{\Wmax}{\xmath{\W_{\max}}}
\newcommand{\Wimp}{\blmath{W}_{\mathrm{imp}}}
\newcommand{\real}[1] {\mathrm{real}\!\left\{#1\right\}}
\newcommand{\realb}[1] {\mathrm{real}\{#1\}}

\newcommand{\abs}[1] {\left| #1 \right|}
\newcommand{\absb}[1] {\big| #1 \big|}
\newcommand{\Half} {\frac{1}{2}}

\newcommand{\reals} {\xmath{\mathbb{R}}}
\newcommand{\complex} {\xmath{\mathbb{C}}}
\newcommand{\RM} {\xmath{\reals^M}}
\newcommand{\CN} {\xmath{\complex^N}}
\newcommand{\eE}{\xmath{\mathbb{E}}}
\newcommand{\fF}{\xmath{\mathbb{F}}}
\newcommand{\FN}{\xmath{\fF^N}}
\newcommand{\baih} {\xmath{\ba_i'}}

\newcommand{\bpsi} {\blmath{\psi}}

\newcommand{\xtz} {\xmath{\tilde{\blmath{x}}_0}}
\newcommand{\xk} {\xmath{\blmath{x}_k}}
\newcommand{\xkk} {\xmath{\blmath{x}_{k+1}}}
\newcommand{\zkk} {\xmath{\blmath{z}_{k+1}}}
\newcommand{\dk} {\xmath{\blmath{d}_k}}
\newcommand{\vk} {\xmath{\blmath{v}_k}}
\newcommand{\vkk} {\xmath{\blmath{v}_{k+1}}}

\renewcommand{\v} {\blmath{v}}

\newcommand{\ba}{\boldsymbol{a}}

\newcommand{\bc}{\boldsymbol{c}}

\newcommand{\br}{\boldsymbol{r}}
\newcommand{\bs}{\boldsymbol{s}}

\newcommand{\bz}{\boldsymbol{z}}

\newcommand{\bleta}{\boldsymbol{\eta}}

\newcommand{\btheta}{\xmath{\bm{\theta}}}

\def\b{{\blmath b}} 

\newcommand{\paren}[1]{\left( #1 \right)}

\long\def\red#1{\bgroup\color{red}#1\egroup}

\newcommand{\of}[1]{\!\paren{#1}}

\newcommand{\Lip} {\xmath{\mathcal{L}}}
\newcommand{\copt}{\xmath{c_{\mathrm{opt}}}}
\newcommand{\cimp}{\xmath{c_{\text{imp}}}}

\newcommand{\cT}{\mathcal{T}}
\definecolor{mich-blue-high}{HTML}{0027CC}
\long\def\blue#1{\bgroup\color{mich-blue-high}#1\egroup}
\long\def\red#1{\bgroup\color{red}#1\egroup}
\newcommand{\ith}{$i$th\xspace}

\makeatletter
\newcommand{\pictslash}[2]{%
  \vcenter{%
    \sbox0{$\m@th#1\varobslash$}\dimen0=.55\wd0
    \hbox to\wd 0{%
      \hfil\pictslash@aux#2\hfil
    }%
  }%
}
\newcommand{\pictslash@aux}[2]{%
    \begin{picture}(\dimen0,\dimen0)
    \roundcap
    \put(0,#1\dimen0){\line(1,#2){\dimen0}}
    \end{picture}%
}
\makeatother % load useful macros

\begin{document}

\title{Poisson Phase Retrieval
\\
in Very Low-count Regimes}

\author{Zongyu~Li,%
~\IEEEmembership{Student Member,~IEEE,}
Kenneth~Lange
% ~\IEEEmembership{Fellow,~IEEE}% <-this % stops a space
and~%
Jeffrey~A.~Fessler,%
~\IEEEmembership{Fellow,~IEEE}
\thanks{Zongyu Li and Jeffrey A. Fessler are with 
Department of Electrical Engineering 
and Computer Science, 
University of Michigan, 
Ann Arbor, MI 48109-2122 
(e-mails: zonyul@umich.edu, 
fessler@umich.edu).}
\thanks{
Kenneth Lange is with Departments of Computational Medicine, 
Human Genetics, and Statistics, 
University of California, 
Los Angeles, CA 90095 
(e-mail: klange@ucla.edu).}% <-this % stops a space
\thanks{Research supported in part by
USPHS grants GM53275 and HG006139, % KL
and by NSF Grant IIS 1838179
and NIH R01 EB022075.
Code for reproducing the results
is available at
\url{https://github.com/ZongyuLi-umich/PPR-low-count}.
}% <-this % stops a space
\thanks{This paper has supplementary downloadable material available at 
\url{http://ieeexplore.ieee.org.}, 
provided by the author. 
The material includes 
experiments with truncated WF 
and derivation of the ADMM algorithm.}
\thanks{The manuscript is accepted in IEEE Transactions on Computational Imaging, doi: 10.1109/TCI.2022.3209936.}
}
\comment{
\markboth{Journal of \LaTeX\ Class Files,~Vol.~14, No.~8, August~2015}%
{Shell \MakeLowercase{\textit{et al.}}: Bare Demo of IEEEtran.cls for IEEE Journals}
}
\maketitle

\begin{abstract}
This paper discusses 
phase retrieval algorithms 
for maximum likelihood (ML) estimation
from measurements following independent Poisson distributions
in very low-count regimes,
\eg, 0.25 photon per pixel.
To maximize the log-likelihood
of the Poisson ML model,
we propose a modified Wirtinger flow (WF) algorithm
using a step size
based on the observed Fisher information.
This approach eliminates all parameter tuning
except the number of iterations.
We also propose 
a novel curvature 
for majorize-minimize (MM) algorithms
with a quadratic majorizer.
We show theoretically that 
our proposed curvature
is sharper than the curvature derived from
the supremum of 
the second derivative of the Poisson ML cost function.
We compare the proposed algorithms 
(WF, MM)
with existing optimization methods,
including WF using other step-size schemes,
quasi-Newton methods such as LBFGS
and
alternating direction method of multipliers 
(ADMM) algorithms,
under a variety of experimental settings.
Simulation experiments 
with a random Gaussian matrix,
a canonical DFT matrix,
a masked DFT matrix and 
an empirical transmission matrix 
demonstrate the following. 
1)
As expected, 
algorithms based on the Poisson ML model
consistently produce higher quality reconstructions
than algorithms derived from Gaussian noise ML models
when applied to low-count data.
Furthermore, 
incorporating regularizers,
such as
corner-rounded anisotropic total variation (TV)
that exploit the assumed properties
of the latent image,
can further improve the reconstruction quality.
2)
For unregularized cases,
our proposed WF algorithm 
with Fisher information for step size
converges faster 
(in terms of cost function and PSNR vs. time)
than other WF methods, \eg,
WF with empirical step size, 
backtracking line search,
and optimal step size for the Gaussian noise model;
it also converges faster than
the LBFGS quasi-Newton method.
3)
In regularized cases,
our proposed WF algorithm
converges faster than 
WF with backtracking line search, 
LBFGS, MM and ADMM.

\comment{and also the fastest 
among all regularized algorithms,
by using a Huber surrogate regularizer.}

%and is even comparable with unregularized algorithms.
% jf: regularized and unregularized have different minimizers
% so we cannot really compare their convergence rates
\comment{
and converges faster than previous WF approaches.
}
\end{abstract}


\IEEEpeerreviewmaketitle
% s,intro
\section{Introduction}
\label{sec:intro}
\IEEEPARstart{P}{hase} retrieval is an inverse problem 
with many applications 
in engineering and applied physics
\cite{jaganathan:15:pra,grohs:20:pru},
including radar \cite{jaming:99:prt},
X-ray  crystallography \cite{millane:90:pri},
astronomical imaging \cite{dainty:87:pra},
Fourier ptychography
\cite{bian:16:fpr,zhang:17:fpm,xu:18:awf,tian:19:fpr}
and coherent diffractive imaging (CDI)
\cite{latychevskaia:18:ipr}. % when citing a book, always give a page or chapter!
In these applications,
the sensing systems
can only measure
the magnitude (or the square of the magnitude)
of the signal,
for example,
optical detection devices 
(\eg, CCD cameras) 
cannot measure the phase of a light wave.
The problem of recovering the original signal 
from only the magnitude
of such linear measurements 
is called phase retrieval.
Mathematically,
the goal
is to recover the unknown signal
$\x \in \FN$
from measurements $\{\yi\}$ that follow
some statistical distribution
\begin{equation}\label{e, yi, p}
\yi \sim p(|\baih\x|^2 + b_i), 
\end{equation}
where $p(\cdot)$
is a probability density function.
Here,
$\baih \in \complex^N$
denotes the $i$th row of the system matrix
$\A \in \complex^{M \times N}$,
where $i=1,\ldots,M$,
%$\x \in \FN$
%denotes the true unknown signal,
and $\bi \in \reals_+$
denotes a known mean background signal
for the $i$th measurement,
e.g., 
as arising from dark current \cite{snyder:95:cfr}.
Here the field
$\fF = \reals$
or
$\fF = \complex$
depending on whether \x is known to be real or complex.

The sensing vectors $\{\baih\}$ 
are often assumed to follow some structure,
\eg, i.i.d. random Gaussian,
or the coefficients of 
discrete Fourier transform (DFT).
For the random Gaussian case,
Candès \etal \cite{candes:13:pea}
showed that
$M \sim \mathcal{O}(N\log N)$
samples are sufficient to recover the signal;
Bandeira \etal \cite{bandeira:13:spi}
posed a conjecture that $M = 4 N - 4$ is necessary 
and sufficient to uniquely recover the original signal
from noiseless measurements.
However, under very low-count regimes with noise,
a much larger $M$ is often needed
to successfully reconstruct the signal.
Additionally,
when \A corresponds to a Fourier transform,
the measurements describe
%for the case that given
only the magnitudes 
of a signal’s Fourier coefficients,
and one usually does not have enough information 
to recover the signal;
while the Fourier transform is injective, 
its point-wise absolute value is not \cite{bandeira:14:prf}.
So a common approach is to create redundancy 
in the measurement process by additional illuminations
of the object using different masks \cite{candes:13:prv}. 
Banderia \etal \cite{bandeira:14:prf}
showed that by using a set of
$\mathcal{O}(\log M)$ random masks
can increase the probability of recovering the signal.

% \vspace*{-2em} % dirty trick
\subsection{Background for Gaussian phase retrieval}
In many previous works,
the measurement vector 
$\y \in \RM$
was assumed to have statistically independent elements
following Gaussian distributions with variance $\sigma^2$:
\begin{equation}\label{e,yi,gaussian}
\yi \sim \mathcal{N}(|\baih\x|^2 + b_i, \sigma^2)
.
\end{equation}
For this Gaussian noise model,
the ML estimate of \x
corresponds to the following
non-convex optimization problem 
\begin{align}
\hat{\x} = \argmin{\x \in \FN} g(\x)
,\ 
g(\x) \defequ \sumi \Big| \yi -\bi 
- \big|\baih \x 
\big|^2\Big|^2
.
\label{g,cost}
\end{align}
To solve \eqref{g,cost}, 
numerous algorithms have been proposed.
One approach reformulates
\eqref{g,cost} by ``matrix lifting''
\cite{candes:13:pea, candes:13:prv, shectman:15:prw},
where a rank-one matrix is introduced
and if the rank constraint is relaxed,
then the transformed problem is convex
and can be solved 
by semi-definite programming (SDP).
The SDP based algorithms can yield
robust solutions 
but can be computational expensive,
especially on large-scale data.
Another approach is
Wirtinger Flow (WF) \cite{candes:15:prv}
and its variants \cite{jiang:16:wfm, cai:16:oro, soltanolkotabi:19:ssr}
that descend the cost function
with a (projected/thresholded/truncated)
Wirtinger gradient 
using an appropriate step size.
In the classic WF algorithm
\cite{candes:15:prv},
the gradient%
\footnote{
If $\x \in \reals^N$,
then all gradients w.r.t. \x in this paper 
should be real and hence
use only the real part of expressions like
\eqref{grad_g}.}
for the Gaussian cost function
\eqref{g,cost} is
\begin{equation}\label{grad_g}
    \nabla g(\x) = 4 \A' \diag\{|\A \x|^2 -\y + \b \}\A \x.
\end{equation}
To descend the cost function,
reference \cite{candes:15:prv}
used a heuristic
where
the step size $\mu$
is rather small for the first few iterations 
and gradually increases 
as the iterations proceed.
The intuition is that 
the gradient is noisy at the early iterations
so a small step size is preferred.
A drawback of this approach is that
one needs to
select hyper-parameters that
control the growth of $\mu$.
An alternative approach
is to perform backtracking for $\mu$
at each iteration \cite{qiu:16:ppr},
\ie,
by reducing $\mu$
until the cost function decreases sufficiently.
This approach guarantees decreasing the cost function 
monotonically
but can increase the compute time of the algorithm
due to the variable number of inner iterations.  
Jiang \etal \cite{jiang:16:wfm} 
derived the optimal step size
for the Gaussian ML cost function \eqref{g,cost}
and showed faster convergence 
than the heuristic step size
when measurements are noiseless
or follow i.i.d. Gaussian distribution.
Cai \etal \cite{cai:16:oro} 
proposed thresholded WF
and showed it can achieve 
the minimax optimal rates of convergence,
but that scheme requires an appropriate selection
of tuning parameters.
Soltanolkotabi \etal \cite{soltanolkotabi:19:ssr}
reformulated the phase retrieval problem 
as a nonconvex optimization problem
and proved that projected Wirtinger gradient descent, 
when initialized in a neighborhood of the desired signal, 
has a linear convergence rate.
However, it can be difficult
to find an initial estimate satisfying 
the conditions mentioned in \cite{soltanolkotabi:19:ssr}.

An alternative to cost function \eqref{g,cost}
(aka intensity model) 
is the magnitude model
that works with the square root of $\y$.
In particular,
\cite{gerchberg:72:paf} 
proposed an algorithm known as 
Gerchberg Saxton (GS) that
introduced a new variable $\btheta$
to represent the phase,
leading to the following joint optimization problem
\begin{align}\label{cost_gs}
    &\hat{\x}, \hat{\btheta} = 
    \argmin{\x \in \FN, \, \btheta \in \CN}
    \|\A \x - 
    \diag\{\sqrt{\max (\y -\b, \blmath{0})}\} \,
    \btheta \|_2^2, \nonumber \\
    &\text{subject to}\quad |\theta_i| = 1,\
    i=1,...,N.
\end{align}
The square root in
\eqref{cost_gs}
is reminiscent of the Anscombe transform
that converts a Poisson random variable
into another random variable
that approximately has a standard Gaussian distribution.
However, that approximation is accurate
when the Poisson mean is sufficiently large
(\eg, above 5),
whereas this paper focuses on the lower-count regime.
The convergence and recovery guarantees of GS were studied in
\cite{netrapalli:15:pru, waldspurger:18:prw}.

In addition to
matrix-lifting, WF, GS and their variants,
several other algorithms
have been proposed to solve phase retrieval problems
under the assumption of the Gaussian measurement noise,
including
Gauss-Newton methods \cite{gao:17:pru},
LBFGS updates 
to approximate the Hessian
in the Newton's method
\cite{li:16:ogd},
majorize-minimize (MM) methods \cite{qiu:16:ppr}, 
alternating direction method of multipliers (ADMM) \cite{liang:18:prv},
and an iterative soft-thresholding with
exact line search algorithm (STELA) \cite{yang:19:pcd}.
It seems unlikely
that any of the many existing methods
for the Gaussian noise case
are optimal
for low-count Poisson noise.

% s,background
\vspace*{-1.3em} % dirty trick
\subsection{Background for Poisson Phase Retrieval}
\label{sec:background}

In many low-photon count applications
\cite{thibault:12:mlr,goy:18:lpc,xu:18:awf,barmherzig:20:lph,vazquez:21:qpr,
lawrence:20:prw,
gnanasambandam:20:ici},
especially in \cite{gnanasambandam:20:ici},
where 0.25 photon per pixel on average is considered,
%however,
a Poisson noise model is more appropriate:
\begin{equation}
\yi \sim \mathrm{Poisson}(|\baih\x|^2 + b_i)
\label{e,yi,poisson}
.\end{equation}
ML estimation of \x
for the model \eqref{e,yi,poisson}
corresponds to
the following optimization problem 
\begin{align}
\hat{\x} =
&
\argmin{\x \in \FN} 
f(\x)
,\quad
f(\x) \defequ \sumi \psi(\baih \x; \yi, \bi)
,\quad \nonumber \\
&\psi(v; y, b) \defequ (|v|^2 + b) - y \log(|v|^2 + b)
.
\label{e,cost}
\end{align}
Here,
$f(\x)$ denotes the negative log-likelihood
corresponding to \eqref{e,yi,poisson},
ignoring irrelevant constants independent of \x,
and the function
$\psi(\cdot;y,b)$
denotes the marginal negative log-likelihood
for a single measurement,
where $v \in \complex$.
Because
$\abs{v}$ is real,
it is helpful to re-write $\psi$
in the form
\(
\psi(v; y,b) = 
\phi(\abs{v}; y,b)
,\)
where
%For \eqref{e,cost},
\begin{equation}\label{h,cost}
    \phi(r;y,b) \defequ (r^2 + b) - y \log(r^2 + b),
    \quad r \in \reals_+
.\end{equation}
One can verify that the function 
$\phi(r;y,b)$
is non-convex over $r \in \reals_+$
when $0 < b < y$.
That property,
combined with the modulus within the logarithm
in \eqref{e,cost},
makes \eqref{e,cost} a challenging optimization problem.
Similar problems
for $b=0$
have been considered previously
\cite{choi:07:prf,
bian:16:fpr,
candes:13:prv,
chen:17:srq,
chang:18:dpp%
}, 
but many optical sensors 
also have Gaussian readout noise
\cite{zhang:17:fpm,kang:20:pen}
so that the mean background signal
is unlikely to be zero.
To accommodate the Gaussian readout noise,
a more precise model would consider
a sum of Gaussian and Poisson noise.
However,
the log likelihood for 
a Poisson plus Gaussian distribution
is complicated,
so a common approximation
is to use a shifted Poisson model
\cite{snyder:93:irf}
that also leads to the cost function in \eqref{e,cost}.
An alternative to the shifted Poisson model 
could be to work with an unbiased inverse transformation
of a generalized Anscombe transform approximation
\cite{Makitalo:13:oio,tian:19:fpr}
or use a surrogate function 
that tightly upper bounds the challenging 
Poisson plus Gaussian
ML objective function 
and apply a majorize-minimize algorithm \cite{fatima:22:pam}.
Algorithms for the 
Poisson plus Gaussian noise model
are interesting topics
for future work.
%but are not under the scope of this paper.

Existing algorithms for the
Poisson phase retrieval are limited in the literature.
Chen \etal \cite{chen:17:srq} proposed 
to solve the Poisson phase retrieval problem
by minimizing a nonconvex functional 
as in the Wirtinger flow (WF) approach;
Bian \etal \cite{bian:16:fpr} used 
Poisson ML estimation and
truncated Wirtinger flow 
in Fourier ptychographic (FP) reconstruction.
Zhang \etal \cite{zhang:17:ana} % "reshaped" WF
consider a scale square root of
\eqref{e,yi,poisson}
for the common case with $\bi = 0$.
Chang \etal \cite{chang:18:tvb}
derived a (TV) regularized ADMM algorithm 
for Poisson phase retrieval 
and established its convergence.
Recently, Fatima \etal \cite{fatima:21:panp} 
proposed a double looped primal-dual majorize-minimize (PDMM) algorithm.

In this paper, we propose novel algorithms
for the Poisson phase retrieval problem
and report empirical comparisons
of the convergence speed and reconstruction quality
of algorithms under a variety of experimental settings.
We presented
a preliminary version of this work
at the 2021 IEEE international conference on image processing (ICIP) \cite{li:21:ppr}. 
We significantly extended this work 
by testing our proposed method 
under more practical experimental settings.
We also added comparisons to related works
such as \cite{jiang:16:wfm, li:16:ogd}.

The main contributions of this paper can be summarized
as follows:
\begin{enumerate}[1)]
\item
We propose a novel method
for computing the step size for the WF algorithm
that can lead to faster convergence
compared to empirical step size \cite{candes:15:prv},
backtracking line search \cite{qiu:16:ppr},
optimal step size derived for 
the Gaussian noise model \cite{jiang:16:wfm},
and LBFGS updates to approximate the Hessian
in Newton's method \cite{li:16:ogd}.
Moreover, our proposed method 
can be computed efficiently 
without any tuning parameter.
\comment{
\item
We propose a novel majorize-minimize (MM) method
using a quadratic majorizer with improved curvature
and proved that our proposed curvature
%is superior to
provides faster convergence
than using
the upper bound of the second derivative of 
the Poisson ML cost function.
}
\item 
We derive a majorize-minimize (MM) algorithm
with quadratic majorizer using a novel curvature.
We show theoretically
that our proposed curvature
is sharper
than the curvature
derived from the upper bound of 
the second derivative of the Poisson ML cost function.

\item
We present numerical simulation results
under random Gaussian, canonical DFT,
masked DFT 
and empirical transmission 
system matrix settings 
for very low-count data,
\eg, 0.25 photon per pixel.
We show that under such experimental settings,
algorithms derived from the Poisson ML model
produce consistently higher reconstruction quality
than algorithms derived from Gaussian ML model,
as expected. % do we have theory?  probably not...
Furthermore, the reconstruction quality
is further improved by incorporating regularizers
that exploit assumed properties of the signal.

\item
We compare the convergence speed 
(in terms of cost function and PSNR vs. time)
of WF with Fisher information 
with other methods for step size
(backtracking line search, optimal Gaussian) 
and LBFGS quasi-Newton method.
We also compare the convergence speed of 
regularized WF with
MM and ADMM \cite{li:21:afp},
using smooth regularizers such as 
corner-rounded anisotropic total variation (TV).
For both cases,
%we show that 
our proposed WF Fisher algorithm converges
the fastest under all system matrix settings.

\end{enumerate}

The rest of this paper is organized as follows.
Section~\ref{sec:methods}
introduces the proposed 
modified WF method
with Fisher information for step size;
and derives the improved curvature for 
the MM algorithm.
Section~\ref{sec:exp_setup}
illustrates implementation details 
of algorithms discussed in Section~\ref{sec:methods}.
Section~\ref{sec:sim}
provides numerical results
using simulated data under
different experimental settings.
Section~\ref{sec:discussion}
and
section~\ref{sec:conclusion}
discuss and conclude 
this paper and provide future directions.

\textit{Notation:}
Bold upper/lower case letters (\eg, \A, \x, \y, \b) 
denote matrices and column vectors, respectively.
Italics (\eg, $\mu, y, b$) denote scalars.
\yi and \bi denote the \ith  
element in vector \y and \b, respectively.
$\reals^N$ and $\complex^N$ denote
$N$-dimensional real/complex normed vector space, respectively.
$(\cdot)^{*}$ denotes the complex  conjugate
and $(\cdot)'$ denotes Hermitian transpose. 
$\diag \{\cdot \}$ is a diagonal matrix 
constructed from a column vector.
%filling into the principal diagonal entries.
Unless otherwise defined,
a subscript denotes outer iterations
and superscript denotes the inner iterations, respectively. 
For example,
$\xk$ denotes the estimate of \x
at the $k$th iteration of an algorithm.
$\varoslash$ denotes element-wise division.
The first and second derivatives
of a scalar function $\psi$
are denoted
$\dot \psi$
and $\ddot \psi$,
respectively.
For gradients associated with complex numbers/vectors,
the notation $\dot{\psi}(\cdot)$ and $\nabla(\cdot)$,
should be considered as an ascent direction,
not as a derivative.
%in real case.

\section{Methods}
\label{sec:methods}

\subsection{Wirtinger flow (WF)}
\label{sec:wf}
This section describes 
the modified WF algorithm 
with proposed step-size approach
based on Fisher information.
To generalize the Wirtinger flow algorithm 
to the Poisson cost function \eqref{e,cost}, 
the most direct approach
simply replaces the gradient
\eqref{grad_g} by \eqref{grad_psi} 
in the WF framework
\cite{zhang:17:ana}
and performs backtracking
to find the step-size $\mu$, 
as in \cite{qiu:16:ppr}.
We propose a faster alternative next.
We treat $0 \log 0$ as $0$ in \eqref{e,cost}
because a Poisson random variable with zero mean
can only take the value $0$.
With this assumption,
one can verify that
$\psi$ has 
the following well-defined ascent direction
(negative of descent direction
\cite{zhang:16:iac}) 
and a second derivative:
\begin{align}\label{grad_psi}
    &\dot{\psi}(v; y, b) = 
    2v \paren{ 1 - 
    \frac{y}{|v|^2 + b}}, \quad v \in \complex.
    \nonumber \\
    &\ddot{\psi}(v;y,b)
    = \mathrm{sign}(v)
    \Bigg( 2 + 2y \frac{|v|^2 - b}{(|v|^2 + b)^2} \Bigg),
    \nonumber \\
    &|\ddot{\psi}(v;y,b)| \le 2 + \frac{y}{4b}.
\end{align}

\subsubsection{Fisher information for Poisson model}
We first make a quadratic approximation
along the gradient direction of the cost function 
at each iteration,
and then apply one step of Newton's method
to minimize that 1D quadratic.
Because computing the Hessian can be 
computationally expensive in large-scale problems,
we follow the statistics literature by
replacing the Hessian by the 
observed Fisher information
when applying Newton's method
\cite{titterington:84:rpe, lange:95:aga}.
Our Fisher approach is based on 
the fact that
the observed Fisher information is the negative
Hessian matrix of the incomplete data log-likelihood functions evaluated at the observed data,
and hence can provide a good approximation to the Hessian with enough data \cite{meng:16:mfc}.
Moreover, the Fisher information matrix
is always positive semi-definite,
and avoids calculation of second derivatives.
Using Fisher information
in gradient-based algorithms
has a long history in statistics
and is central to Fisher's method of scoring
\cite{titterington:84:rpe,osborne:92:fmo,hudson:94:fmo,lange:95:aga}.
%For more explanation, see \cite{meng:16:mfc}. % is this reference really relevant?

Specifically,
we first approximate
the 1D line search problem
associated with \eqref{e,cost} 
by the following Taylor series 
\begin{align}\label{quad_approx}
    \mu_k &= 
    \argmin{\mu \in \reals} f_k(\mu),
    \nonumber \\
    f_k(\mu) &\defequ 
    f(\xk - \mu \nabla f(\xk))
    % \nonumber \\
    %& % f(\xk + \mu \nabla f(\xk))
    \approx 
    f(\xk) 
    % \nonumber \\&
    - \|\nabla f(\xk)\|_2^2 \, \mu
    \nonumber \\ & \
    + \frac{1}{2}\nabla f(\xk)' 
    \nabla^2 f(\xk) \nabla f(\xk) \mu^2,
\end{align}
where one can verify 
that the
minimizer is
\begin{equation}\label{sol_muk}
    \mu_k = \frac{\|\nabla f(\xk)\|_2^2}
    {\real{\nabla f(\xk)'\nabla^2 f(\xk) \nabla f(\xk)}}.
\end{equation}
We next approximate
the Hessian matrix
$\nabla^2 f(\x)$
using the observed Fisher information matrix:
\begin{align}\label{fisher_f}
%\MoveEqLeft
    \nabla^2 f(\x) &\approx \I(\x, \b)  \\
    & \defequ
    \mathbb{E}_{\y} 
    \Big[  \nabla^2 f(\x; \y, \b) \Big| \x, \b \Big] 
    \nonumber \\
    &= 
    \mathbb{E}_{\y} 
    \Big[ 
    \big( \nabla f(\x;\y, \b) \big)
    \big( \nabla f(\x;\y, \b) \big)'
    \Big| \x, \b
    \Big] \nonumber \\
    &=
    \A' \mathbb{E}_{\y}
    \Big[
    \big( \dot\psi\subdot(\v;\y, \b) \big)
    \big( \dot\psi\subdot(\v;\y, \b) \big)'
    \Big| \v, \b
    \Big] \A \nonumber
.\end{align}
%where $\psi_.(\cdot)$ denotes element-wise application
%of the function $\psi$ to its first argument
%(as in the Julia language).
%
Here the dot subscript notation
$\dot\psi\subdot(\v;\y, \b)$ denotes element-wise application
of the function $\dot\psi$ to its arguments
(as in the Julia language),
so the gradient
$ \dot\psi\subdot(\v;\y, \b) $
is a vector in $\complex^M$.
One can verify that
the marginal Fisher information
for a single term $\psi(v;y,b)$
is
\begin{align}\label{fisher_psi}
    \bar{I}(v,b) &= \mathbb{E}_{y}
    \Big[ 
%    \dot{\psi}(y;v,b) \cdot \dot{\psi}(y;v,b)^*
    \absb{ \dot{\psi}(v;y, b) }^2
    \Big| v,b
    \Big] \nonumber \\
    &= \frac{4 |v|^2 }{|v|^2 + b}, 
    \quad v \in \complex,\ b > 0.
\end{align}
Substituting
\eqref{fisher_psi} into \eqref{fisher_f}
using the statistical independence
of the elements of the gradient vector,
and then substituting
\eqref{fisher_f}  into \eqref{sol_muk}
yields the simplified step-size expression
\begin{equation}\label{sol_muk_fisher}
    \mu_k \defequ \frac{\|\nabla f(\xk)\|_2^2}
    {\dk' \, \D_1 \, \dk} \in \reals_+,
\end{equation}
where
$\dk \defequ \A \nabla f(\xk)$
and $\D_1 \defequ \diag \{\bar{I}\subdot(\A \xk,\b)\}$.
(Careful implementation avoids redundant matrix-vector products.)

This approach removes all tuning parameters
other than number of iterations.
In addition,
using the observed Fisher information
leads to a larger step size than
using the best Lipschitz constant of \eqref{e,cost},
\ie, $\max_i \{2 + y_i /(4 b_i)\}$
when $b_i >0$,
hence accelerating convergence.

To facilitate fair comparisons in subsequent sections,
we also derive a Fisher information step size
for the Gaussian noise model here.
%Similarly,
The marginal Fisher information
for the scalar case of the
Gaussian cost function \eqref{g,cost} is 
\begin{align}\label{fisher_gau}
    \bar{I}(v,b) &= \mathbb{E}_{y}
    \Big[\big|4|v|(|v|^2 - b - y)\big|^2
    \Big| v, b \Big]
    \\ \nonumber
    &= 16 |v|^2 (|v|^2 + b)
.\end{align}
Substituting
\eqref{fisher_gau} into \eqref{sol_muk_fisher},
one can also derive
a convenient step size
$\mu_k$
for the WF algorithm
for the Gaussian model 
\eqref{g,cost}
using its observed Fisher information
to approximate the exact Hessian.
We used such step size in our experiment
as will be discussed in Section~\ref{sec:sim}.

\subsubsection{WF with regularization}
To potentially improve 
the reconstruction quality, 
one often adds
%an additional term
a regularizer or penalty
% in  the unregularized
to the
Poisson log-likelihood cost function, 
leading to a cost function of the form
\begin{equation}
\label{e,Phi}
\Psi(\x) = f(\x) + \beta R(\x),
\end{equation}
where
$R : \fF^N \mapsto \reals_+$
is a regularizer
and $\beta \geq 0$ denotes the regularization strength.
The general methods in the paper
are adaptable to many regularizers,
but for simplicity we focus
on regularizers that are based on the assumption
that $\T \x$
is approximately sparse,
for a $K \times N$ matrix \T.
\comment{When \T is prox-friendly,
like the orthogonal discrete wavelet transform (ODWT),
we consider the (non-smooth) regularizer
\(
R(\x) = \| \T \x \|_1
,\)
and we focus on algorithms
based on proximal operators
such as
MM and ADMM.}
%are more suitable than WF,
%because WF requires a well-defined gradient.
In particular,
we used the corner-rounded
anisotropic finite-difference matrix
(aka total variation (TV)) 
for regularization.
Because the WF algorithm
requires a well-defined gradient,
we replaced the $\ell_1$ norm term
with a Huber function
regularizer of the form
\begin{align}\label{alg,huber}
R(\x) =
& \bm{1}' h\subdot(\T \x; \alpha)
= \min_{\z}
\frac{1}{2} \|\T \x - \bz \|_2^2 + \alpha \|\bz\|_1,
\nonumber \\&
h(t;\alpha)
\defequ \left\{
    \begin{array}{ll}
        \frac{1}{2}|t|^2, & |t|<\alpha,\\
        \alpha |t|- \frac{1}{2}\alpha^2, 
        & \text{otherwise}, 
    \end{array}
    \right.
\end{align}
which involves solving for \z analytically in terms of \x.
This smooth regularizer is suitable
for gradient-based methods like WF
and for quasi-newton methods like
LBFGS, as well as for 
versions of MM and ADMM.
We refer to \eqref{alg,huber} as ``TV regularization''
even though it is technically (anisotropic) ``corner rounded'' TV.

For the smooth regularizer \eqref{alg,huber},
we majorize the Huber function $h(t)$
using quadratic polynomials 
with the optimal curvature
using the ratio
$\dot{h}(z)/z$
\cite[p.~184]{huber:81},
so that the step size $\mu_k$ becomes
\begin{align}\label{sol_muk_fisher_reg}
&\mu_k 
\defequ \frac{\|\nabla \tilde{f}(\xk)\|_2^2}
    {\nabla \tilde{f}(\xk)' 
    \, (\A'\D_1 \A + \beta \T' \D_2 \T)\, \nabla \tilde{f}(\xk)},
\nonumber \\
&\nabla \tilde{f}(\xk) \defequ \nabla f(\xk) 
+ \beta \T' \dot{h}\subdot(\T \x; \alpha),
\nonumber \\
&\D_2 \defequ \diag \{\min{\!}\subdot(\alpha \varoslash |\T \xk|, 1)\},
\end{align}
where $\varoslash$ denotes element-wise division.

\subsubsection{Truncated Wirtinger flow}

To potentially reduce the error
in gradient estimation
due to noisy measurements,
Chen \etal \cite{chen:17:srq}
proposed a truncated Wirtinger flow (TWF)
approach that uses
only those measurements 
satisfying a threshold criterion
to calculate the Wirtinger flow gradient.
In particular,
the threshold criterion \cite{bian:16:fpr}
is defined as
\begin{equation}\label{meth,twf}
    \big|\yi - |\baih \x|^2\big|
    \le a^h \frac{\big\|\y - |\A \x|^2\big\|_1}{M}
    \cdot
    \frac{|\baih \x|^2}{\|\x\|_2},
\end{equation}
where $a^h$ is a user-defined parameter 
that controls the threshold value.
When $a^h$ is chosen appropriately,
$\yi$ values that do not satisfy \eqref{meth,twf}
will be truncated
when calculating the gradient,
to try to reduce noise.
However,
we did not use gradient truncation
in our experiments
(Section~\ref{sec:sim})
because
we did not observe any improvement 
on the cost function value at convergence
for various setting of $a^h$
compared to WF (shown in the supplement),
which is consistent with results 
in \cite{bian:16:fpr}.
Furthermore,
we found that the
TWF can instead be computationally inefficient
because it requires
computing the truncated indices 
in each iteration,
especially when both the iteration number 
and $M$ are large. 

\subsubsection{Summary}
Algorithm \ref{alg:wfls} summarizes the 
Wirtinger flow algorithm for the Poisson model that
uses the observed Fisher information for the step size
and the optional gradient truncation for noise reduction.

\begin{algorithm}[ht!]
\SetAlgoLined
\SetInd{0.5em}{0.5em}
 \textbf{Input:} $\A, \y, \b, \x_0, \T, \beta$ and 
 $n$ 
 (number of iterations)
 \\
 \comment{
 Initialize: $\x_0 \leftarrow$ 
 random Gaussian vector
 \\}
 \For{$k=0,...,n-1$}{
  \uIf{\textup{gradient is truncated}}{$\nabla \tilde{f}(\xk) = \A_{\cT}' \dot\bpsi\subdot([\A \xk]_{\cT}; \y_{\cT}, \b_{\cT})$  \\ 
    $+ \beta \T' \dot{h}\subdot(\T \xk)$}
  \Else{$\nabla \tilde{f}(\xk) = \A' \dot\bpsi\subdot(\A \xk; \y, \b)
    + \beta \T' \dot{h}\subdot(\T \xk)$ }
    \uIf{\textup{cost function is regularized}}{$\mu_k \leftarrow$ 
    Computed by \eqref{sol_muk_fisher_reg}}
    \Else{ $\mu_k \leftarrow$ 
    Computed by \eqref{sol_muk_fisher}
    }
    $\xkk = \xk - \mu_k \nabla f(\xk)$
 }
 \textbf{Output:} $\x_n$
 \caption{Wirtinger flow for the Poisson model}
 \label{alg:wfls}
\end{algorithm}

\subsection{Majorize-minimize (MM)}

This section introduces
our proposed MM algorithm
with a quadratic majorizer
using a novel curvature formula
for the Poisson phase retrieval problem.

A majorize-minimize (MM) 
algorithm \cite{hunter:04:ato}
is a generalization of
the expectation-maximization
(EM) algorithm
that solves an optimization problem 
by iteratively constructing 
and solving simpler surrogate optimization problems.
Quadratic majorizers are 
very common in MM algorithms 
because they have closed-form solutions 
and are well-suited to conjugate gradient methods.

\comment{We start with deriving the following upper bound 
for the magnitude of the second-order ascent direction 
of $\psi$:
\begin{align}\label{curv_psi}
    &\ddot{\psi}(v;y,b)
    = \mathrm{sign}(v)
    \Bigg( 2 + 2y \frac{|v|^2 - b}{(|v|^2 + b)^2} \Bigg),
    \nonumber \\
    &|\ddot{\psi}(v;y,b)| \le 2 + \frac{y}{4b}.
\end{align}}

\begin{figure}[hbt!]
    \centering
    \includegraphics[width = \linewidth]{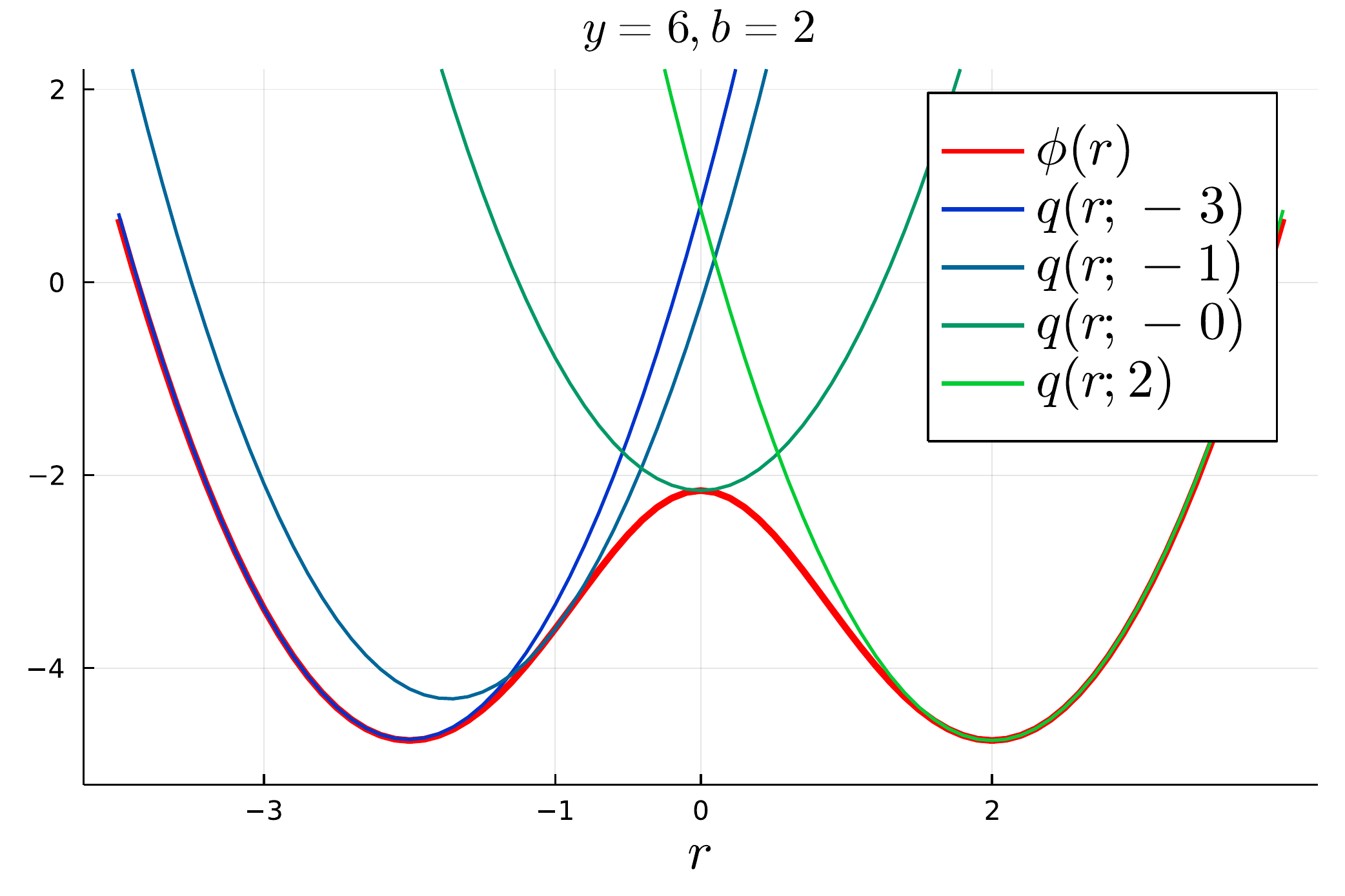}
    \caption{Quadratic majorizers for the non-convex 
    Poisson log-likelihood function $\phi(r;y,b)$ when $y=6$ and $b=2$. }
    \label{fig:quad_mm}
\end{figure}

\begin{figure}[hbt!]
    \centering
    \subfloat[]{\includegraphics[width=0.5\linewidth]{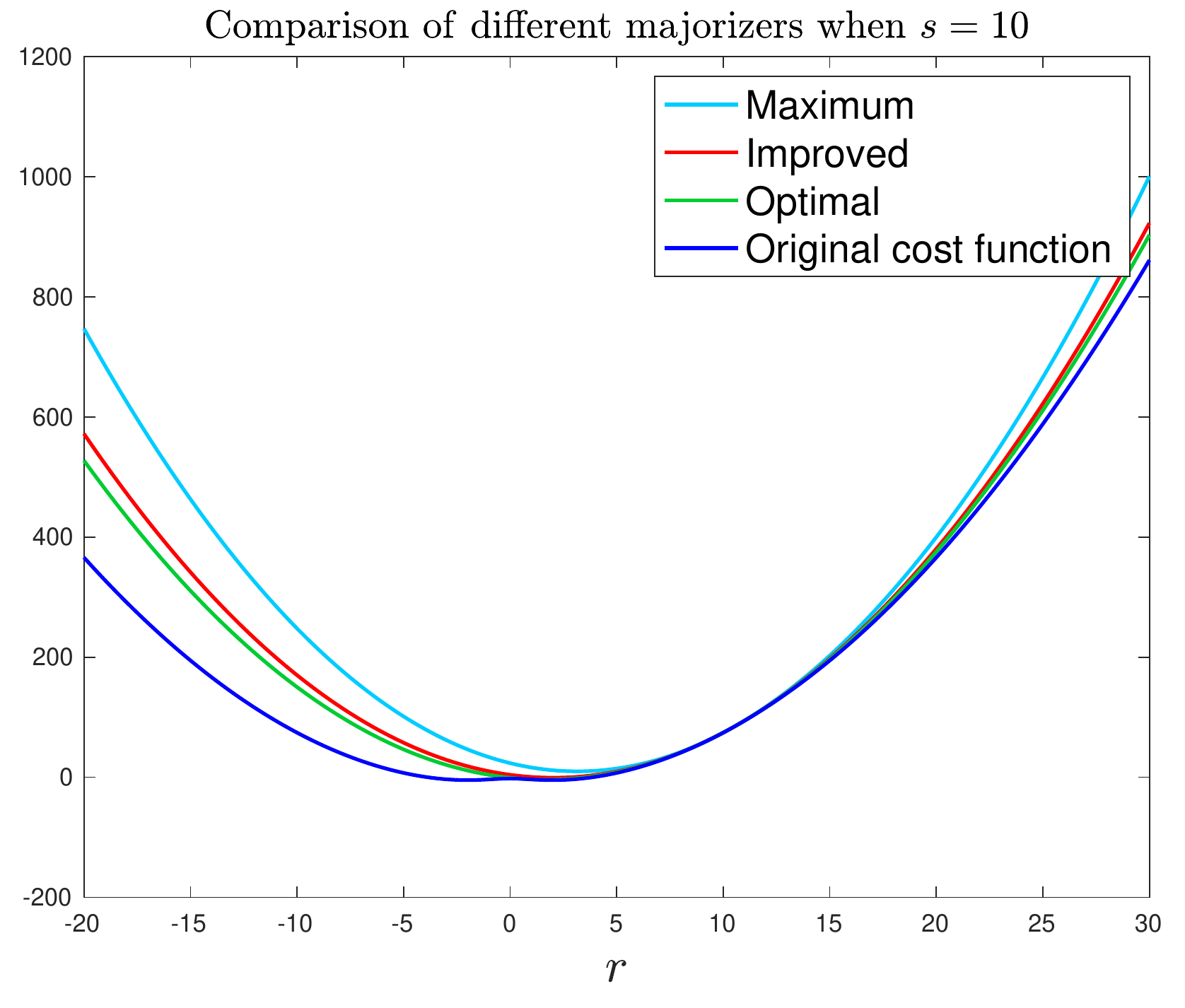}}
    \subfloat[Zoom in around $r= 0.5$.]{\includegraphics[width=0.5\linewidth]{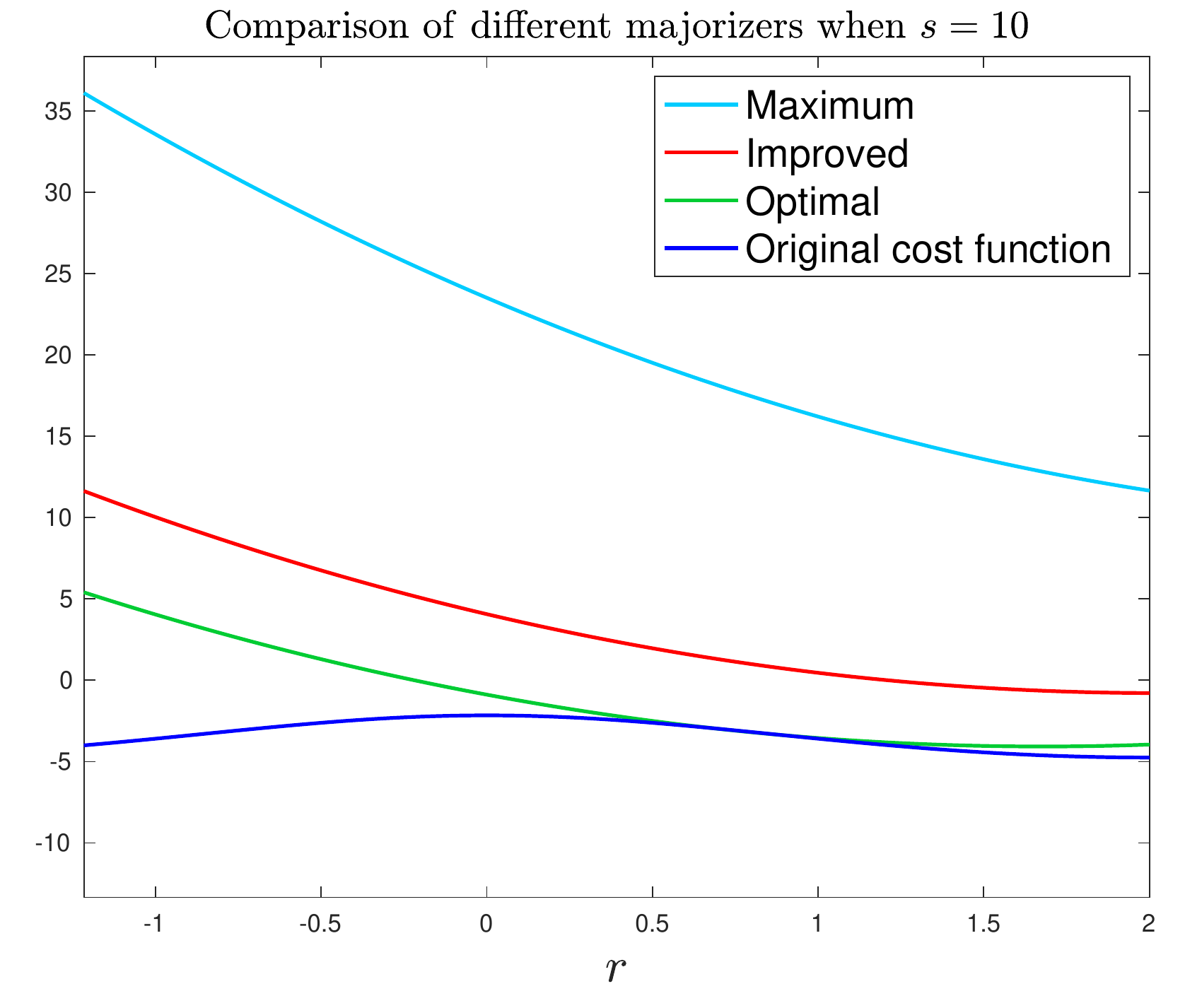}}
    \caption{Comparison of quadratic majorizers with 
    maximum, improved and the optimal curvatures,
    for $y=6$ and $b=2$, visualized around $r = 0.5$.
    All three curves touch at the point $r=s=10$ by construction.
    }
    \label{fig:mm_compare}
\end{figure}

The bounded curvature property
derived in \eqref{grad_psi}
enables us to 
derive an MM algorithm \cite{bohning:88:moq}
with a quadratic majorizer for \eqref{e,cost}.
\fref{fig:quad_mm}
illustrates that
%Without too much effort, 
one can construct 
a quadratic majorizer 
on \reals for \eqref{h,cost}.
With a bit more work
to generalize to $\complex^N$, 
a quadratic majorizer
for the
Poisson ML cost function \eqref{e,cost} 
has the form
\begin{align}\label{quad_mm_form}
q(\x;\xk) &\defequ f(\xk) + \real{ (\x - \xk)' \A' \dot\bpsi\subdot(\A\xk; \y, \b) }
\nonumber \\
&+ \Half (\x - \xk)' \A' \W \A (\x - \xk)
,\end{align}
where \W denotes a diagonal curvature matrix. 
From \eqref{grad_psi}, 
one choice of \W
uses the maximum of $\ddot \psi$:
\begin{equation}\label{quad_mm_max}
    \Wmax \defequ \diag \{ 2 + \y /(4\b ) \}
\in \reals^{M \times M}.
\end{equation}
However, \Wmax is suboptimal
because the curvature
of a quadratic majorizer of $\psi(v;\cdot)$
varies with $v = [\A \xk]_i$.
For example,
when $|v| \rightarrow \infty$,
then \eqref{e,cost} is dominated by 
the quadratic term 
having curvature = 2;
so if $y$ is large  
and $b$ is small,
then \Wmax
can be much greater than 
the optimal curvature 2.
Thus, 
instead of using \Wmax
to build majorizers,
we propose to use 
the following improved curvature: 
\begin{align}\label{quad_mm_imp}
    \Wimp &\defequ\diag \{ \bc\subdot(\A \xk; \y, \b) \} 
        \in \reals^{M \times M}, \nonumber \\
    c(s; y, b) &\defequ
    \left\{
        \begin{array}{ll}
            \ddot{\psi}\Big(\frac{b+
            \sqrt{b^2+b|s|^2}}{|s|}; y, b \Big),
            & s \ne 0, \\
            2, & s = 0.
        \end{array}
    \right.
\end{align}
One can verify 
$\lim_{s \rightarrow 0} c(s; y, b) =2 $ 
so \eqref{quad_mm_imp} 
is continuous over $s \in \complex$.
The next subsection
proves that
\eqref{quad_mm_imp}
provides a majorizer in 
\eqref{quad_mm_form}
and is an improved curvature compared to \Wmax,
though it is not necessarily the sharpest possible
\cite{deleeuw:09:sqm};
the sharpest (optimal) curvature $\copt(s)$ in real case can be expressed as
\begin{equation}\label{eq:opt_curv}
    \copt(s)=
    \sup_{r \ne s}
    \frac{2\Big(\phi(r)-\phi(s)-\dot{\phi}(s)(r-s)\Big)}{(r-s)^2} 
,
\end{equation}
where $\phi(\cdot)$ is the marginal Poisson cost function
defined in \eqref{h,cost}.
However,
%one can verify that
\eqref{eq:opt_curv} usually
does not have a closed-form solution
due to its transcendental derivative;
while our $\Wimp$ has a simpler form 
and is more efficient to compute.
\fref{fig:mm_compare} visualizes 
the quadratic majorizer with different curvatures
and the original Poisson cost function \eqref{e,cost}.
We find the optimal curvature 
numerically
by first discretizing $r$ and 
then finding the supremum
over all discrete segments.

For the ML case 
where constraints or regularizers are absent,
the quadratic majorizer \eqref{quad_mm_form} 
associated with \eqref{quad_mm_max}
or \eqref{quad_mm_imp}
leads to the following MM update:
\begin{align}\label{quad_mm_inv}
    \xkk &= \argmin{\x \in \FN} 
    q(\x;\xk)
    \nonumber \\ 
    &= \xk - (\A'\W\A)^{-1} \A' \dot\bpsi\subdot(\A\xk; \y, \b)
.\end{align}
If $\x \in \reals^N$,
then the MM update for \xkk is
\[
\xk - (\real{\A'\W\A})^{-1} \realb{\A' \dot\bpsi\subdot(\A\xk; \y, \b)}
.
\]
When $N$ is large,
the matrix inverse operation 
in \eqref{quad_mm_inv} 
is impractical,
so we
%an alternative is to 
run a few inner iterations
of conjugate gradient (CG)
to descend the quadratic majorizer
and hence descend the original cost function. 

\subsubsection{Proof of the proposed curvature}
For simplicity,
we drop the subscript $i$
and irrelevant constants
and focus on the negative log-likelihood
for real case for simplicity
as in \eqref{h,cost}.

One can generalize
the majorizer derived here
for \eqref{h,cost}
to the complex case
by taking the magnitude 
and some other minor modifications.

First, we consider some simple cases: 
\begin{itemize}
    \item If $y=0$, then \eqref{h,cost} is a quadratic function,
    so no quadratic majorizer is needed.
    \item If $b=0$ and $y > 0$
    then \eqref{h,cost} has unbounded 2nd derivative
    so no quadratic majorizer exists.
    \item If $b=0$ and $r=0$,
    then $y$ must be zero
    because a Poisson random variable with zero mean
    can only take the value $0$.
    Thus again quadratic majorizer is not needed.
\end{itemize}
So hereafter we assume that $y>0,\ b>0$.
Under these assumptions, 
the derivatives of \eqref{h,cost} are:
\begin{align}
    \dot{\phi}(r) &= 2r \paren{ 1-\frac{y}{r^2+b}},
    \label{eq:grad_h}
    \\
    \ddot{\phi}(r) &= 2+2y\frac{r^2-b}{(r^2+b)^2},
    \label{eq:curv_h}
    \\
    \phi^{(3)}(r) &= \frac{2yr(3b-r^2)}{(r^2+b)^3},
    \label{eq:grad3_h}
\end{align}
where $\phi^{(3)}(r)$ denotes the third derivative.
Clearly, $\dot{\phi}(r)$ is convex on
$(-\infty, -\sqrt{3b}]$ and $[0,\sqrt{3b}] $,
and concave on
$[-\sqrt{3b},0]$ and $[\sqrt{3b},+\infty)$,
based on the sign of $\phi^{(3)}(r)$.

A quadratic majorizer of $\phi(\cdot)$ 
at point $s$ has the form:
\begin{equation}\label{eq:H}
    \Phi(r;s)=\phi(s)+\dot{\phi}(s)(r-s)+\frac{1}{2}c(s)(r-s)^2
.\end{equation}
The derivative of this function
(w.r.t.~$r$) is:
\begin{equation}\label{eq:grad_H}
    \dot{\Phi}(r;s)=c(s)(r-s)+\dot{\phi}(s).
\end{equation}
By design, this kind of quadratic majorizer satisfies
$\Phi(s;s)=\phi(s)$ and $\dot{\Phi}(s;s)=\dot{\phi}(s)$.
From \eqref{eq:grad3_h}, 
we note that $r^2=3b$ is a maximizer of $\ddot{\phi}$
so the maximum curvature is:
\begin{equation}\label{eq:max_curv}
    \ddot{\phi}(r) \le 
    2y\frac{2b}{(4b)^2}+2=2+\frac{y}{4b}
.\end{equation}

\textbf{Proposition:}
$\Phi(r;s)$ defined in \eqref{eq:H} 
is a majorizer of $\phi(r)$
when
$ c(s) = \cimp(s)
,$
where:
\begin{equation}\label{eq:cs_curv}
    \cimp(s) \triangleq \left\{
    \begin{array}{ll}
        \ddot{\phi}\of{u(s)}, & s \ne 0,\\
        \displaystyle \lim_{s \rightarrow 0}
        \ddot{\phi}\of{u(s)}, & s = 0, 
    \end{array}
    \right.
\end{equation}
where
\begin{equation}\label{eq:g}
    u(s) \defequ \frac{b+\sqrt{b^2+bs^2}}{s}
.\end{equation}
By construction, the proposed curvature $c(s)$ 
is at most the max curvature 
given in \eqref{eq:max_curv}.

\textbf{Proof:}
Because of the symmetry of $\ddot{\phi}(r)$,
it suffices to 
prove the proposition for $s\ge 0$
without loss of generality.
First we consider some trivial cases:
\begin{enumerate}
    \item If $s=0$, one can verify 
    $\lim_{s \rightarrow 0}\ddot{\phi}\of{u(s)}=2$.
    In this case, $\Phi(r;s)$ is simply 
    \begin{align}\label{p,s=0}
    \Phi(r;0) &= \phi(0) + \frac{1}{2} c(0) r^2 \\
    &= r^2 + b - y\log(b) \nonumber \\
    &\ge r^2 + b - y\log(r^2+b) = \phi(r). \nonumber
    \end{align}
    \item If $s=\sqrt{3b}$, one can verify
    \begin{equation}\label{p, s3b}
    \ddot{\phi}(g(\sqrt{3b})) = 2 + \frac{y}{4b}
,\end{equation}
which equals the maximum curvature.
\end{enumerate}

Hereafter, we consider only
$s>0$ and $s \ne \sqrt{3b}$. 

To proceed, it suffices to prove 
\begin{equation}\label{p,dhH}
    \begin{split}
        \forall r \in (-\infty,s],\ \dot{\phi}(r) \ge \dot{\Phi}(r;s), \\
        \forall r \in [s,+\infty),\ \dot{\phi}(r) \le \dot{\Phi}(r;s)
    ,\end{split}
\end{equation}
because if \eqref{p,dhH} holds,
then $\forall \tilde{r} < s$:
\begin{align}
\Phi(s;s) - \Phi(\tilde{r};s)
    &= \int_{\tilde{r}}^s \dot{\Phi}(r;s) dr 
    \nonumber\\
    &\le \int_{\tilde{r}}^s \dot{\phi}(r) dr 
    = \phi(s) - \phi(\tilde{r})
\label{p,dhH,l}
,\end{align}
and $\forall \tilde{r}>s$:
\begin{align}
    \Phi(\tilde{r};s) - \Phi(s;s)
    &= \int_s^{\tilde{r}} \dot{\Phi}(r;s) dr 
    \nonumber\\
    &\ge \int_s^{\tilde{r}} \dot{\phi}(r) dr 
    = \phi(\tilde{r}) - \phi(s) 
\label{p,dhH,g}
.\end{align}
Together with $\Phi(s;s) = \phi(s)$,
we have shown that 
\eqref{p,dhH} implies
$\Phi(r;s) \ge \phi(r),\, \forall r \in \reals$.

Substituting
$\dot{\Phi}(r;s) = c(s)(r-s) + \dot{\phi}(s)$
into \eqref{p,dhH},
one can verify that showing
\eqref{p,dhH} becomes showing
\begin{equation}\label{p,dch}
    \cimp(s) \ge \frac{\dot{\phi}(r) - \dot{\phi}(s)}{r-s},\quad 
    \forall r \in \reals, \ r \ne s.
\end{equation}
Furthermore, 
when $s>0$,
the parabola $\Phi(\cdot;s)$
is symmetric about its minimizer:
%the axis of symmetry for $H(r;s)$ 
% this is a *very* long sentence so try to reword to clarify and split into >=2 sentences!
\begin{align}
    \delta &= \delta(s) \defequ 
    \argmin{r}
    \ \Phi(r;s)
    = s - \frac{\dot{\phi}(s)}{\cimp(s)}
    \nonumber\\
    &=
    \frac{s \, \ddot{\phi}(u(s))-\dot{\phi}(s)}
    {\ddot{\phi}(u(s))} \ge 0
    \label{p,symm,a}
.\end{align}
This minimizer is nonnegative because
$\dot{\phi}(s) \le 2s$
and
\begin{align}\label{p,ddh,gs}
    \cimp(s) = \ddot{\phi}(u(s)) &= 2+
    \frac{y s^2 (b+\sqrt{b^2+bs^2})}
    {b(b+s^2+\sqrt{b^2+b s^2})^2} \nonumber \\
    &\ge 2.
\end{align}
Thus, if $\phi(r)\le \Phi(r;s)$ 
when $r\ge 0$, we have
$\phi(-r)=\phi(r)\le \Phi(r;s) \le \Phi(-r;s) = \Phi(r+2\delta;s)$,
so it suffices to prove \eqref{p,dch}
only for $r \ge 0$,
which simplifies \eqref{p,dch} to showing
\begin{equation}\label{p,dch,ge0}
    \cimp(s) \ge \frac{\dot{\phi}(r) - \dot{\phi}(s)}{r-s},\quad 
    \forall r \ge 0, \ r \ne s.
\end{equation}
In short, if \eqref{p,dch,ge0} holds, 
then
$\Phi(r;s)\ge \phi(r), \ \forall r \in \reals$.

To prove \eqref{p,dch,ge0}, 
we exploit a useful property of $\cimp(s)$.
Under geometric view,
$\cimp(s)$ defines (the ratio of)
an affine function connecting points 
$(u(s), \dot{\phi}(u(s)))$ and $(s,\dot{\phi}(s))$
is tangent to $\dot{\phi}(r)$ at point $r = u(s)$,
so that one can verify 
\begin{equation}\label{p,c,tan}
    \ddot{\phi}(u(s)) = \cimp(s) = \frac{\dot{\phi}(u(s))-\dot{\phi}(s)}{u(s)-s},
    \quad u(s) \ne s
.\end{equation}

The reason why $u(s) \ne s$
is that one can verify 
$u(s)=s$ implies $s = \sqrt{3b}$
for $s>0$
that has already been proved above.

Let $\xi(r) = (\dot{\phi}(r) - \dot{\phi}(s))/(r-s)$,
where $r\ge 0$ and $r \ne s$, plugging in 
$\dot{\phi}(r)$ and $\dot{\phi}(s)$ yields:
\begin{equation}\label{p,f}
    \xi(r) = 2 + \frac{2y(sr-b)}{(s^2+b)(r^2+b)}.
\end{equation}
Differentiating w.r.t.~$r$ leads to:
\begin{equation}\label{p,df}
    \dot{\xi}(r) = \frac{2y}{s^2+b} \cdot 
    \frac{-sr^2+2br+bs}{(r^2+b)^2},
\end{equation}
where one can verify the positive root 
of $-sr^2+2br+bs = 0$ is $u(s)$ that is given by \eqref{eq:g}.

Together with $\dot{\xi}(r) > 0$ when 
$r \in (0,u(s))$ 
and 
$\dot{\xi}(r) < 0$ when 
$r \in (u(s), \infty)$,
we have 
\eqref{p,dch,ge0}
holds because
$\xi(r)$ achieves 
its maximum at $\xi(u(s))$:
\begin{equation}\label{p,f,c}
    \xi(r) \le \xi(u(s)) = \cimp(s).\quad \blacksquare
\end{equation}

\subsubsection{Regularized MM}

For the regularized cost function \eqref{e,Phi},
one can use the quadratic majorizer
\eqref{quad_mm_form}
as a starting point.
If the regularizer is prox-friendly,
\comment{
The gradient of the majorizer $q(\x;\xk)$
has (best) Lipschitz constant
\(
\Lip = \| \A'\W\A \|_2 
\)
that can be precomputed when
\(\W = \Wmax\),
e.g., by the power method.
When
$\W = \Wimp$,
the Lipschitz constant changes each iteration,
so instead
we used the inequality
% \cite{chun:20:cao}:
$
\|\A'\W\A\|_2
\le
\|\A'\A\|_2\|\W\|_2 
,$
%so that 
%$\Lip = \max \{|\A' \W \A| \blmath{1}_N\}$. 
and seek for 
an upper bound for 
$\|\A'\A\|_2$ that is simpler to compute.
}%
then
the minimization step of an MM algorithm
for the regularized optimization problem
is
\begin{equation}\label{quad_mm_l1}
    \xkk = \argmin{\x \in \FN}
    q(\x;\xk)
    + \beta \|\T \x \|_1.
\end{equation}
To solve \eqref{quad_mm_l1},
one can apply proximal gradient methods
\cite{daubechies:04:ait,beck:09:afi,kim:18:aro}.
We can use the
proximal optimized gradient method (POGM)
with adaptive restart
\cite{kim:18:aro}
that provides faster worst-case convergence bound
than the fast iterative shrinkage-thresholding algorithm (FISTA)
\cite{beck:09:afi}.

For non-proximal friendly regularizers,
we can ``smooth'' it using the Huber function
\eqref{alg,huber},
leading to the optimization problem of the form
\begin{equation}\label{quad_mm_huber}
\xkk = \argmin{\x \in \FN} 
q(\x;\xk) + \beta 1'h\subdot(\T \x;\alpha),
\end{equation}
and we use nonlinear CG for this minimization,
with step sizes based on Huber's quadratic majorizer.

Algorithm \ref{alg:mm} summarizes 
our MM algorithm with quadratic majorizer 
using the improved curvature
\eqref{quad_mm_imp}.

\begin{algorithm}[ht!]
\SetAlgoLined
\SetInd{0.5em}{0.5em}
 \textbf{Input:} $\A, \y, \b, \x_0$ and 
 $n$ 
 (number of iterations)
 \\
 \comment{
 Initialize: $\x_0 \leftarrow$ 
 random Gaussian vector
 \\}
 \For{$k=0,...,n-1$}{
    Build $q(\x;\xk)$ \eqref{quad_mm_form} using $\Wimp$ \eqref{quad_mm_imp}
    \\
    \uIf{\textup{cost function is regularized}}{
        \uIf{$\T$ \textup{is prox-friendly}}{
        Update $\xk$ by \eqref{quad_mm_l1} using POGM
        }
        \Else{
        Update $\xk$ by \eqref{quad_mm_huber} using CG
        }
    }
    \Else{
    Update $\xk$ by \eqref{quad_mm_inv} or CG
    }
 }
 \textbf{Output:} $\x_n$
 \caption{MM algorithm for the Poisson model}
 \label{alg:mm}
\end{algorithm}
\section{Implementation Details}
\label{sec:exp_setup}

This section introduces 
the implementation details
of algorithms discussed in 
the previous section
and our experimental setup
for the numerical simulation
(Section~\ref{sec:sim}).
We ran all algorithms 
on a server with
Ubuntu 16.04 LTS operating system
having Intel(R) Xeon(R) 
CPU E5-2698 v4 @ 2.20GHz 
and 187 GB memory.
All elements in the measurement vector \y
were simulated to follow independent Poisson distributions
per \eqref{e,yi,poisson}.
All algorithms were implemented 
in Julia v1.7.3.
All the timing results 
presented in Section~\ref{sec:sim} 
were averaged across 10 independent test runs.

\subsection{Initialization}
\label{subsec:init}

Luo \etal \cite{luo:19:osi} 
proposed the optimal initialization strategy
under random Gaussian system matrix setting
with Poisson noise.
Since this paper focuses on very low-count regimes,
the scale factor $\kappa$ in \cite{luo:19:osi} 
is a very small number
so that
$y - \kappa \approx y$.
Therefore,
we used \xtz,
the leading eigenvector
of $\A'\diag\{\y \varoslash (\y + 1)\}\A$
(instead of $\A'\diag\{(\y - \kappa) 
\varoslash (\y + 1)\}\A$)
as an initial estimate of \x.

To accommodate signals of arbitrary scale,
we scaled that leading eigenvector
using a nonlinear least-square (LS) fit:
\begin{equation}\label{init,scal}
\hat{\alpha}
= \argmin{\alpha \in \reals}
\|\y-\b-|\alpha \A \xtz|^2\|_2^2
= \frac{\sqrt{(\y - \b)' |\A\xtz|^2}}{\|\A\xtz\|_4^2}
.\end{equation}
Finally, our initial estimate
is the element-wise absolute value
of $\hat{\alpha} \x_0$ 
if \x is known
to be real and nonnegative;
and is $\hat{\alpha} \x_0$ otherwise.

\subsection{Ambiguities}
To handle the global phase ambiguity,
\ie,
all the algorithms 
can recover the signal only to 
within a constant phase shift
due to the loss of global phase information,
before quantitative comparison,
we corrected the phase of $\hat{\x}$ by 
\begin{equation}\label{e,phase,correct}
\hat{\x}_{\textup{corrected}} 
\defequ 
\mathrm{sign}\paren{\langle \hat{\x}, \x \rangle}
\hat{\x}.
\end{equation}

\subsection{System matrix and True signals}
\subsubsection{System matrix}
We investigated 4 different choices
for the system matrix \A: 
complex random Gaussian matrix
(having $80000$ rows), 
canonical DFT 
(with reference image),
masked DFT matrix
(with 20 masks)
and 
a transmission matrix (ETM)
that is acquired empirically
through physical experiments
\cite{chandra:17:phasepack, metzler:17:cis}.

For the canonical DFT,
we used a reference image
as used
%a common approach % not sure how "common" it is safer not to say this
in holographic coherent 
diffraction imaging (HCDI)
\cite{barmherzig:20:lph},
specifically,
the measurements follow
\begin{equation}\label{e,canon,dft}
\y \sim \textup{Poisson}
(|\mathcal{F}\{
[\x, \blmath{0}, \mathcal{R}]
\}|^2 + \b)
,
\end{equation}
where $\mathcal{F}$
denotes discrete Fourier transform (DFT)
and
$\mathcal{R}$ denotes a known reference image.
This paper uses the reference image
shown in \fref{fig:ref,image,canon,dft},
taken by screen shot from
\cite{barmherzig:20:lph}.

\begin{figure}[hbt!]
    \centering
    \includegraphics[width=0.6\linewidth]{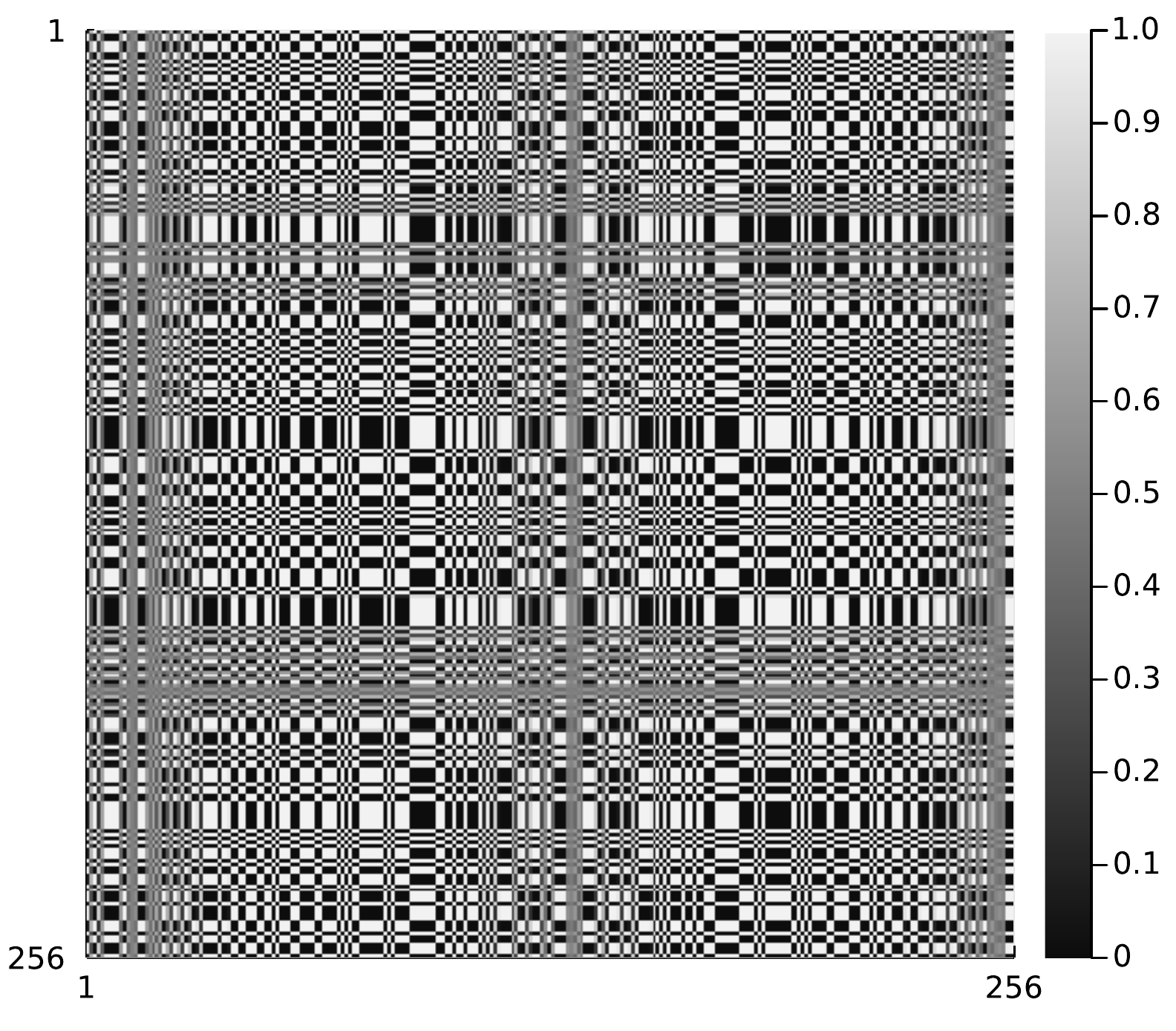}
    \caption{Reference image
    from \cite{barmherzig:20:lph}
    used in HCDI
    and our canonical DFT experiments.}
    \label{fig:ref,image,canon,dft}
\end{figure}

For the masked DFT case,
%in particular,
the measurement vector \y 
in the Fourier phase retrieval problem
has elements with means given by
\begin{equation}
    \eE[ y[\tilde{n}] ] = % there is noise correct? correct!
    \Bigg|\sum_{n=0}^{N-1}
    x[n] e^{- \imath 2\pi n \tilde{n}/
    \tilde{N}}\Bigg|^2 + \, b[\tilde{n}],
\end{equation}
where $\tilde{N} = 2N-1$
(here we consider the over-sampled case),
and $\tilde{n} = 0,...,\tilde{N}-1$.
After introducing redundant masks,
the measurement model becomes
\begin{equation}
    \eE[ y_l[\tilde{n}] ]
    =\Bigg| \sum_{n=0}^{N-1}
    x[n] D_l[n] e^{-\imath 2\pi n \tilde{n}/
    \tilde{N}}\Bigg|^2 + 
    b_l[\tilde{n}],
\label{e,masks}
\end{equation}
where
$\eE[\y_l] \in \mathbb{R}^{\tilde{N}}$
for $i=1,\ldots,L$
and 
$D_l$ denotes the $l$th of $L$ masks.
Our experiment used
$L = 21$ masks
to define the  overall system matrix 
$\A \in \complex^{L\tilde{N}\times N}$,
where the first mask 
has full sampling
and the remaining 20
have sampling rate 0.5 
with random sampling patterns.

We scaled each system matrix % right?  all cases?
by a constant factor
such that the average count of measurement vector $\y$ is 0.25,
and the background count $b$ is set to be 0.1.

\subsubsection{True images}
\begin{figure}[ht!]
    \centering
    \subfloat[]{\includegraphics[width=0.5\linewidth]{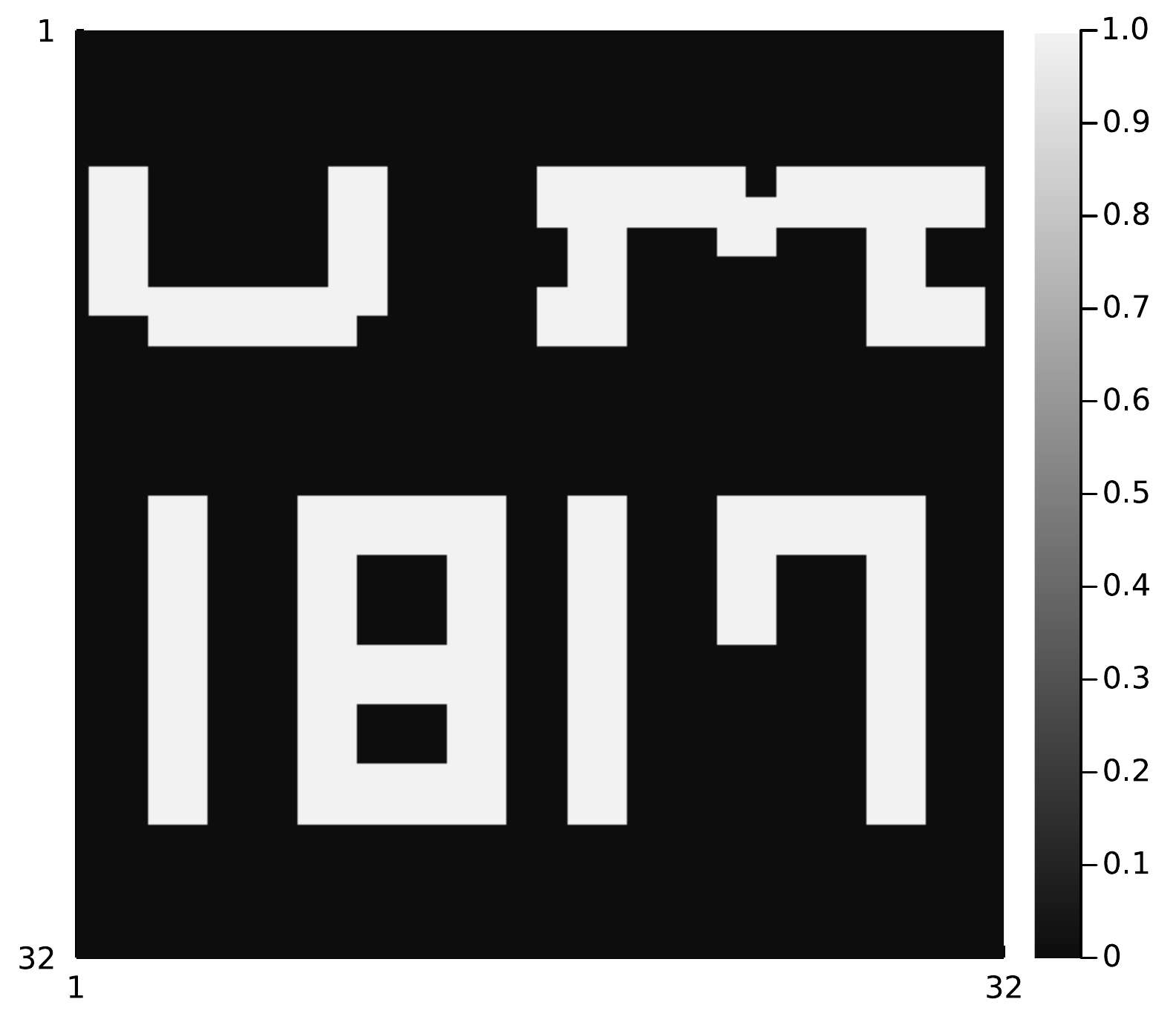}}
    \subfloat[\cite{marchesini:17:pgb}]{\includegraphics[width=0.5\linewidth]{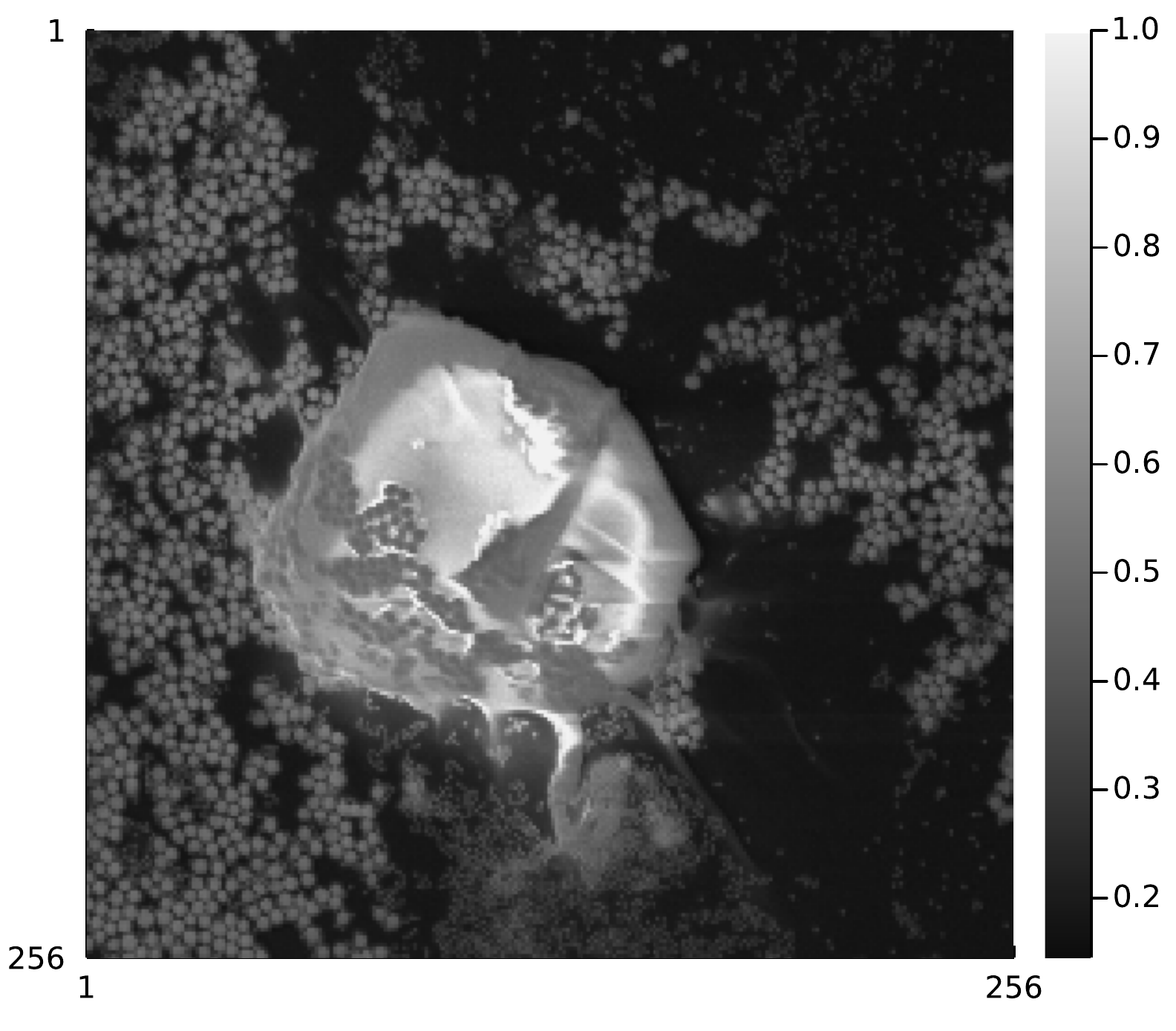}}
    \\
    \subfloat[\cite{barmherzig:20:lph}]{\includegraphics[width=0.5\linewidth]{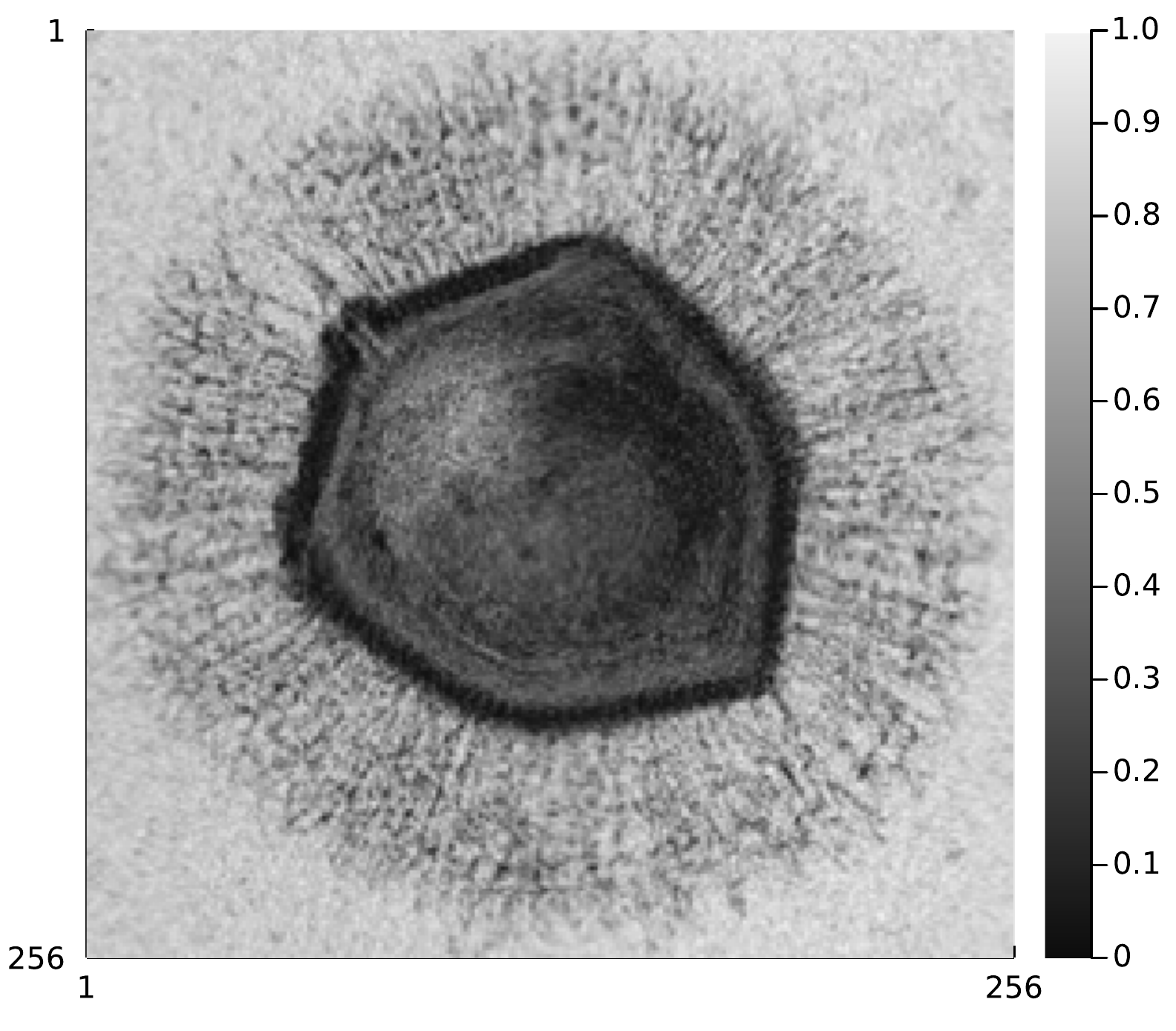}}
    \subfloat[\cite{metzler:17:cis}]{\includegraphics[width=0.5\linewidth]{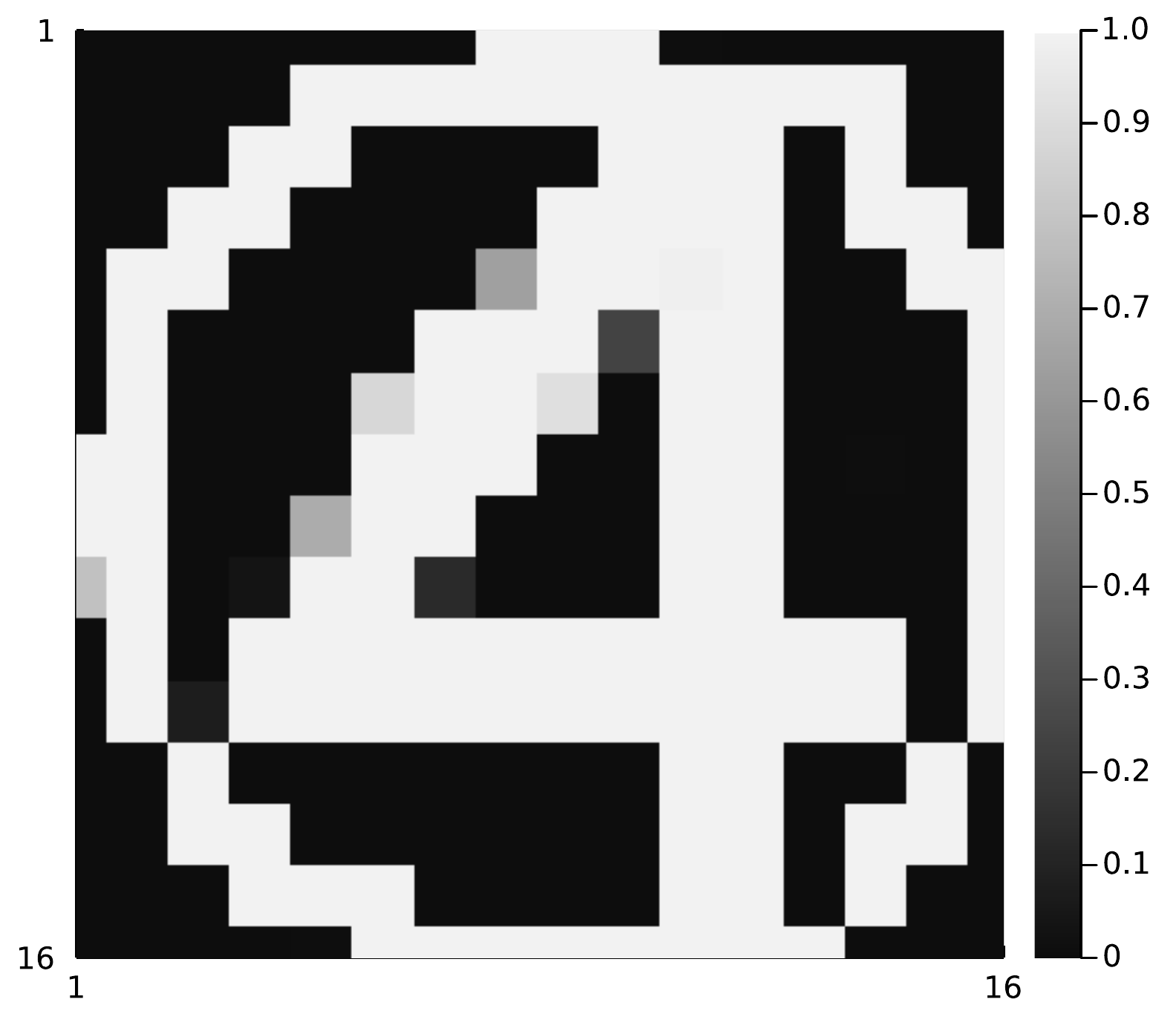}}
    \\
    \subfloat[Real part of (d).]{\includegraphics[width=0.5\linewidth]{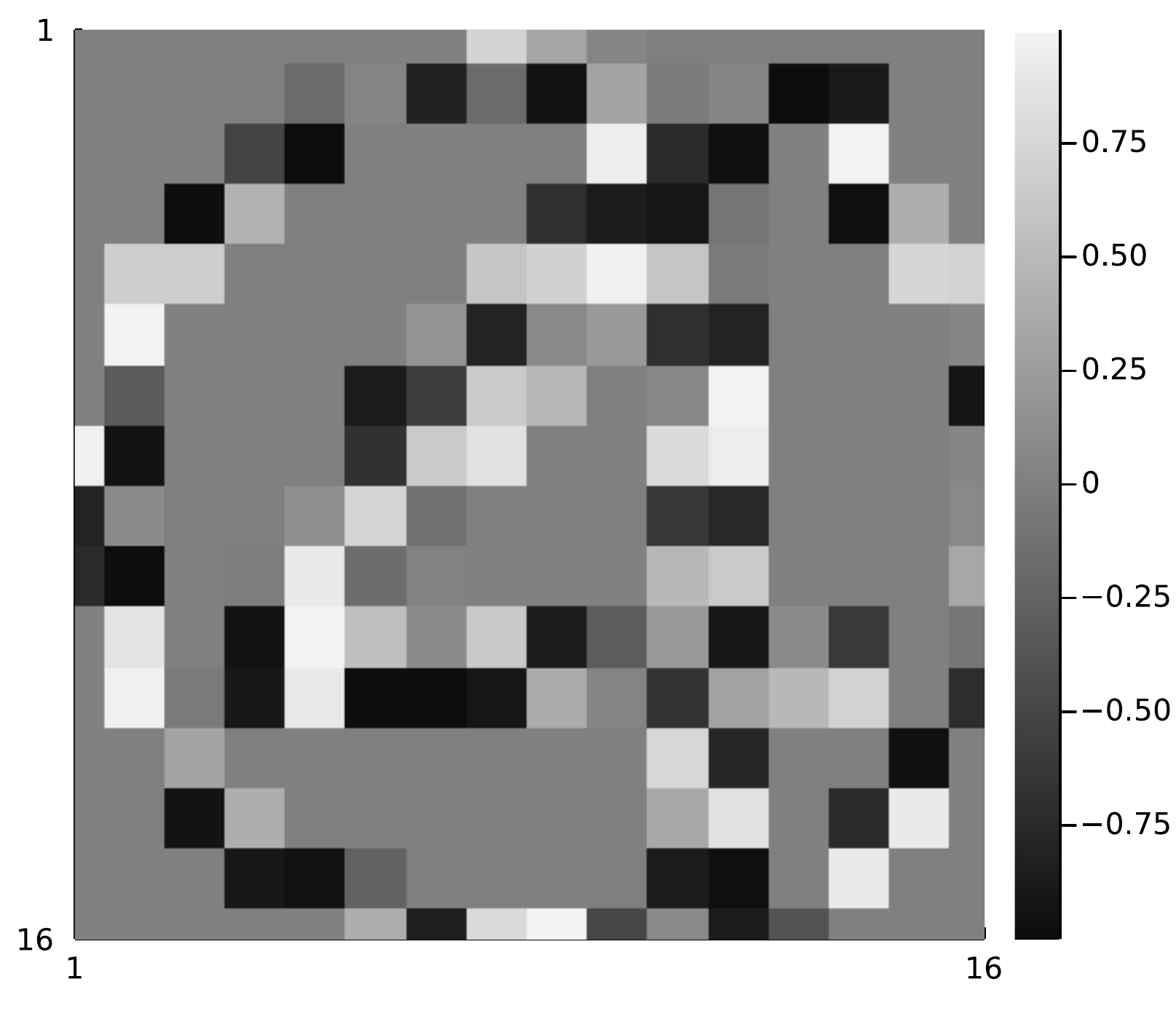}}
    \subfloat[Imaginary part of (d).]{\includegraphics[width=0.5\linewidth]{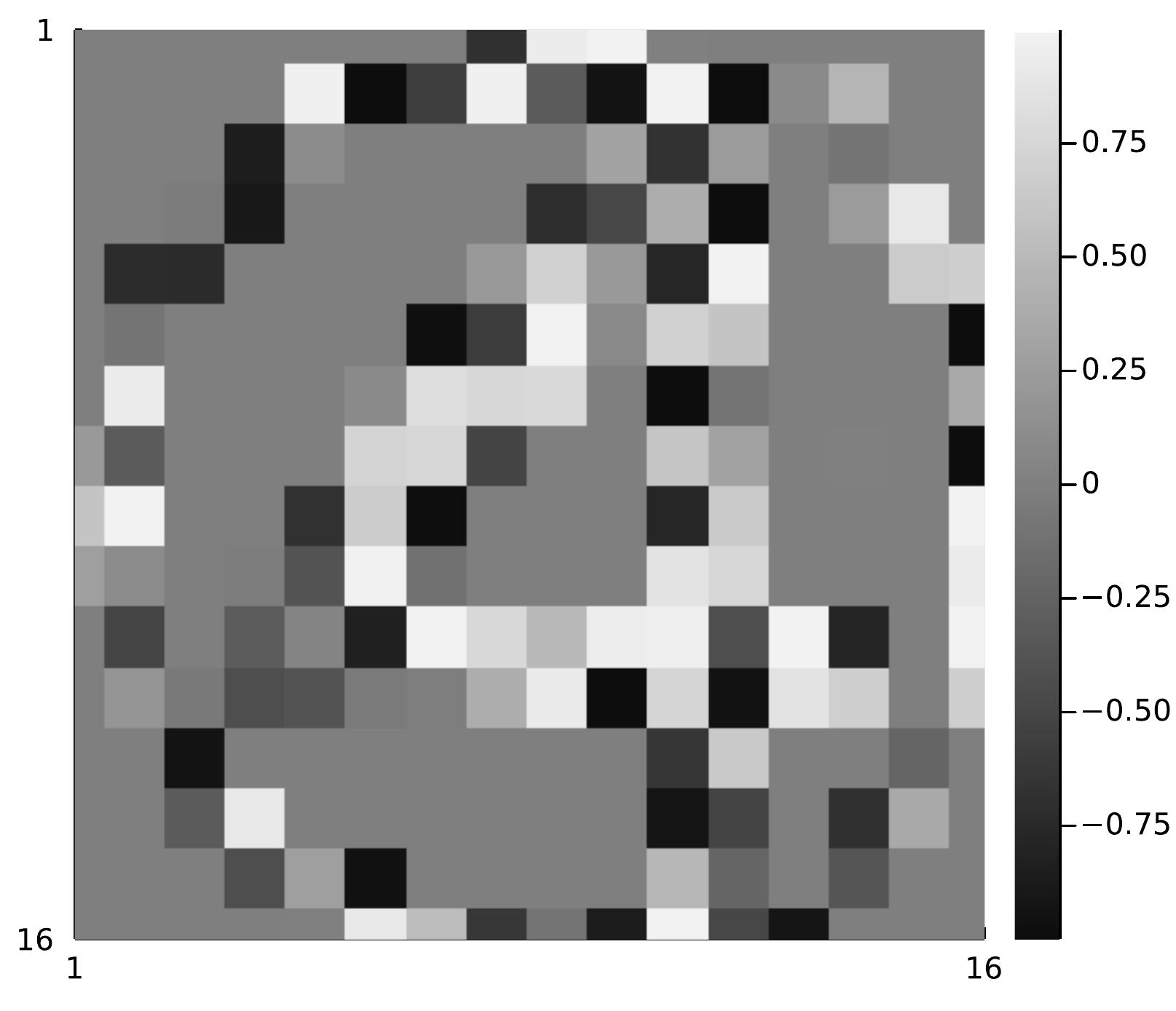}}
    \caption{True images used in the simulations.
    Subfigure (d) shows the magnitude of a complex image.}
    \label{fig:xtrue}
\end{figure}

We considered 4 images 
as the true images in our experiments
\fref{fig:xtrue} shows such images;
(b) is from \cite{marchesini:17:pgb},
(c) is from
\cite{barmherzig:20:lph}, % right?, yes
(d)-(f) are from
\cite{metzler:17:cis}.
We used subfigure 
(a) for experiments with random Gaussian system matrix,
(b) for masked DFT matrix,
(c) for canonical DFT matrix
and (d) for empirical transmission matrix,
respectively.

\section{Numerical Simulations Results}
\label{sec:sim}
\subsection{Convergence speed of WF with Fisher information}
\label{subsec:wf,fisher}

This section
compares convergence speeds,
in terms of cost function vs. time 
and PSNR vs. time,
between 
WF using our proposed Fisher information for step size,
and
empirical step size \cite{candes:13:prv},
backtracking line search \cite{qiu:16:ppr},
the optimal step size for the Gaussian ML cost function \cite{jiang:16:wfm},
and LBFGS quasi-Newton 
to approximate the Hessian in Newton's method \cite{li:16:ogd}.
The LBFGS algorithm was from the ``Optim.jl'' Julia package
\cite{mogensen:18:oam}.
\begin{figure}[ht!] % awkward to put this in middle of paragraph, but ok
    \centering
    \subfloat[Random Gaussian system matrix.]
    {\includegraphics[width=0.8\linewidth]{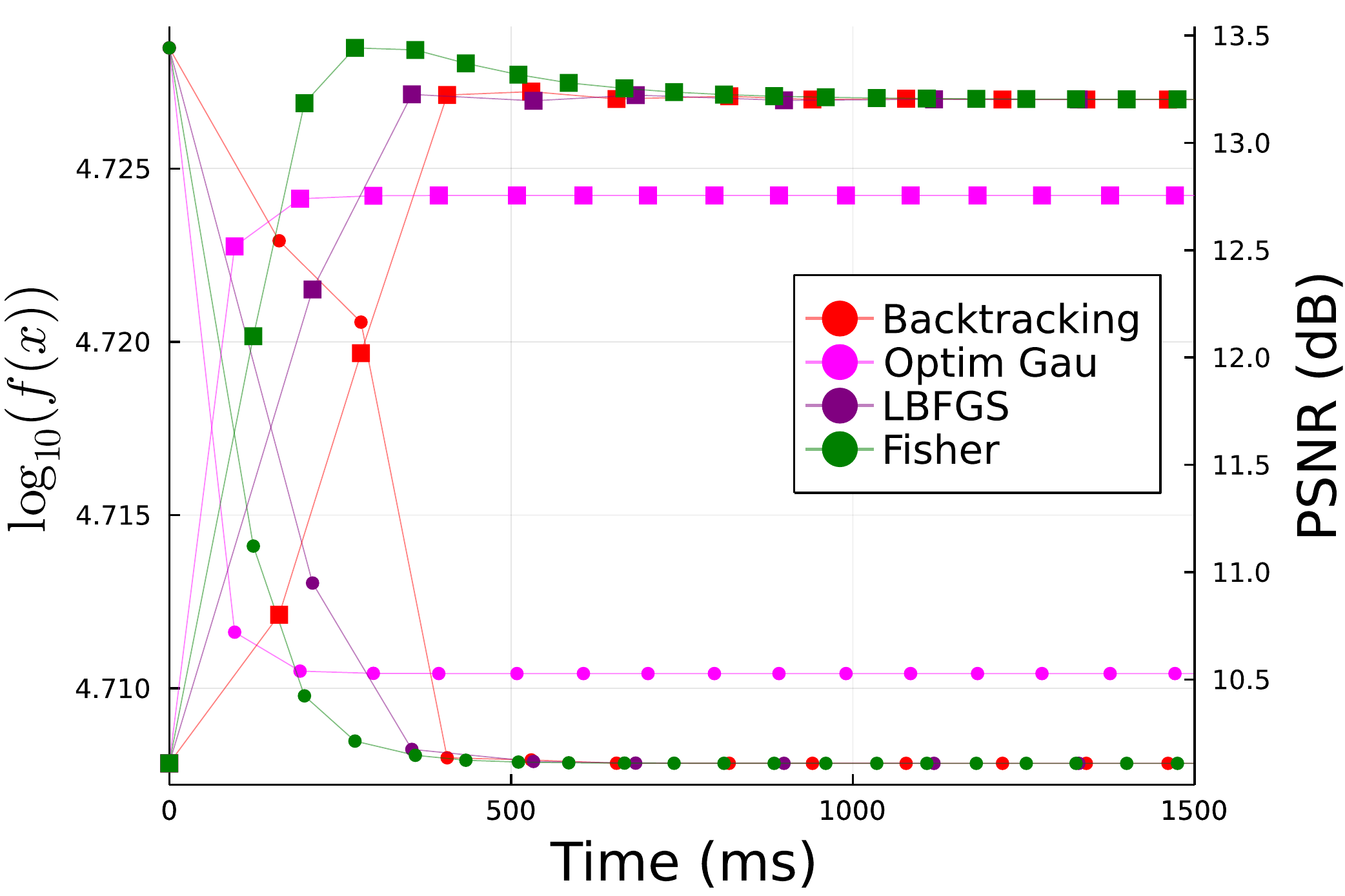}}
    \\
    \subfloat[Masked DFT system matrix.]
    {\includegraphics[width=0.8\linewidth]{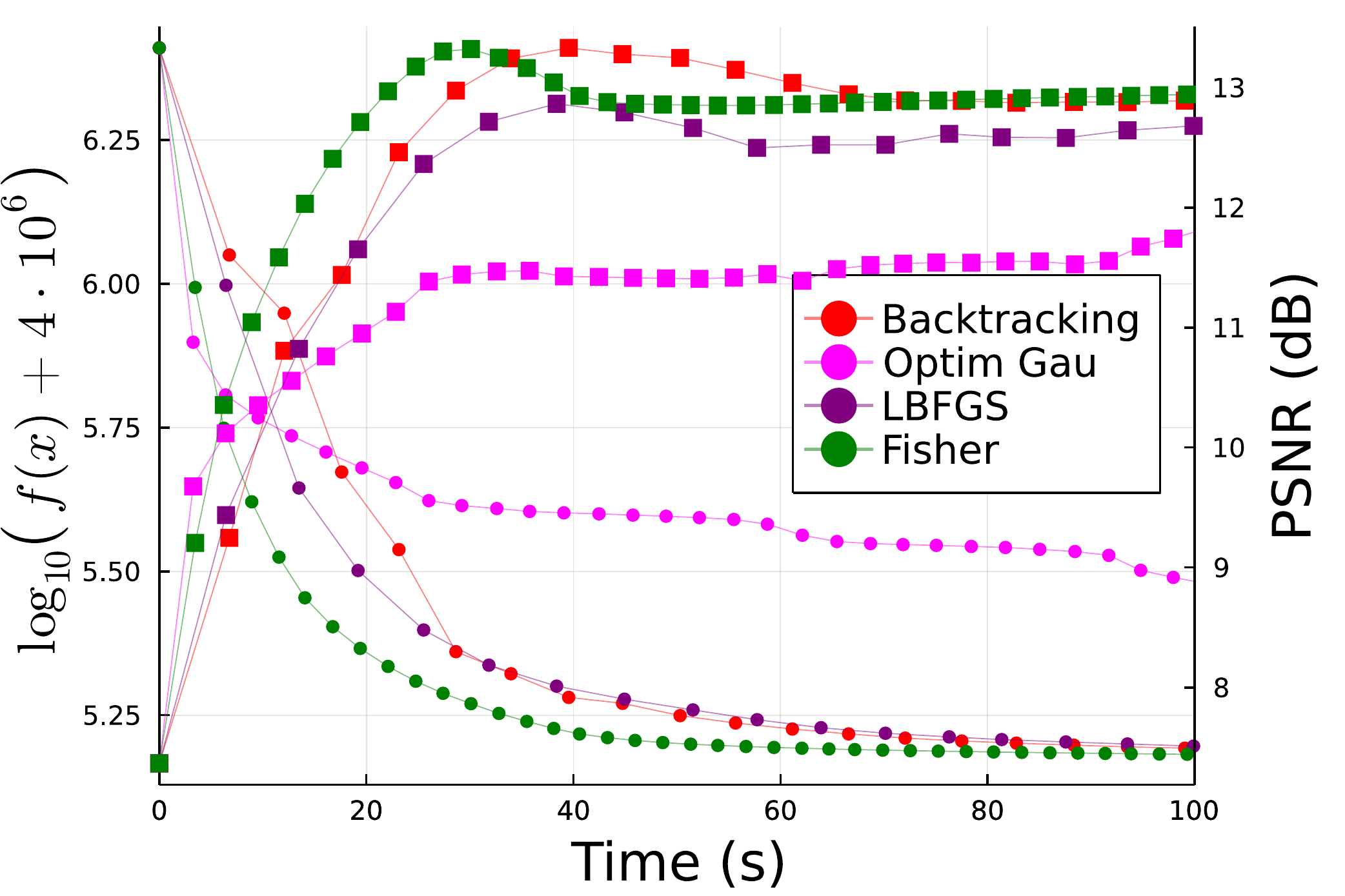}}
    \\
    \subfloat[Canonical DFT system matrix.]{\includegraphics[width=0.8\linewidth]{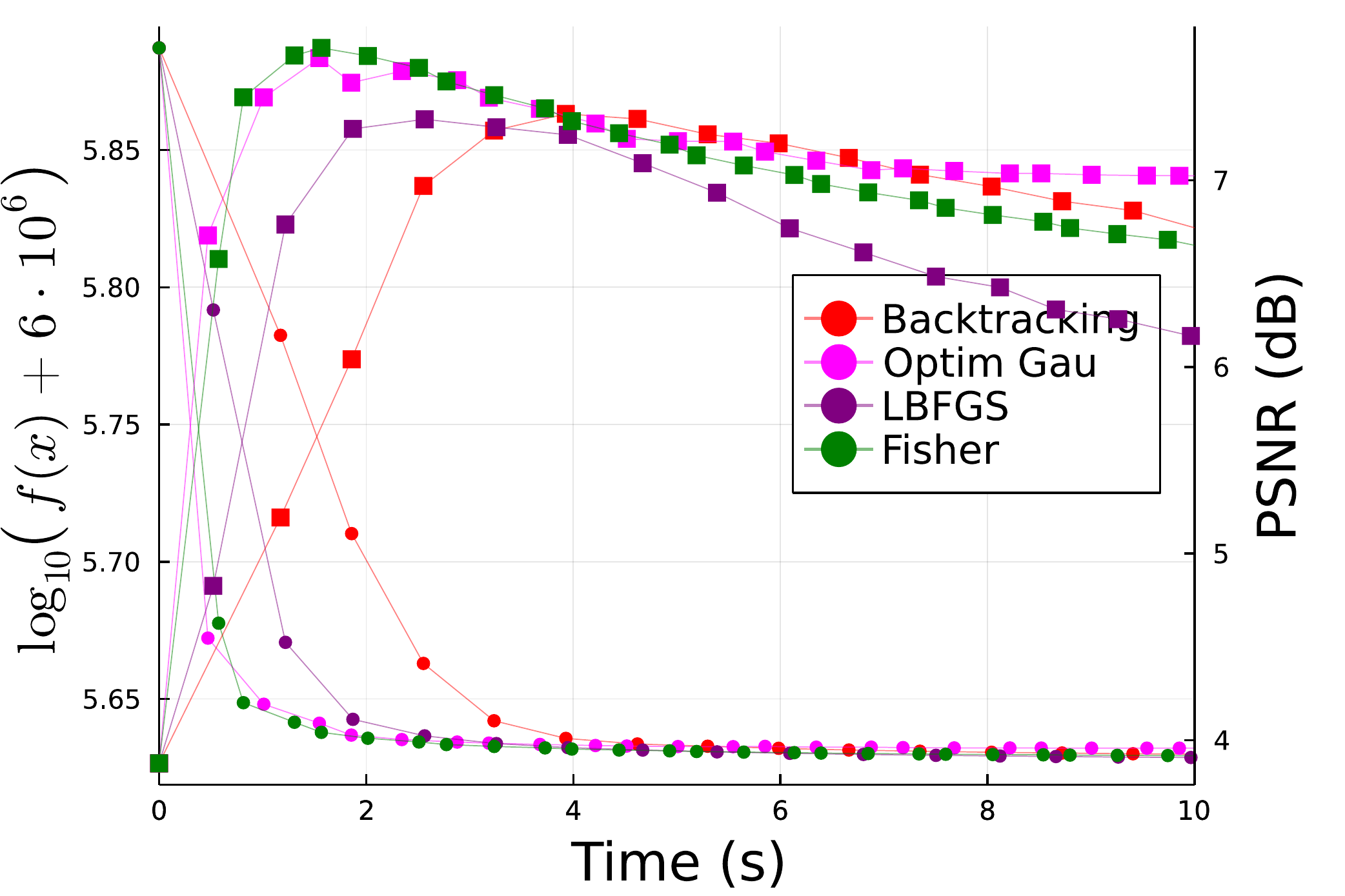}}
    \\
    \subfloat[Empirical transmission matrix.]{\includegraphics[width=0.8\linewidth]{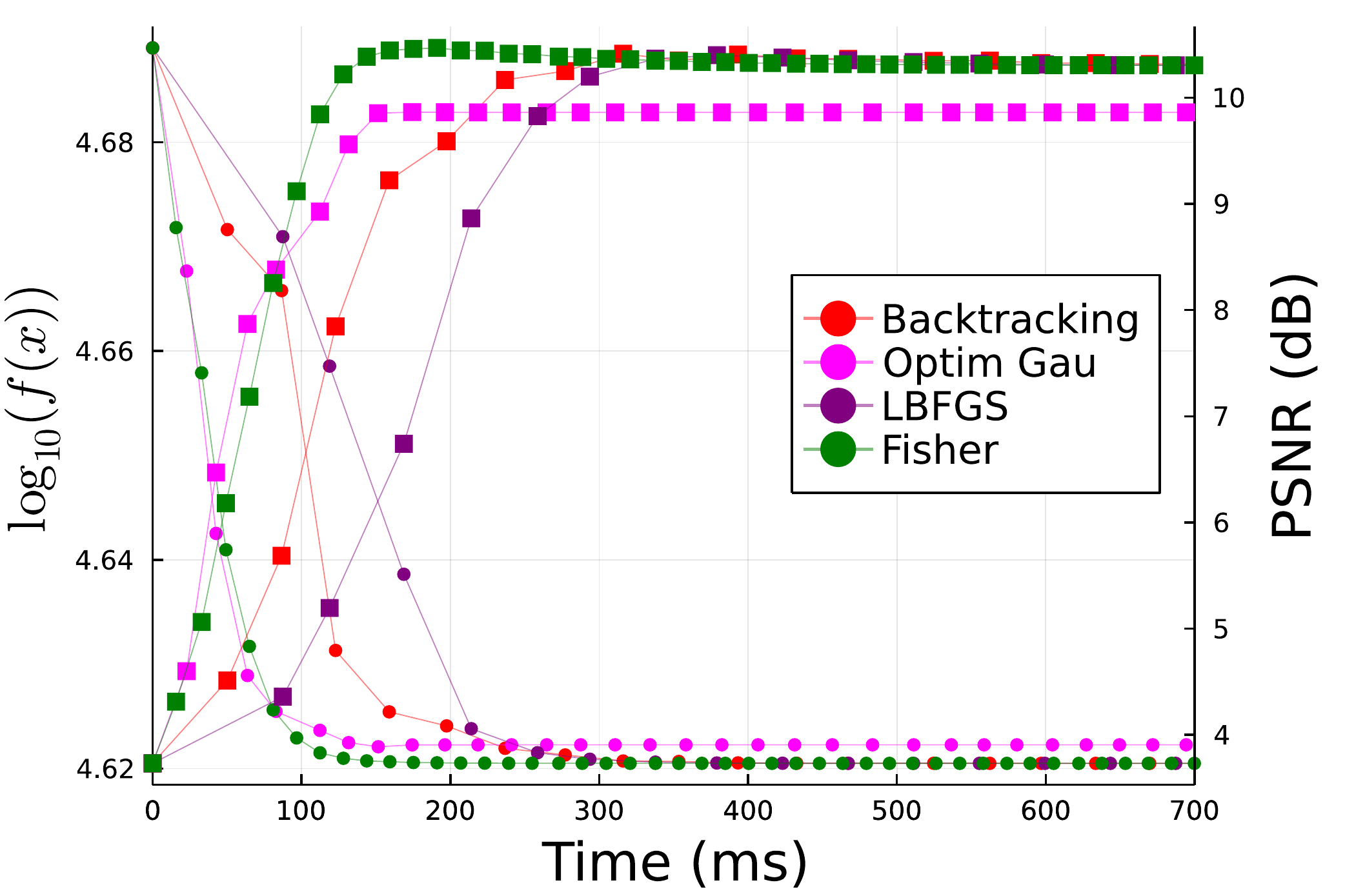}}
    \caption{
    Comparison of convergence speed for various WF methods
    and LBFGS under different system matrix settings.
    The ``Optim Gau'' curve is WF using the curvature from
    \cite{jiang:16:wfm}
    that is optimal for Gaussian noise.
    The circle marker corresponds to the cost function
    and the square marker corresponds to PSNR.}
    \label{fig:wf,fisher}
\end{figure}
\begin{figure*}[ht!]
    \centering
    \subfloat[]{\includegraphics[width=0.25\linewidth]{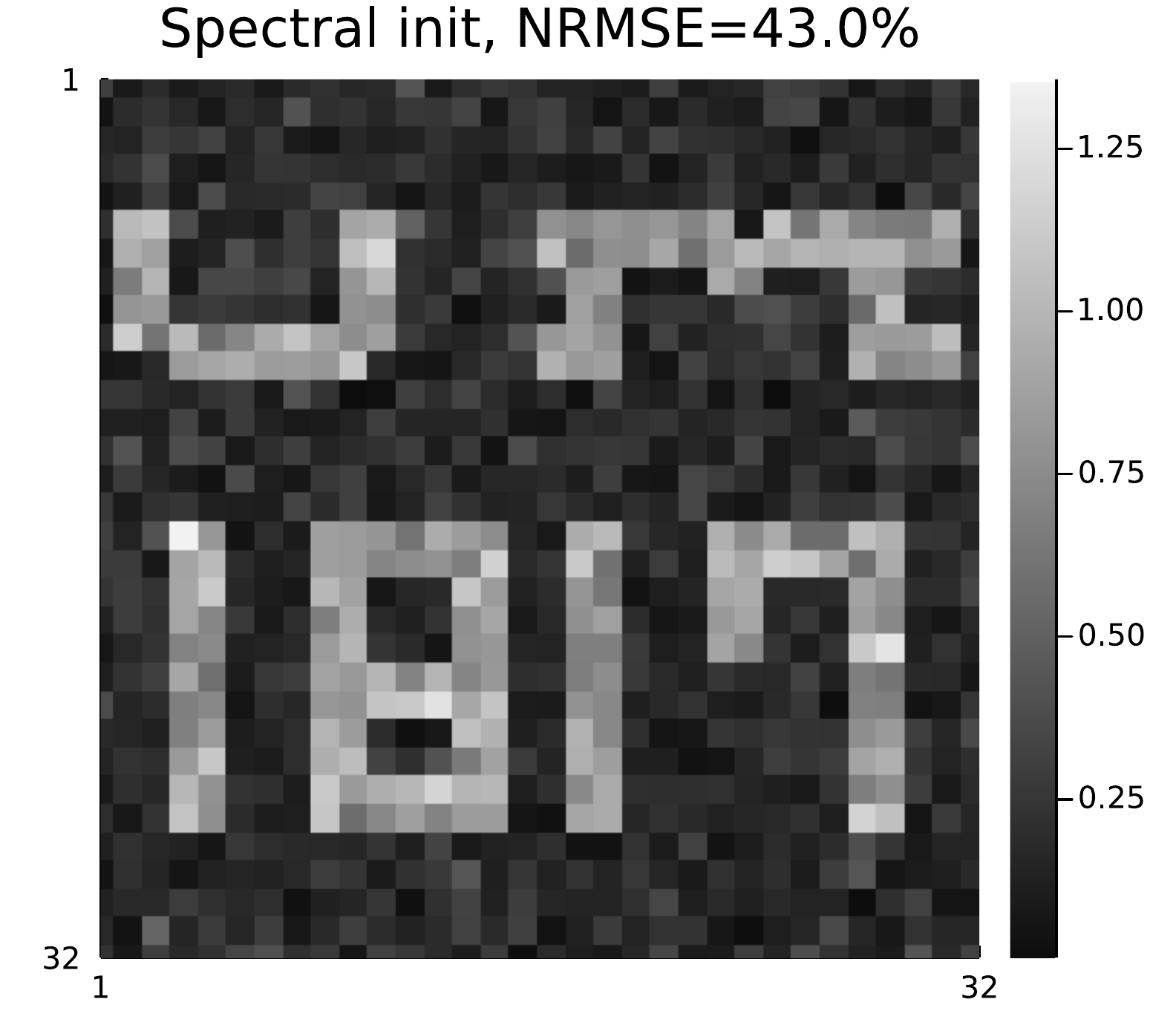}}
    \subfloat[]{\includegraphics[width=0.25\linewidth]{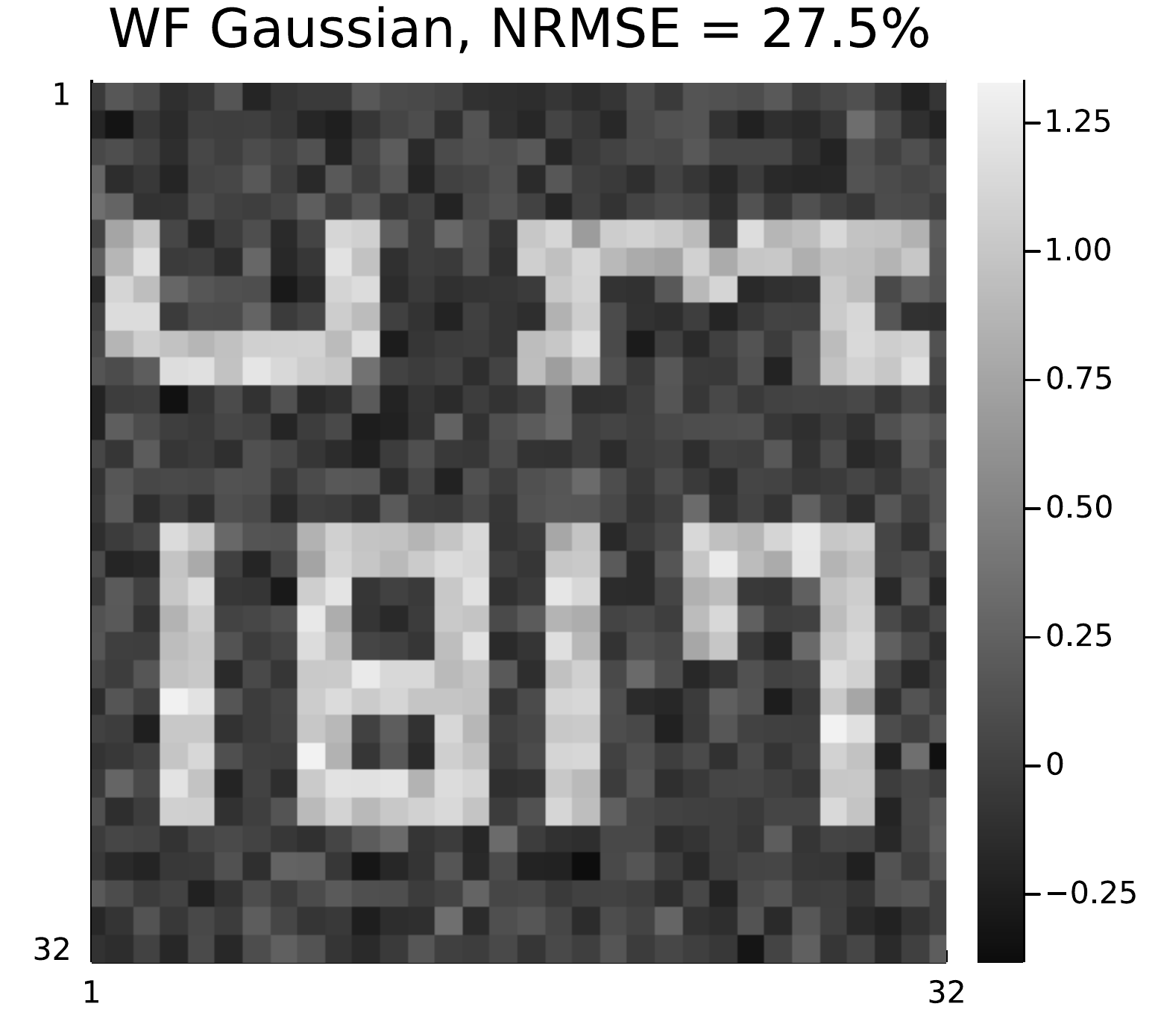}}
    \subfloat[]{\includegraphics[width=0.25\linewidth]{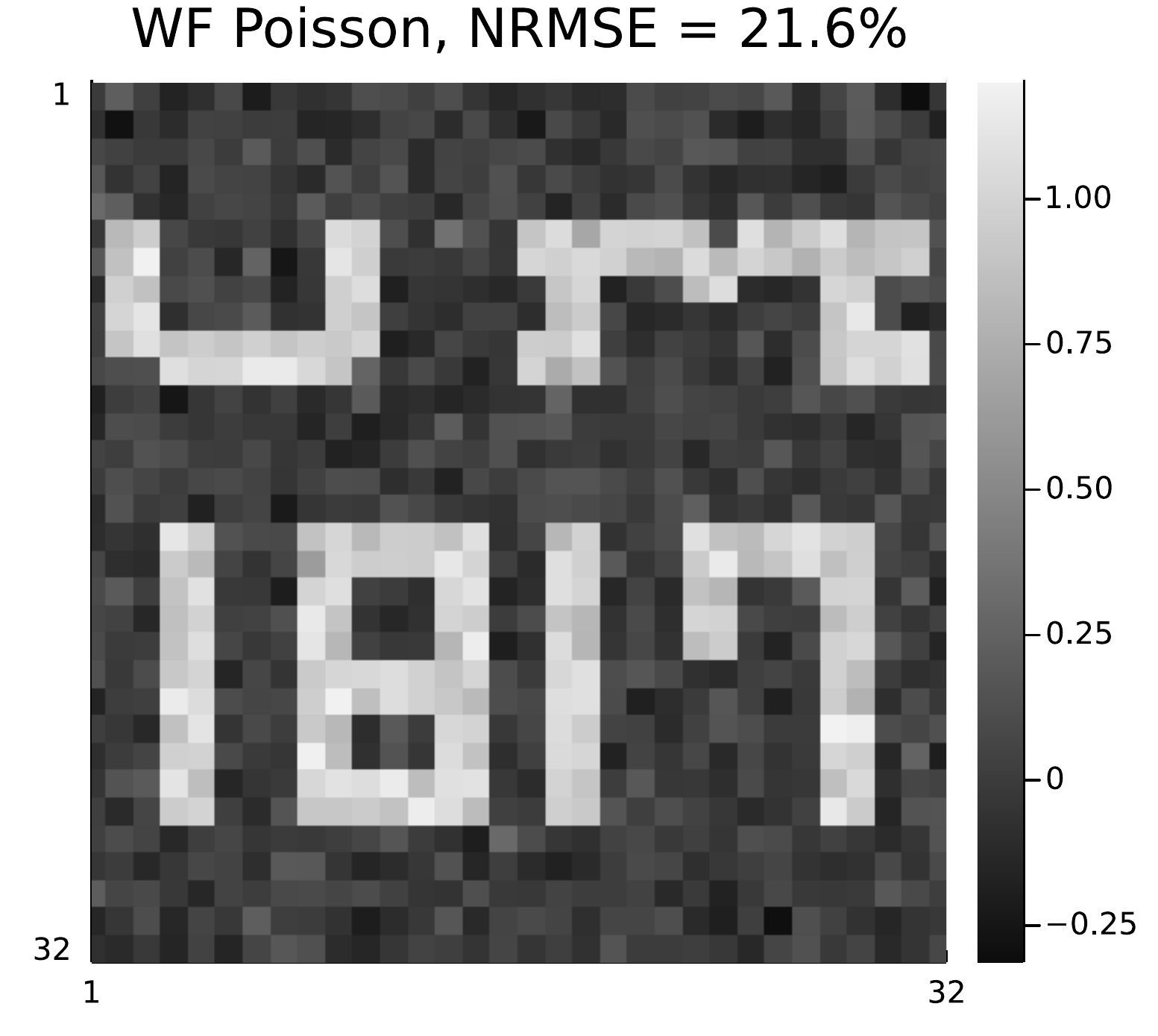}}
    \subfloat[]{\includegraphics[width=0.25\linewidth]{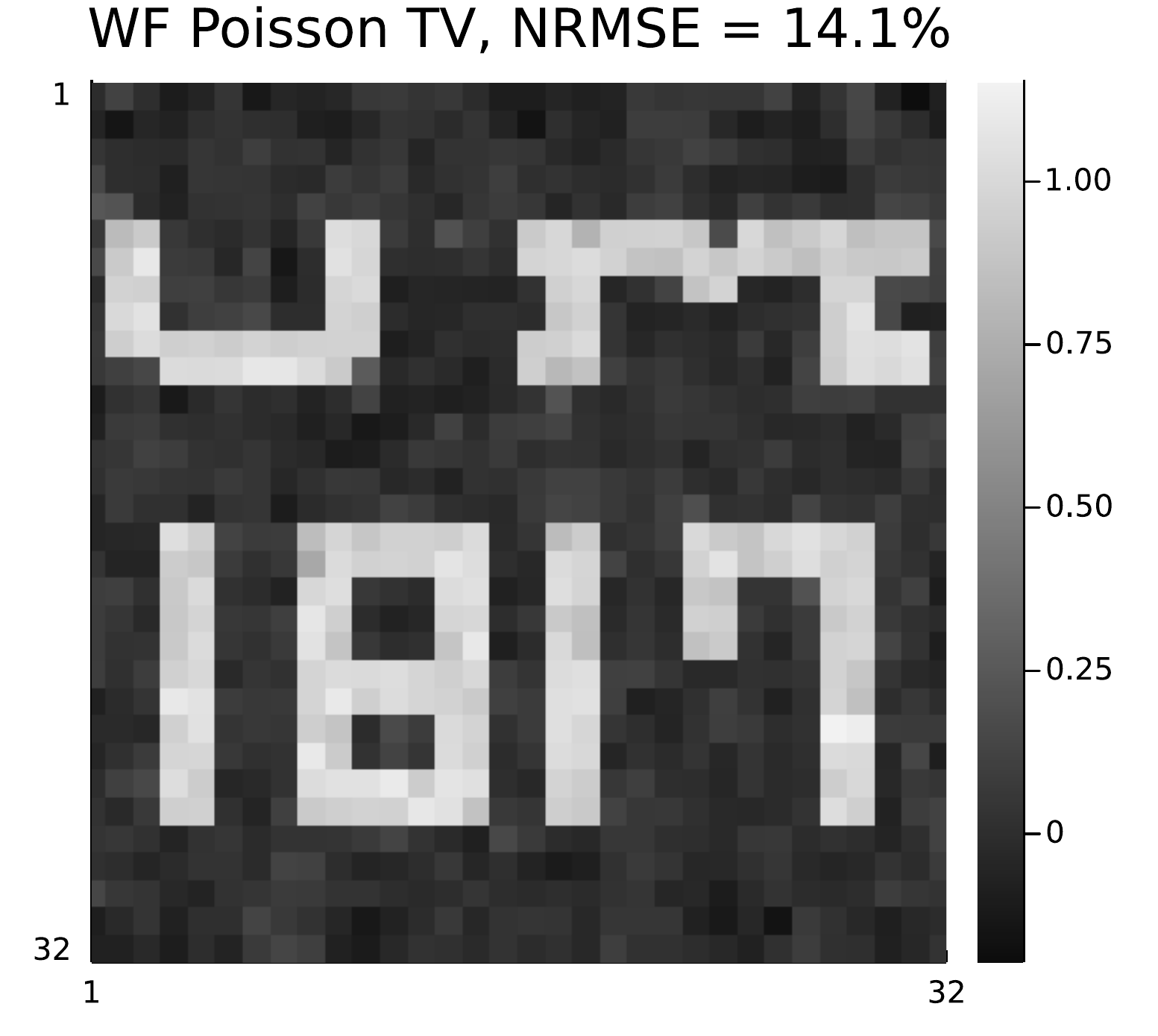}}
    \\
    \subfloat[]{\includegraphics[width=0.25\linewidth]{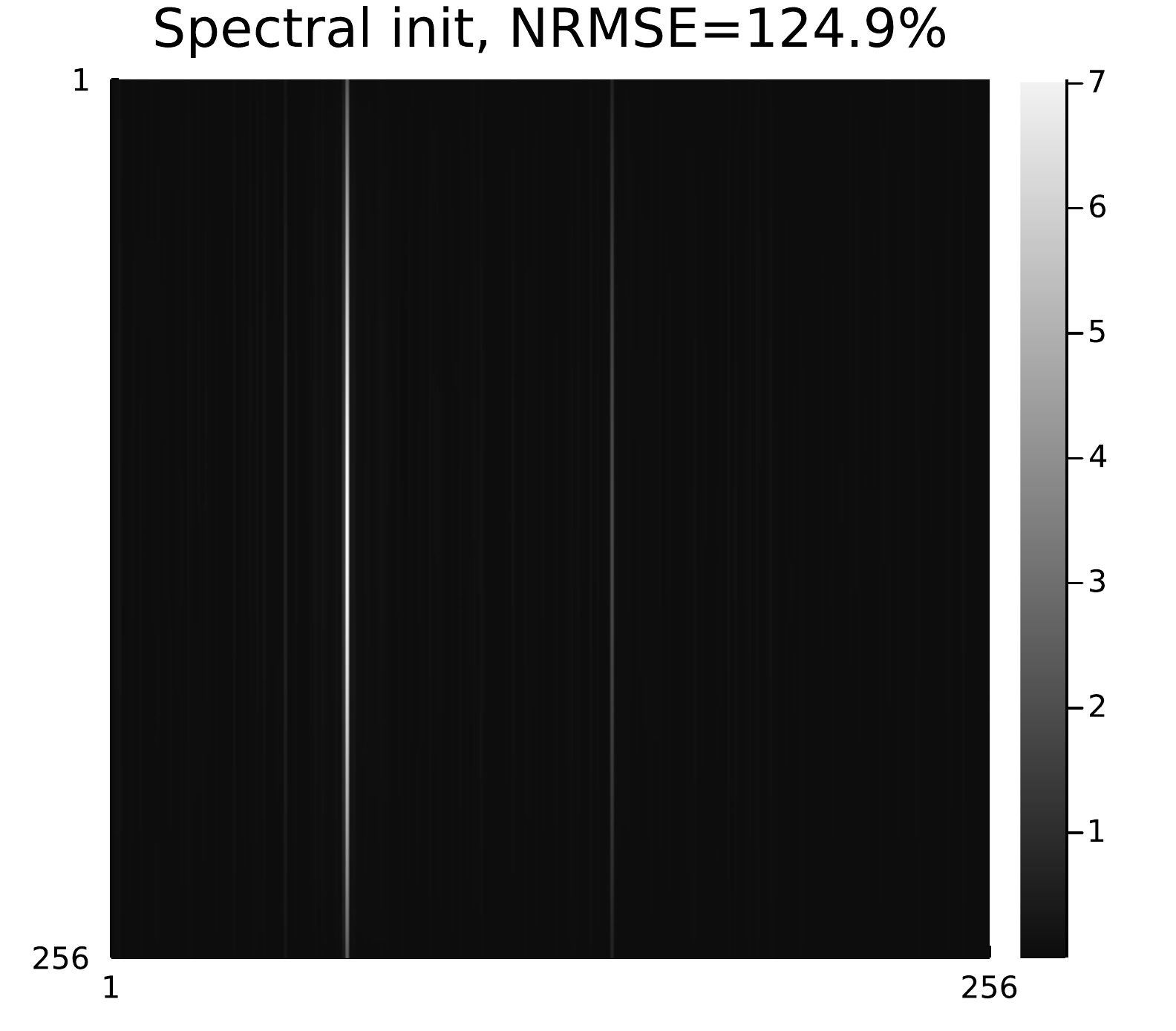}}
    \subfloat[]{\includegraphics[width=0.25\linewidth]{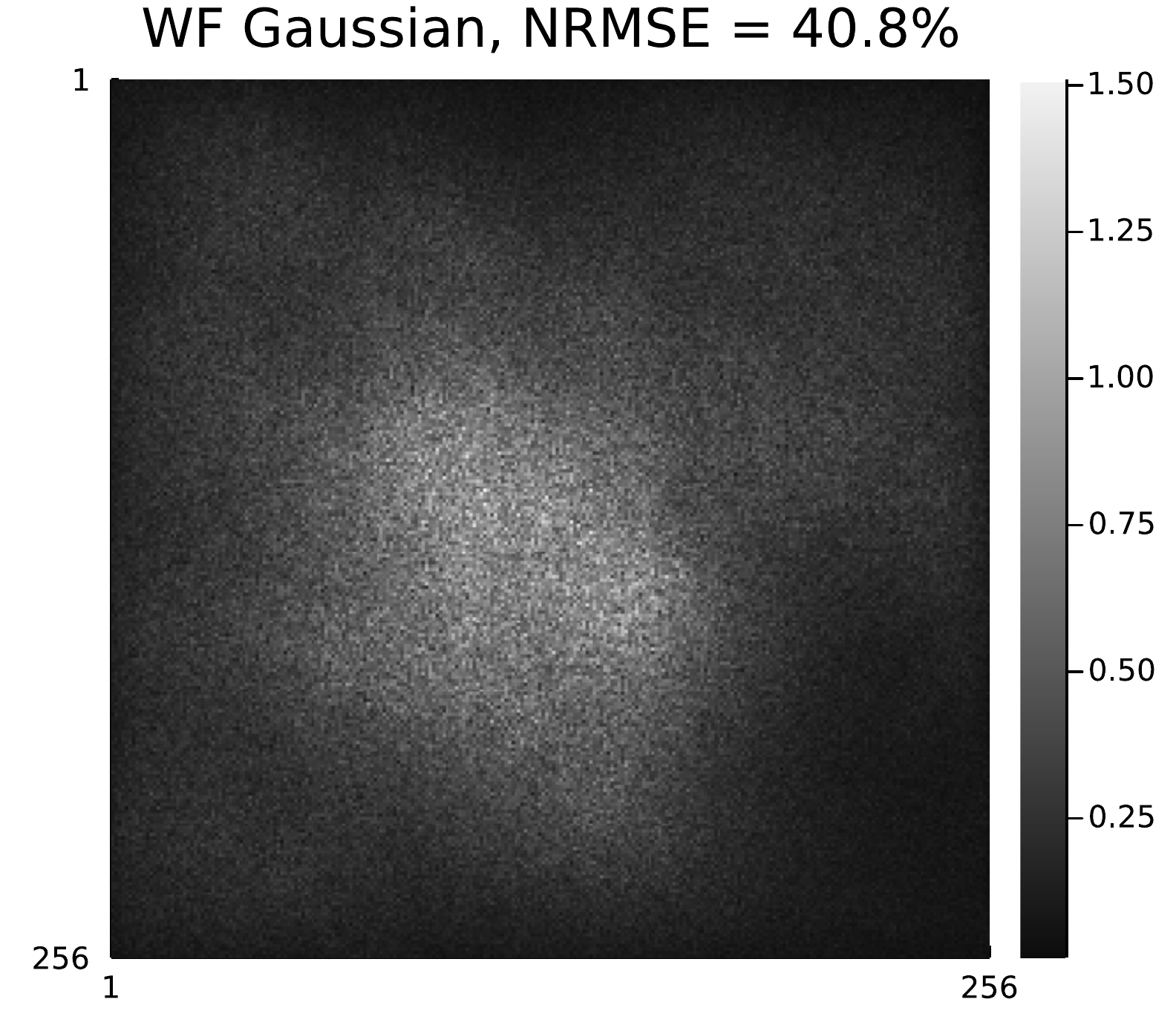}}
    \subfloat[]{\includegraphics[width=0.25\linewidth]{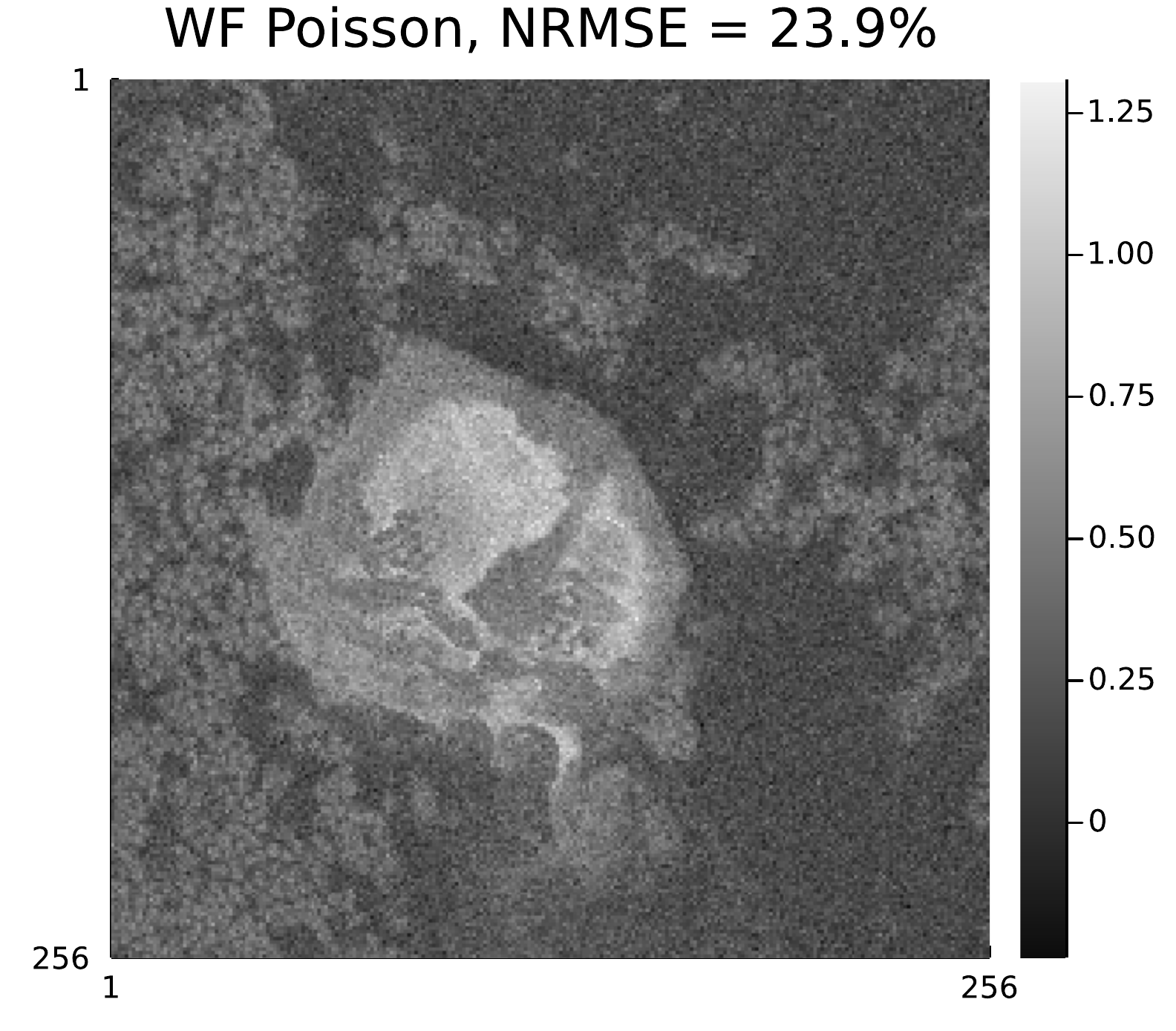}}
    \subfloat[]{\includegraphics[width=0.25\linewidth]{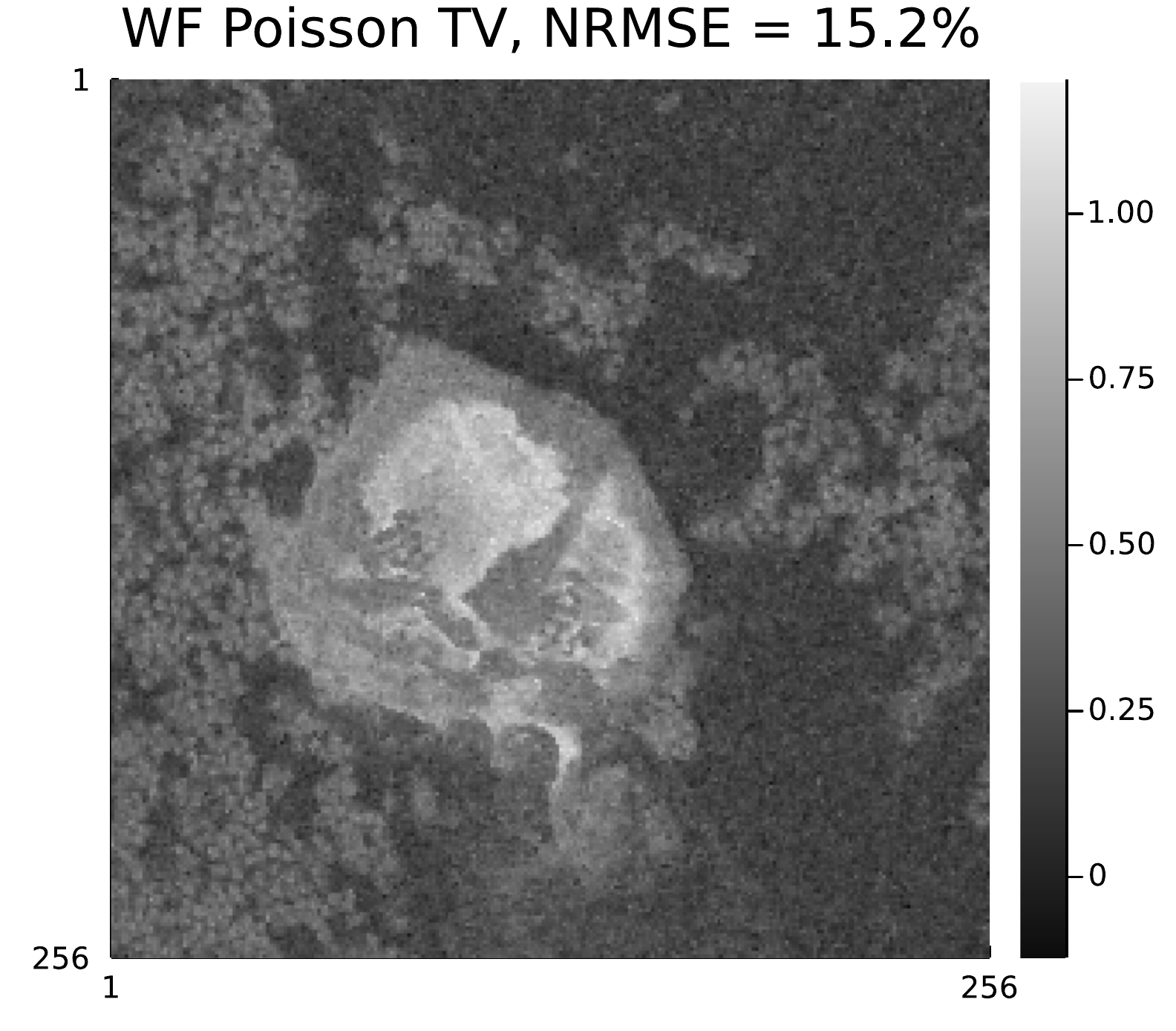}}
    \\
    \subfloat[]{\includegraphics[width=0.25\linewidth]{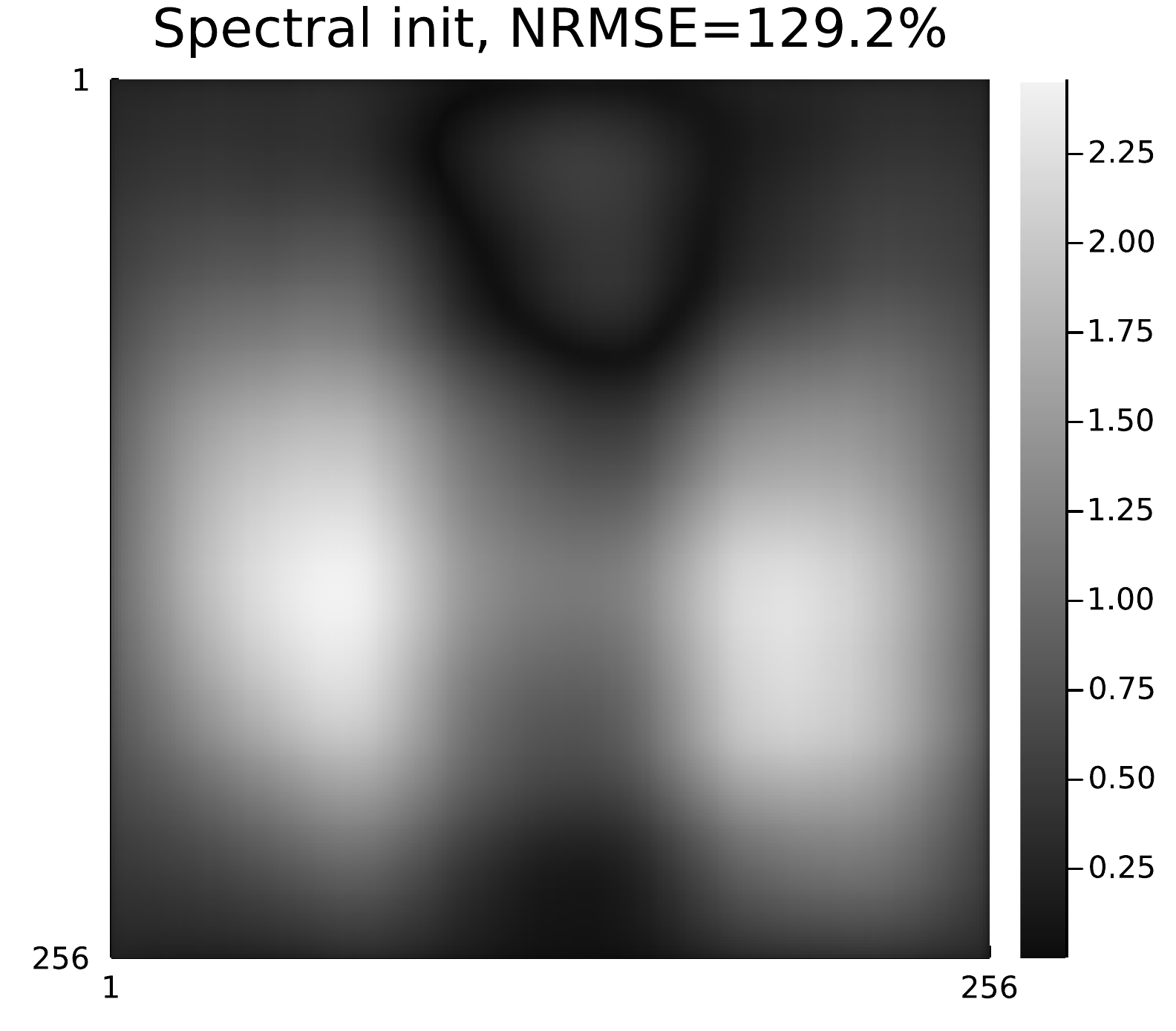}}
    \subfloat[]{\includegraphics[width=0.25\linewidth]{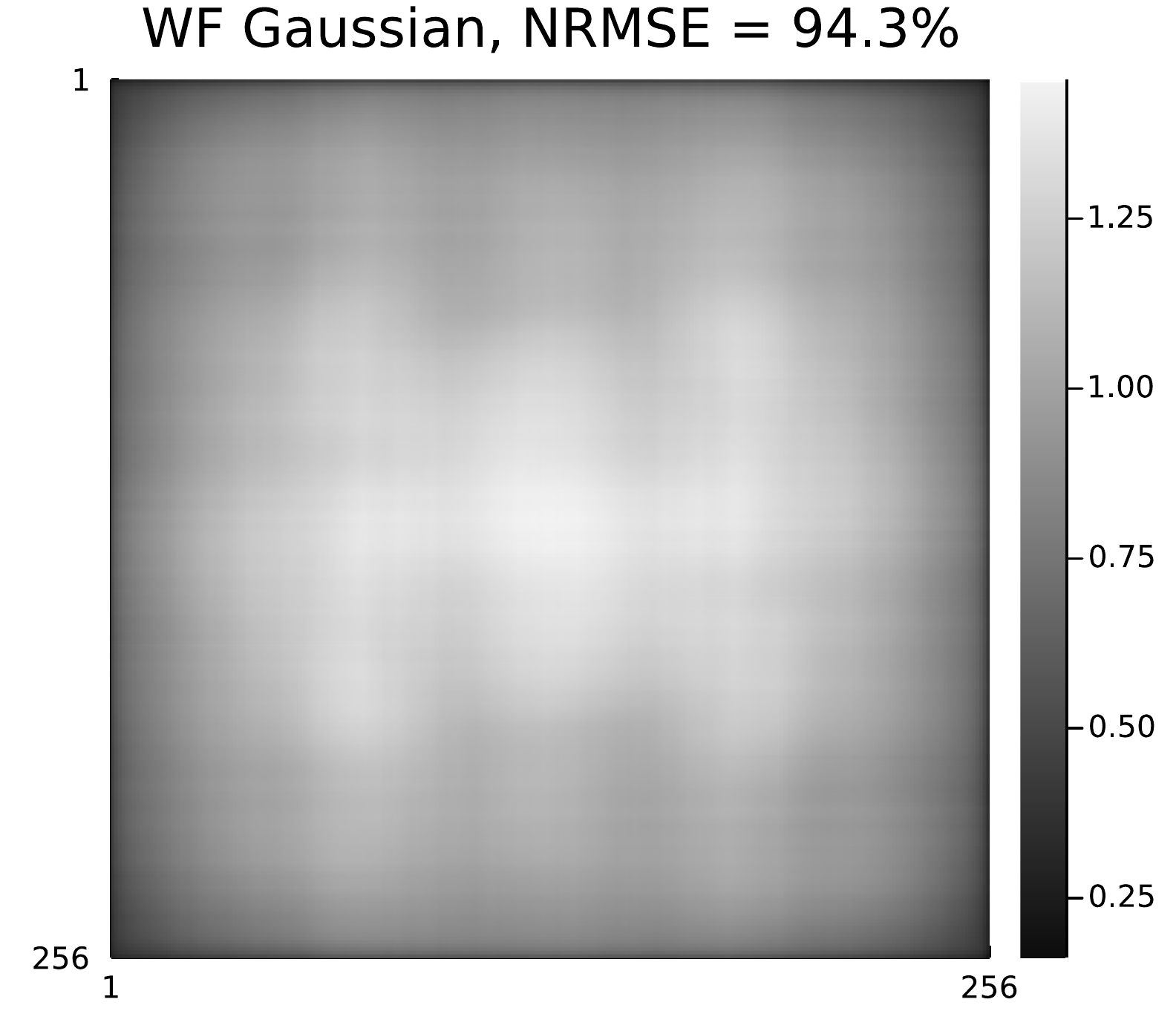}}
    \subfloat[]{\includegraphics[width=0.25\linewidth]{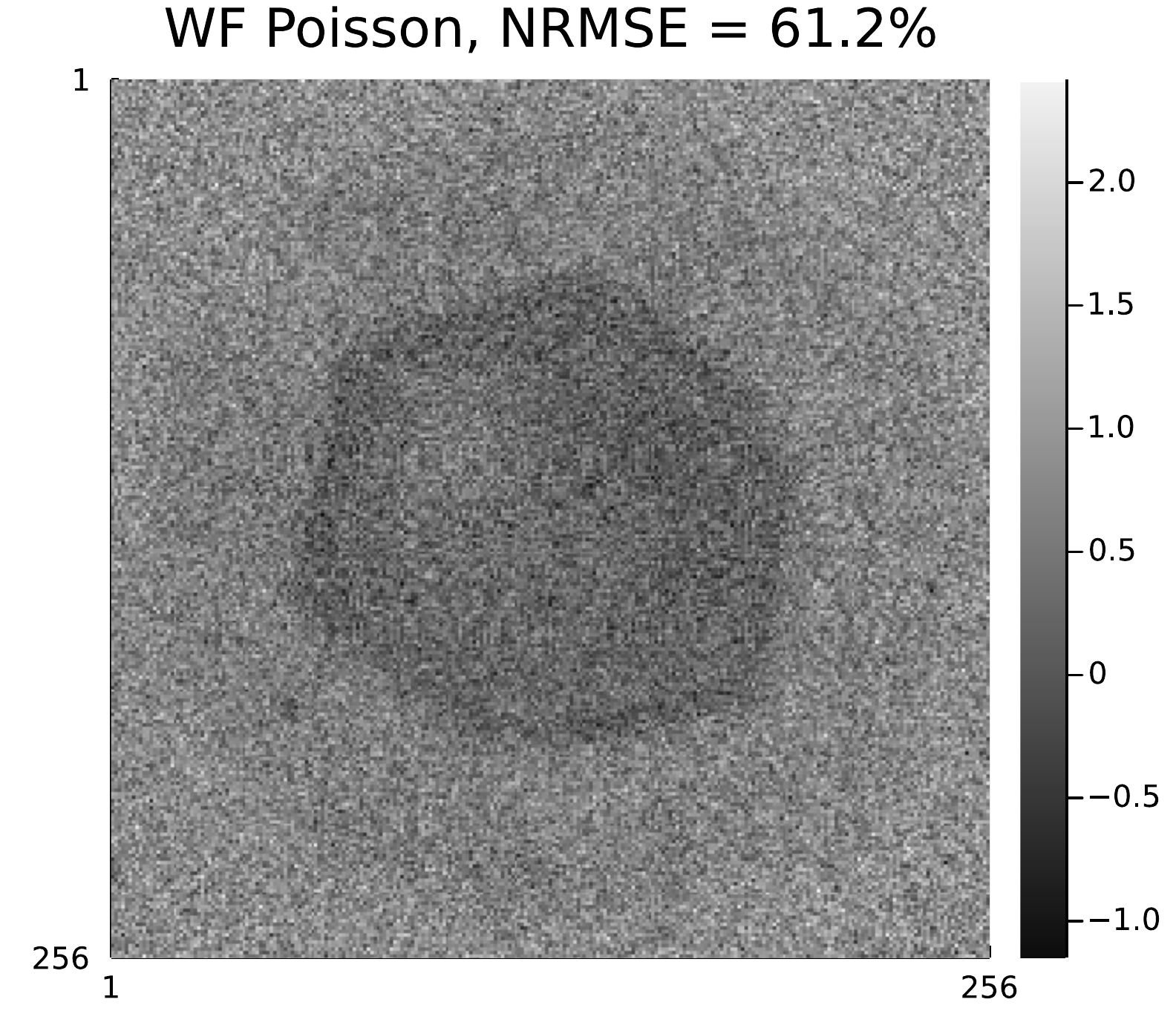}}
    \subfloat[]{\includegraphics[width=0.25\linewidth]{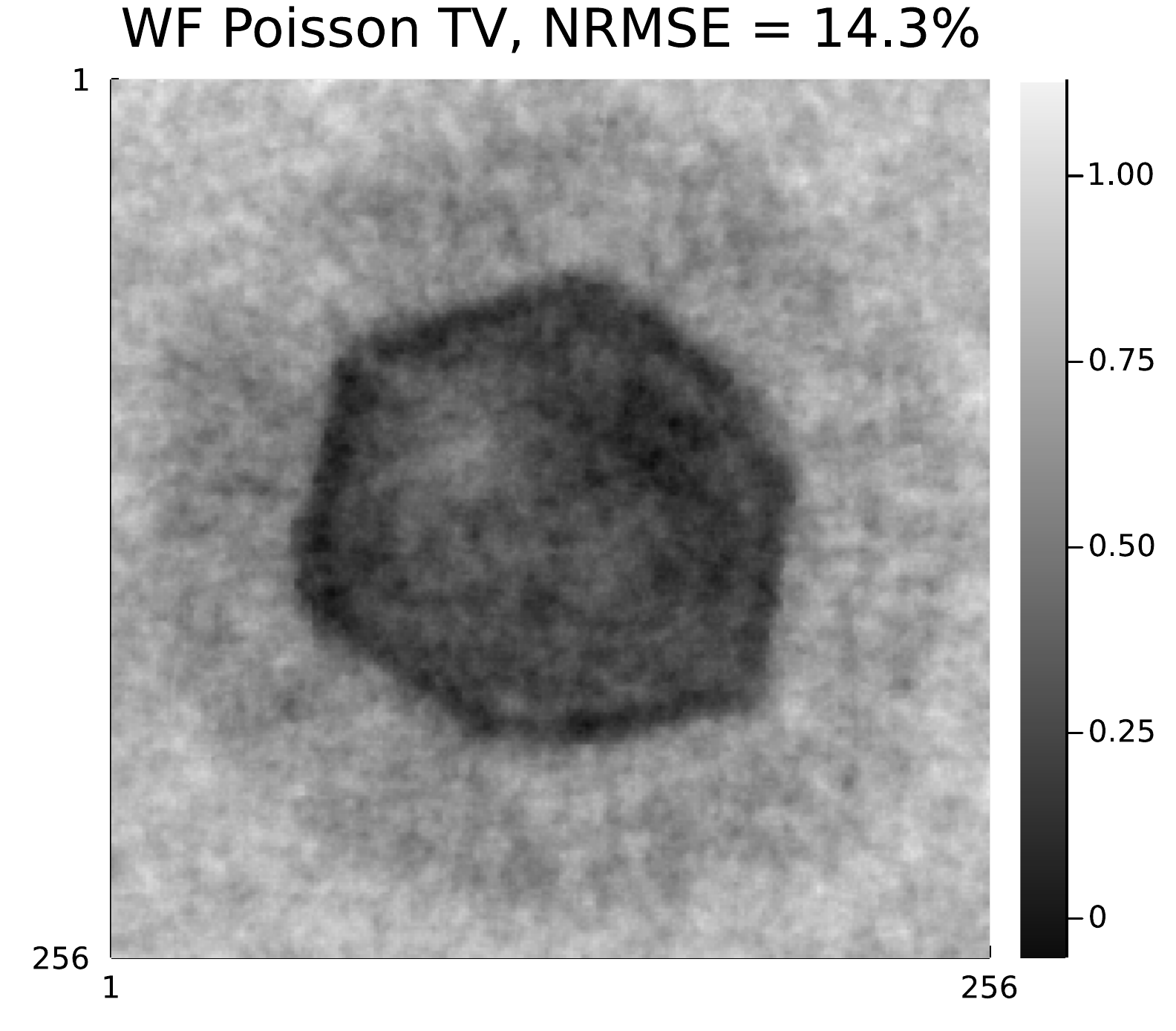}}
    \\
    \subfloat[]{\includegraphics[width=0.25\linewidth]{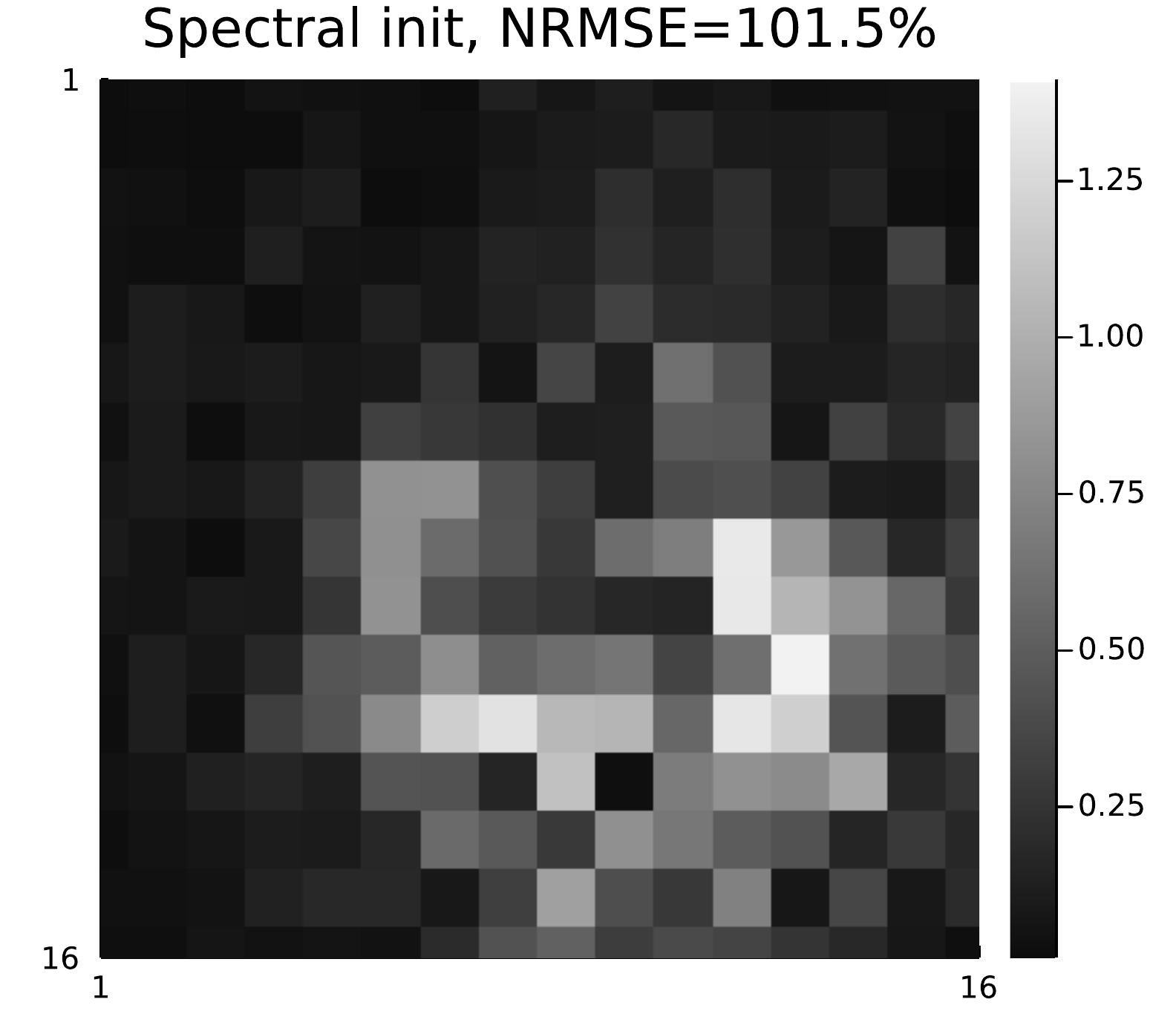}}
    \subfloat[]{\includegraphics[width=0.25\linewidth]{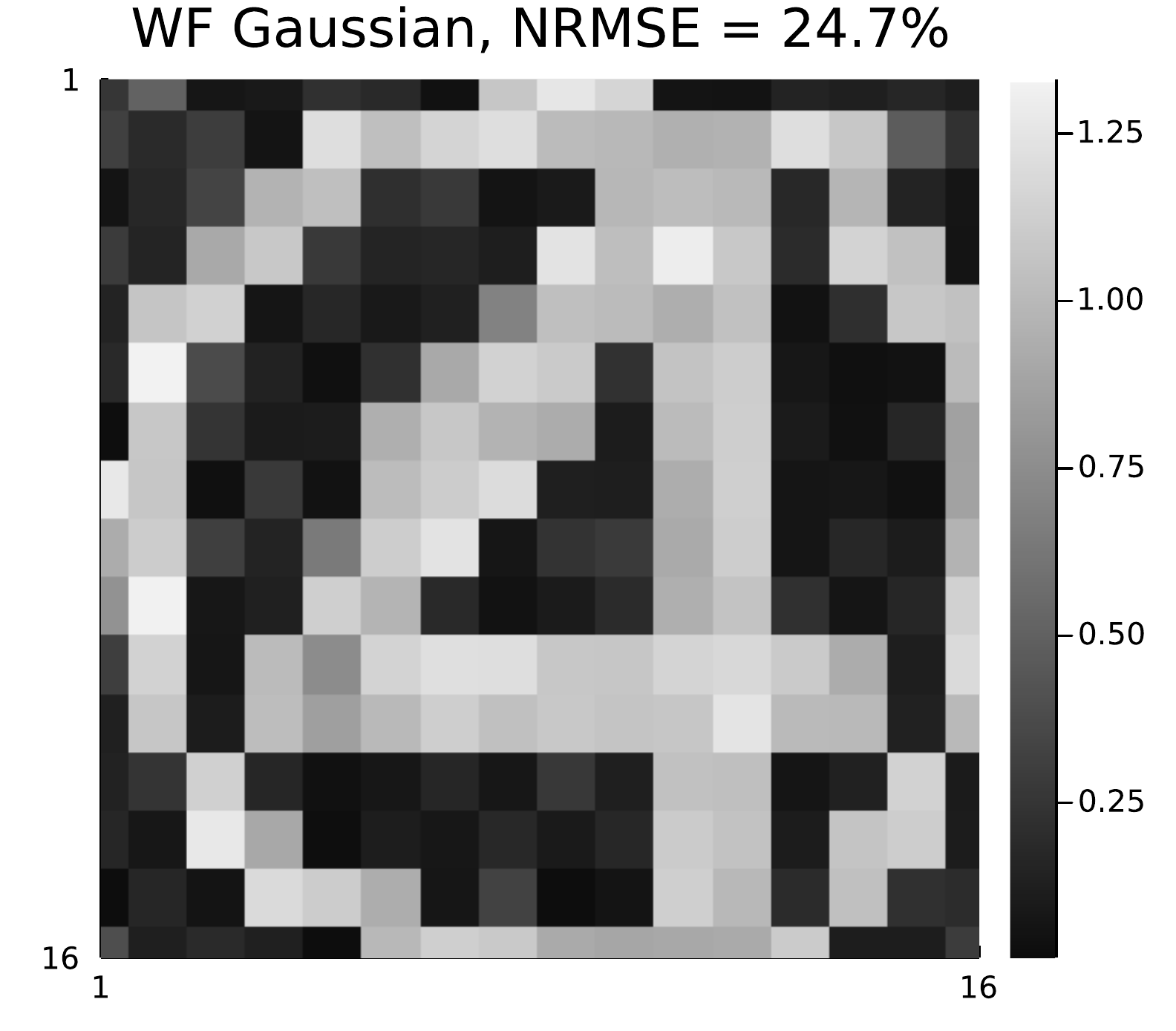}}
    \subfloat[]{\includegraphics[width=0.25\linewidth]{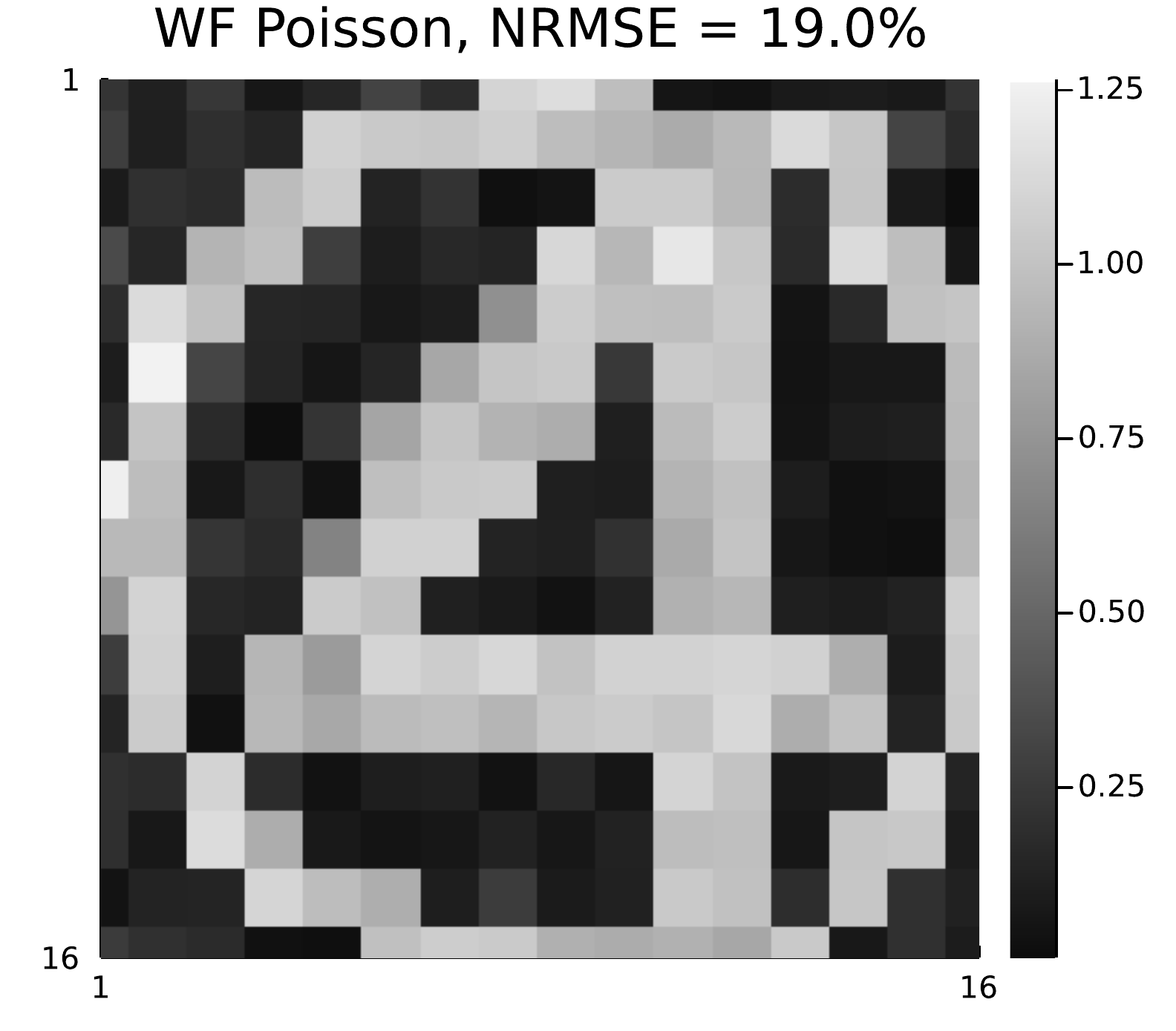}}
    \subfloat[]{\includegraphics[width=0.25\linewidth]{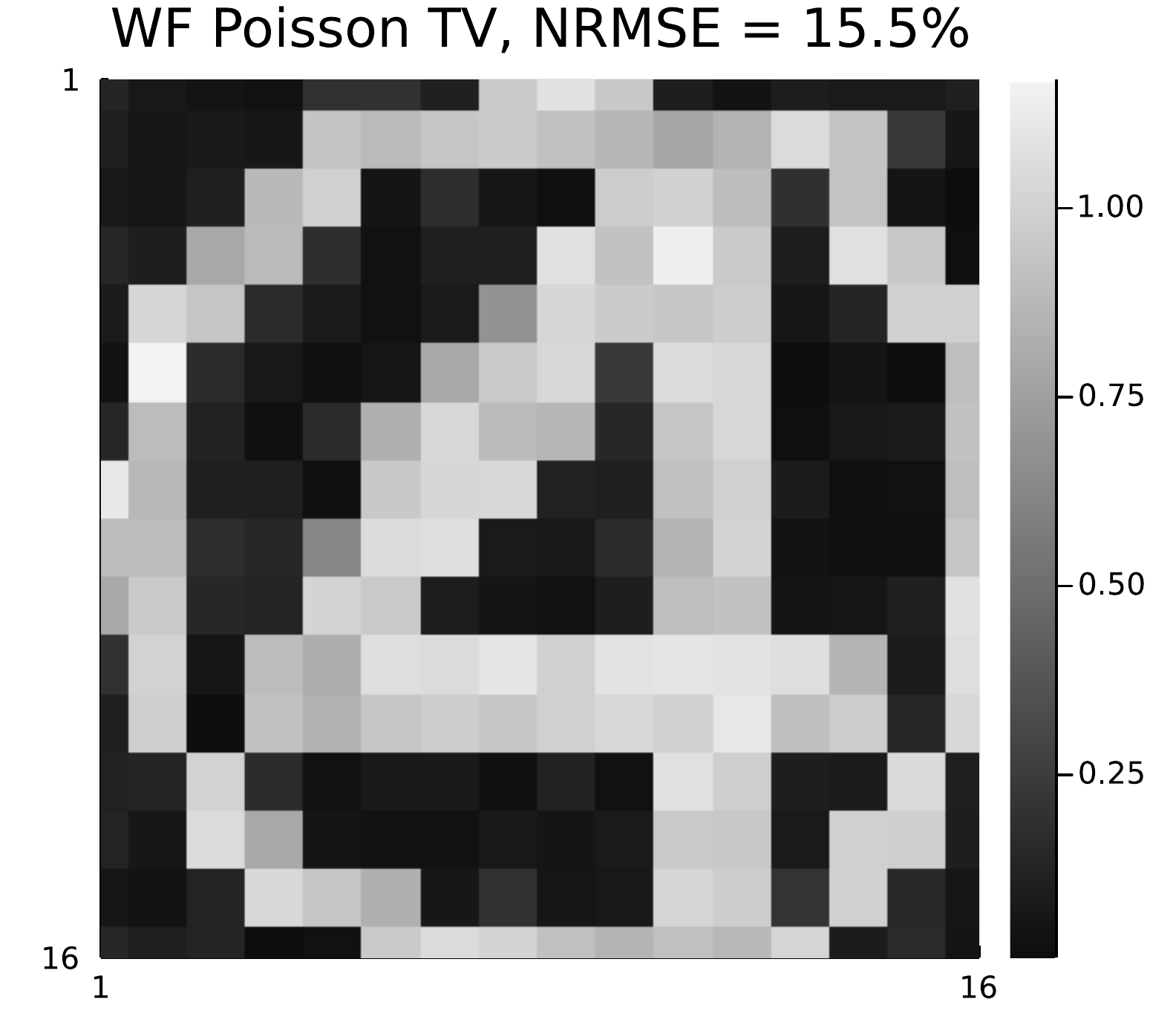}}
    \caption{Reconstruction quality comparison
    between four methods (left to right):
    the optimal Poisson 
    spectral initialization \cite{luo:19:osi},
    the WF Gaussian method,
    the WF Poisson method,
    and WF Poisson with TV regularization. 
    System matrices:
    (a)-(d) random Gaussian; 
    (e)-(h) masked DFT; 
    (i)-(l) canonical DFT with reference image;
    (m)-(p) ETM.
    Magnitude of complex images shown.
    All WF algorithms used the proposed Fisher information for step size.
    }
    \label{fig:poi-vs-gau}
\end{figure*}
\fref{fig:wf,fisher}
shows that,
for all system matrix choices,
WF with Fisher information converged faster
(in terms of decreasing the cost function) than all other methods;
the LBFGS algorithm had comparable convergence speed
as WF with backtracking line search.
We found that WF with the empirical step size 
did not converge using hyper-parameters in \cite{candes:13:prv}
so we excluded those results in \fref{fig:wf,fisher}. 
The backtracking approach, although
slower than Fisher approach per wall-time, 
is faster per-iteration.
However, the step size found by 
backtracking line search
could be sensitive to hyper-parameter choices.
For the WF algorithm 
with optimal step size 
(derived based on Gaussian noise model \cite{jiang:16:wfm}),
we conjectured that
it reached a non-stationary point that
has larger cost function value 
than those of other methods,
as expected.

In terms of PSNR, 
we found that in random Gaussian, 
masked DFT and empirical transmission cases,
WF with Fisher information 
increased the PSNR faster than all other methods;
WF with optimal Gaussian step size
led to lower PSNR, perhaps again due to 
reaching a sub-optimal minimizer.
However, for the canonical Fourier case,
we found that all methods started decreasing PSNR
after several iterations.
%this can be due to
The algorithms may be more sensitive to noise
in the canonical Fourier matrix setting,
especially in the very low-count regime
considered here.
%Also we conjectured that
Apparently
WF with optimal Gaussian step size 
overfits the noise more slowly
due % which could be attributed
to its sub-optimal step size
under Poisson noise.

\subsection{Comparison of Poisson and Gaussian algorithms}
\label{subsec:poi-vs-gau}

This section 
compares the reconstruction quality,
\ie, the NRMSE to the true signal,
between WF derived from
the Gaussian ML cost function \eqref{g,cost},
and WF derived from
the Poisson ML cost function \eqref{e,cost}
as well as regularized WF
under different system matrix settings.
We used corner-rounded TV regularizer
with $\beta = 32$ and $\alpha = 0.1$
in the regularized WF algorithm.

\fref{fig:poi-vs-gau} shows that
algorithms derived from the Poisson model
yielded consistently better reconstruction quality
(lower NRMSE)
than algorithms derived from the Gaussian model,
as expected.
Furthermore, by incorporating regularizer that 
exploits the assumed property of the true signal,
the NRMSE was further decreased.
Spectral initialization worked well 
in random Gaussian matrix setting, but 
not for other system matrices, 
as expected from its theory.
The WF Gaussian approach
failed to reconstruct in masked and canonical DFT system matrix setting.
Since incorporating appropriate regularizers 
helps algorithms 
yield higher quality reconstructions,
a question is naturally raised
about
which regularized algorithm
converges the fastest.
The next subsection
presents such comparisons.

\subsection{Convergence speed of 
regularized Poisson algorithms}

As discussed in Section~\ref{sec:methods},
many algorithms can be modified to accommodate regularizers.
%Here in this section,
We compared the convergence speeds
of
regularized Poisson algorithms 
(WF Fisher, WF backtracking, LBFGS, MM and ADMM \cite{li:21:afp}),
with a smooth regularizer 
(corner-rounded TV),
%and then compared their convergence speed
under different system matrix settings.
Based on \fref{fig:wf,fisher},
we did not run simulations of 
regularized WF with empirical step size
and with Gaussian optimal step size,
due to their non-converging trend
and sub-optimal solution, respectively.
For all other algorithms,
we set the regularization parameters
to be 
$\beta = 32$ and $\alpha = 0.1$ 
(defined in \eqref{e,Phi} and \eqref{alg,huber}).

\begin{figure}[ht!]
    \centering
    \subfloat[Random Gaussian system matrix.]
    {\includegraphics[width=0.8\linewidth]{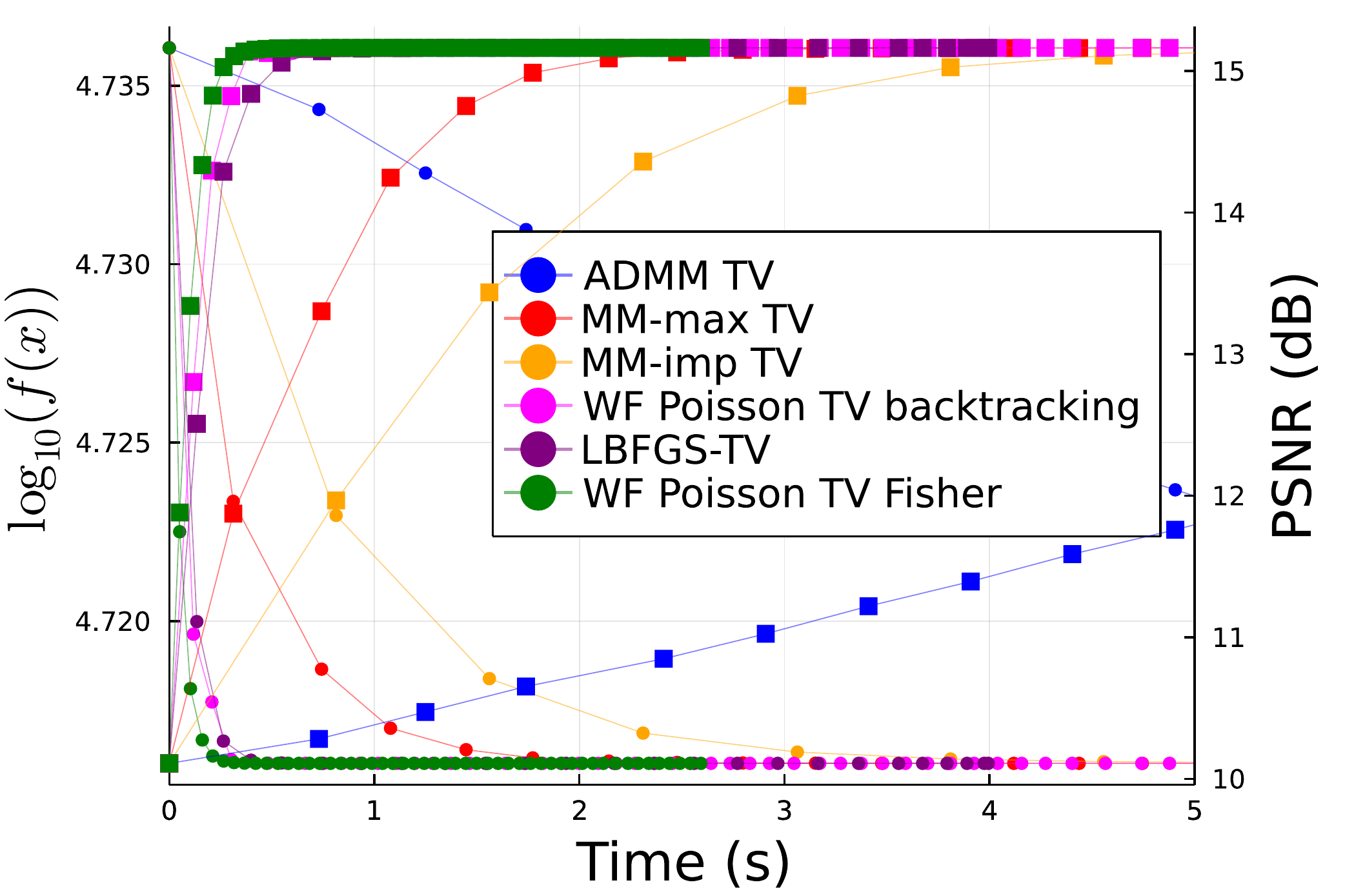}}
    \\
    \subfloat[Masked DFT system matrix.]
    {\includegraphics[width=0.8\linewidth]{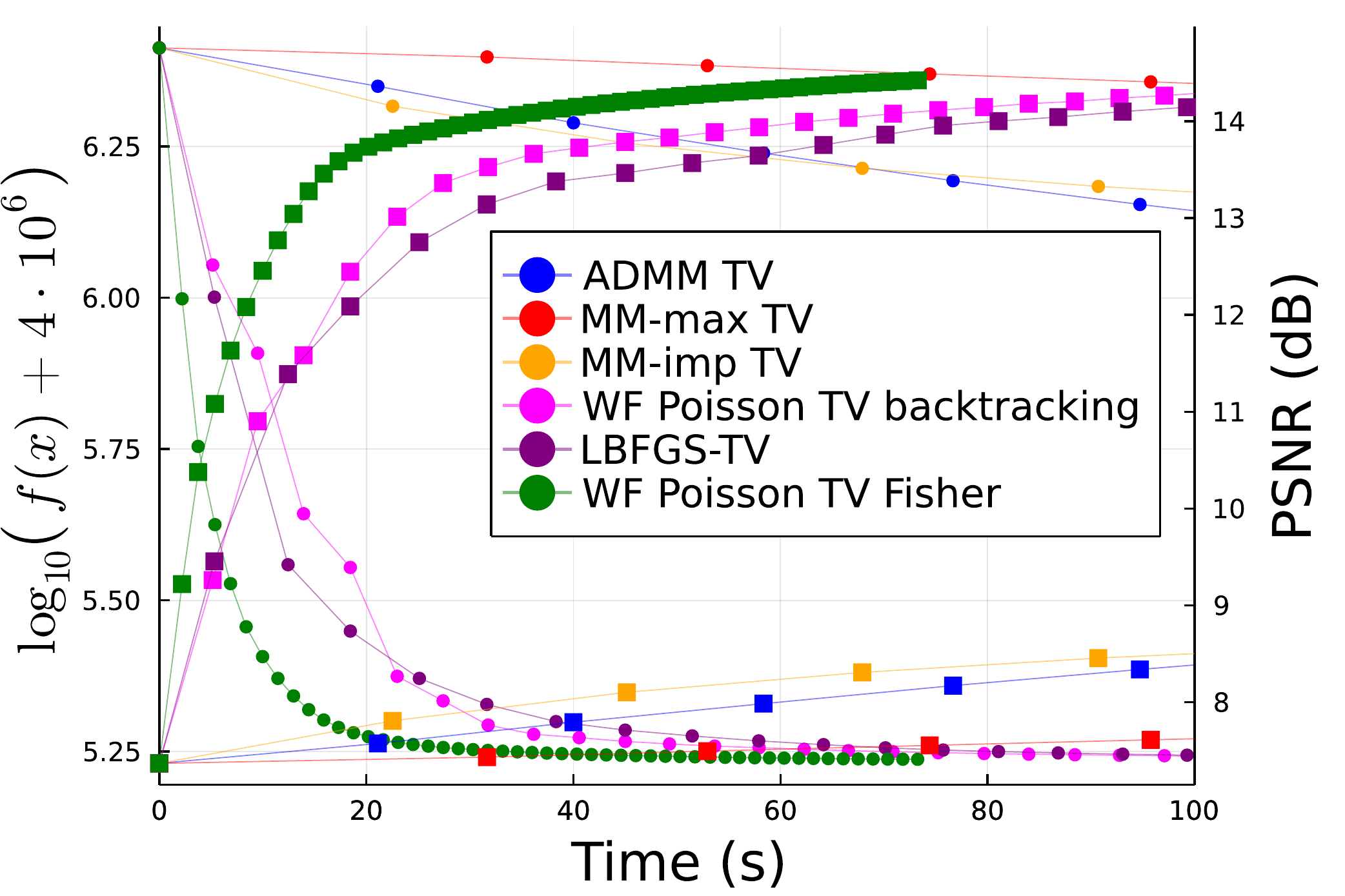}}
    \\
    \subfloat[Canonical DFT system matrix.]
    {\includegraphics[width=0.8\linewidth]{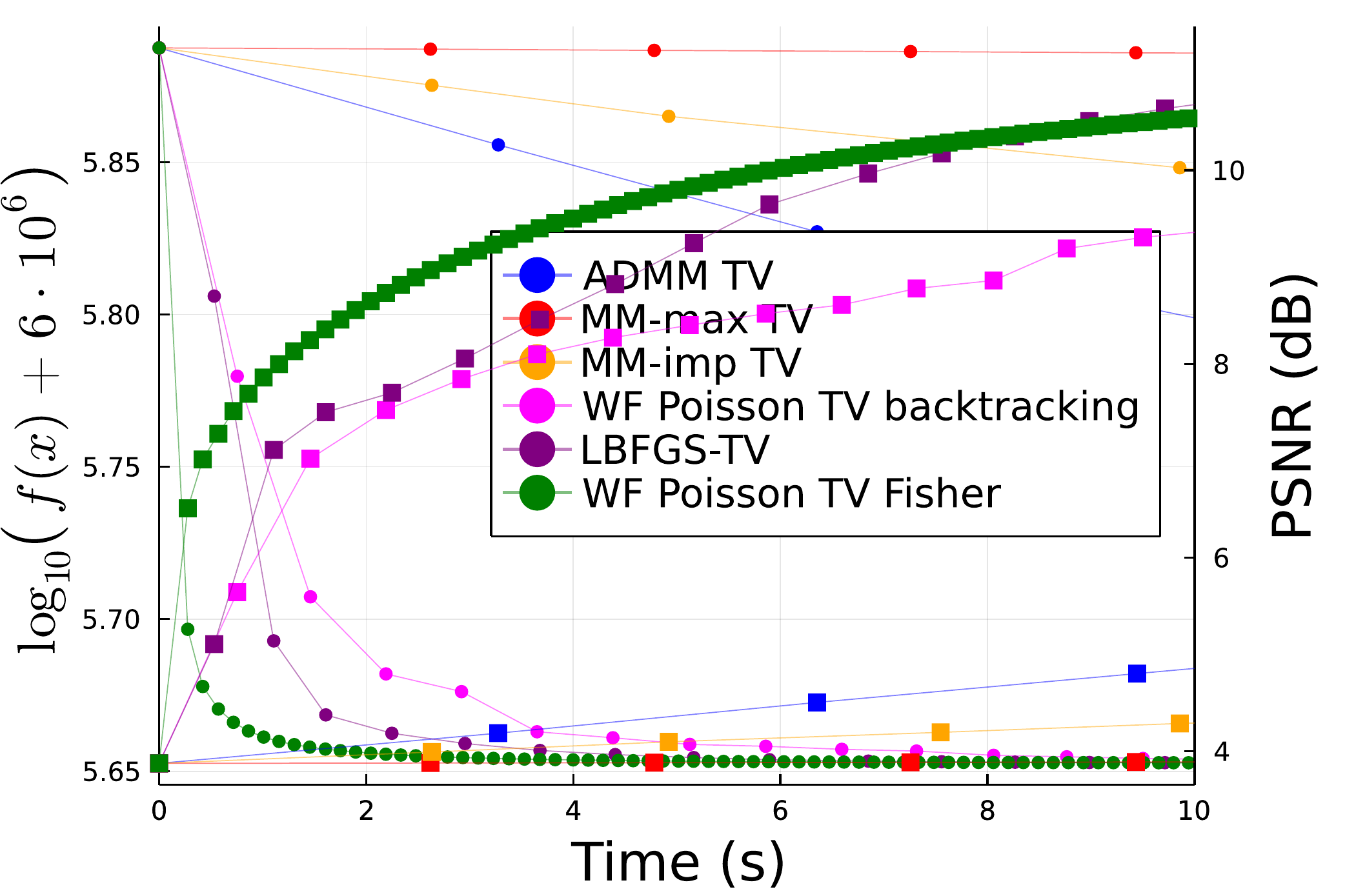}}
    \\
    \subfloat[Empirical transmission matrix.]
    {\includegraphics[width=0.8\linewidth]{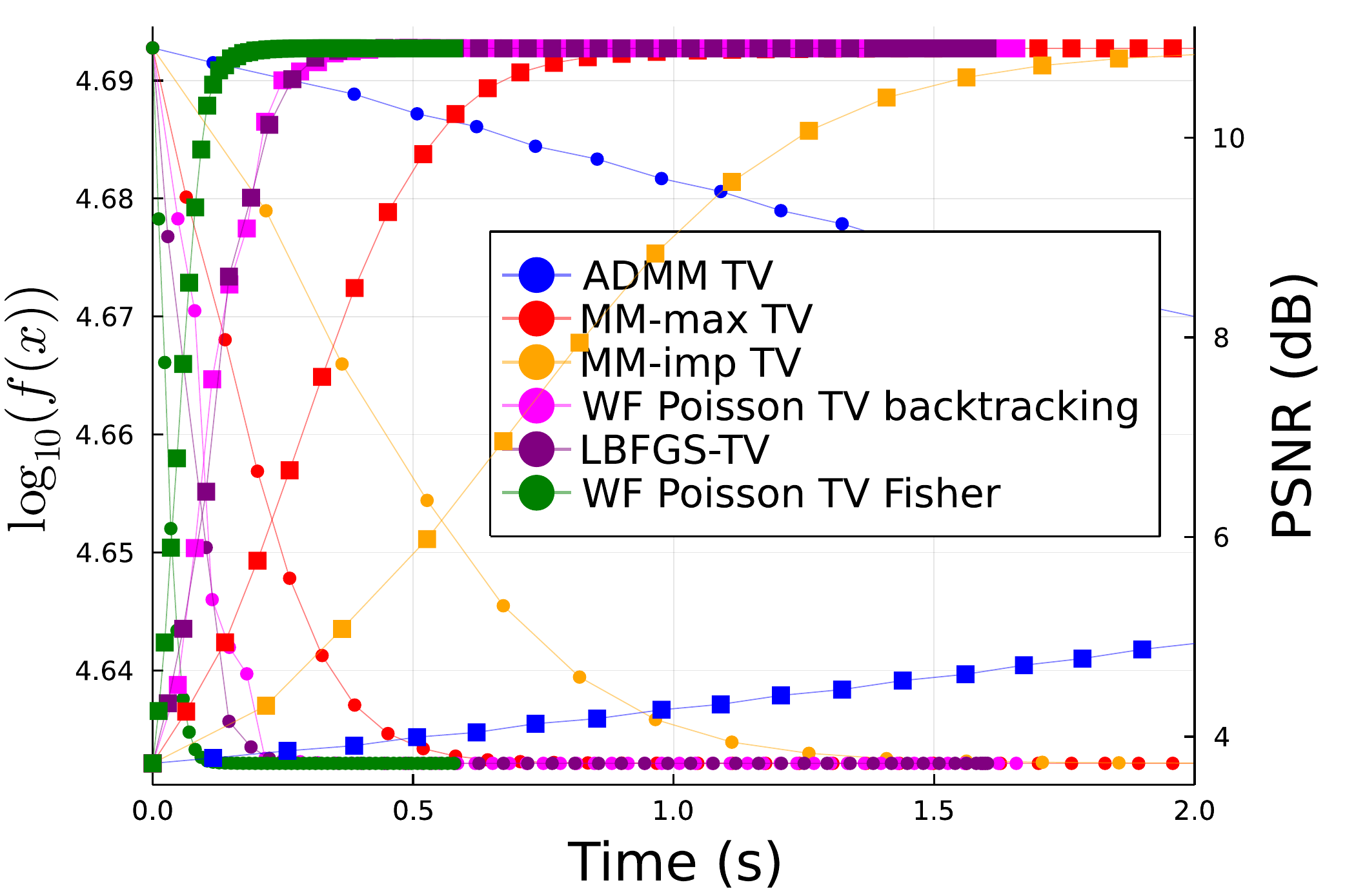}}
    \caption{Comparison of convergence speed
    of variant algorithms with corner-rounded TV regularizer. The circle marker corresponds to cost function and the square marker corresponds to PSNR.
    }
    \label{fig:reg,tv,speed}
\end{figure}

\fref{fig:reg,tv,speed} 
shows that the regularized WF with 
our proposed Fisher information
for step size converged the fastest 
compared to other methods
under all different system matrices.
The LBFGS again had a comparable convergence speed as
WF using backtracking line search.
The MM algorithm with improved curvature, 
was slower in wall-time
due to extra computation per iteration,
but was faster per iteration 
due to its sharper curvature.
In masked and canonical Fourier case, however,
MM with improved curvature was faster than 
the maximum curvature in wall-time comparison,
which can be attributed to large magnitude 
low frequency components in the coefficients 
of the Fourier transform.

\section{Discussion}
\label{sec:discussion}

\begin{figure}[hbt!]
    \centering
    \includegraphics[width=0.8\linewidth]{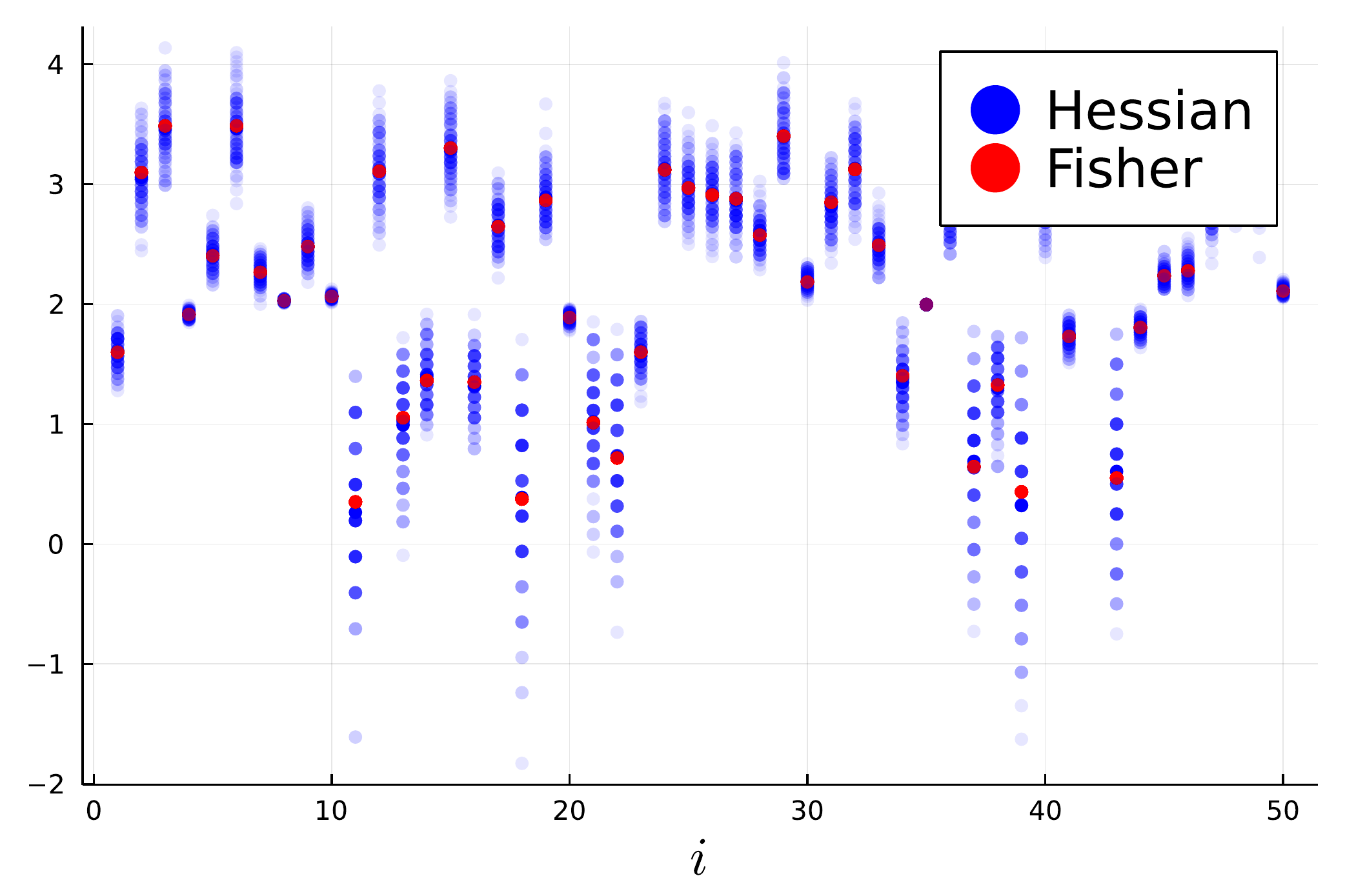}
    \caption{Visualization of the marginal Hessian \eqref{grad_psi} 
    and the marginal observed Fisher information \eqref{fisher_psi}.
    The horizontal axis denotes the $i$th element in the marginal Hessian/Fisher.
    Data were simulated with a random Gaussian matrix and 100 independent realizations.}
    \label{fig:fisher-vs-hessian}
\end{figure}

Current methods for phase retrieval
mostly focus on ML estimation for Gaussian noise;
fewer algorithms were derived for Poisson noise
\cite{chen:17:srq, bian:16:fpr, chang:18:tvb}.
Here we proposed a novel WF algorithm
and an MM algorithm
and then did an empirical study 
on the convergence speed
as well as reconstruction quality
of several Poisson phase retrieval algorithms.
In our proposed WF algorithm,
we first replaced the gradient term 
in Gaussian WF \eqref{grad_g}
with its Poisson counterpart \eqref{grad_psi}.
Then we did a quadratic approximation
of the cost function
and
applied one iteration
of Newton's method
to define an ``optimal'' step size.
We then proposed to use
the observed Fisher information
to approximate the Hessian
when computing the step size,
which is a common method 
in computational statistics.
Moreover, 
the Fisher information matrix
is guaranteed to be 
positive semi-definite
and is more computationally efficient
compared to the Hessian.
To further illustrate 
our proposed method
of using Fisher information
to approximate the Hessian,
\fref{fig:fisher-vs-hessian}
visualizes these two matrices (in marginal forms).
%The Fisher information values
%provide good approximations
%to the typical Hessian values
%while retaining nonnegativity.

As shown in \fref{fig:fisher-vs-hessian}, 
the Hessian is noisy
and can have some negative elements.
Such undesirable features 
can lead to
unstable step size calculations.
%a ill-posed 
%non-convex optimization problem;
In contrast, the elements in Fisher information matrix
are non-negative and less noisy.
We ran some experiments and found that
when the background counts \bi are large,
using the noisy Hessian to calculate the step size
can lead to divergence of the cost function,
due to the negative values in the marginal second derivative.
Setting such negative values 
in the second derivative to zero is a possible solution,
but we found that approach
led to slower convergence
than using the Fisher information.
One potential alternative to our approach is to use
the empirical Fisher information,
but that may be suboptimal
since the empirical Fisher information
does not generally capture second-order information \cite{kunstner:19:lot}.

\comment{
% our step size calculation does not change the cost function!
% so we cannot make any claims about the landscape i think!
% I agree, we are making approximations to the cost function...
so that the geometric landscape 
of the cost function is benign
and potentially allows algorithms 
to find a global minimizer 
more efficiently \cite{sun:16:aga}.
}

To accommodate our WF algorithm with non-smooth regularizers, 
\eg, $\|\T \x\|_1$,
we used a Huber function to
approximate the $\ell_1$ norm
with a quadratic function
around zero,
so that the Wirtinger gradient 
is well-defined everywhere.
A limitation of this paper
is that we did not consider other regularizers
in our experiments,
though
our algorithms
can be generalized to handle
other smooth regularizers 
with minor modifications.
One drawback of TV regularization 
is that it assumes piece-wise uniform 
latent images so it lacks generalizability 
to other kinds of images, 
One way to address this is to 
train deep neural networks \cite{remez:18:caf, zha:22:snl}
with a variety of images,
potentially leading to better generalizability.

\comment{
\blue{When preparing for this paper,
we started with investigating 
algorithms for Poisson phase retrieval
and accelerating some of them;
we did not know which method work the best
(\eg, had the fastest convergence rate)
until we tested them.
Although experiment results showed 
that our proposed WF with Fisher information
converged the fastest than other methods,
we keep the MM algorithm with improved curvature
in the main body
as part of novelty and contribution of this paper.
}
}
\section{Conclusion}
\label{sec:conclusion}

This paper proposed and compared algorithms
based on ML estimation
and regularized ML estimation
for phase retrieval from Poisson measurements,
in very low-photon count regimes,
\eg, 0.25 photon per pixel.
We proposed a novel method
that used the Fisher information
to compute the step size in the WF algorithm;
this approach eliminates all parameter tuning 
except the number of iterations.
We also proposed a novel MM algorithm
with improved curvature compared to the one
derived from the upper bound of 
the second derivative 
of the cost function.

Simulation results experimented on 
random Gaussian matrix, 
masked DFT matrix,
canonical DFT matrix
and an empirical transmission matrix
showed that:
1) For unregularized algorithms,
the WF algorithm using 
our proposed Fisher information for step size 
converged faster than 
using empirical step size,
backtracking line search,
optimal step size for Gaussian noise model
and LBFGS.
Moreover, 
our proposed Fisher step size can be computed efficiently
without any tuning parameter.
2) As expected, algorithms derived from the Poisson noise model
produce consistently better reconstruction quality
than algorithms derived from the Gaussian noise model
for low-count data.
Furthermore,
by incorporating regularizers that exploit
the assumed properties of the true signal,
the reconstruction quality can be further improved.
3) For regularized algorithms
with smooth corner-rounded TV regularizer,
WF with Fisher information 
converges faster than 
WF with backtracking line search,
LBFGS, MM and ADMM.

Future work includes
precomputing and tabluting the optimal curvature 
for the quadratic majorizer,
establishing sufficient conditions 
for global convergence,
investigating algorithms with other kind of regularizers
(\eg, deep learning methods \cite{remez:18:caf, zha:22:snl}),
investigating sketching methods % Will Grissom says it helps...
for large problem sizes
\cite{luo:22:ris},
and testing Poisson phase retrieval algorithms
under a wider variety of experimental settings.

\section*{References}
% \raggedright

\printbibliography[heading=none]
% \clearpage
% \newpage

\comment{ % for the appendix we don't need different numbering i think!
\setcounter{equation}{0}
\setcounter{figure}{0}
\renewcommand{\thefigure}{M.\arabic{figure}}
\renewcommand{\theequation}{M.\arabic{equation}}

\clearpage
\newpage
\section*{\blue{Supplement}}
\setcounter{equation}{0}
\setcounter{figure}{0}
\renewcommand{\thefigure}{S.\arabic{figure}} % supplement needs different numbering for clarity
\renewcommand{\theequation}{S.\arabic{equation}}
% s,supp
\comment{
\onecolumn
\section*{Supplemental Materials
\\
for\\
Algorithms for Poisson Phase Retrieval
\\
Z. Li, K. Lange, J. Fessler
}
}
\setcounter{subsection}{0}
\subsection{WF with gradient truncation}

Reference \cite{bian:16:fpr} shows that
under Poisson noisy measurement setting,
the relative error (RE) is almost
monotonically decreasing as
the tuning parameter $a^h$ increases.
Here we found a similar trend.
\fref{fig:trun-vs-notrun} shows that
under all three different system matrix settings,
the Poisson ML cost function values of TWF
is also monotonically decreasing 
as $a^h$ increases;
and are consistently higher 
than those of non-truncated WF.
Moreover, we found that
TWF can significantly
increase the computational time
compared to the non-truncated WF
for random Gaussian and ETM matrix settings;
for masked DFT matrix setting,
both versions of WF showed comparable running time,
presumably due to the efficient implementation 
of fast Fourier transform (FFT).
Based on the results presented in \fref{fig:trun-vs-notrun}, 
we did not use TWF for the rest simulations 
in this paper.
And we will focus on incorporating regularization
for potential reconstruction 
quality improvement.
\begin{figure}[hbt!]
    \centering
    \subfloat[Cost function value (logarithm) of WF vs. TWF.]{\includegraphics[width=0.9\linewidth]{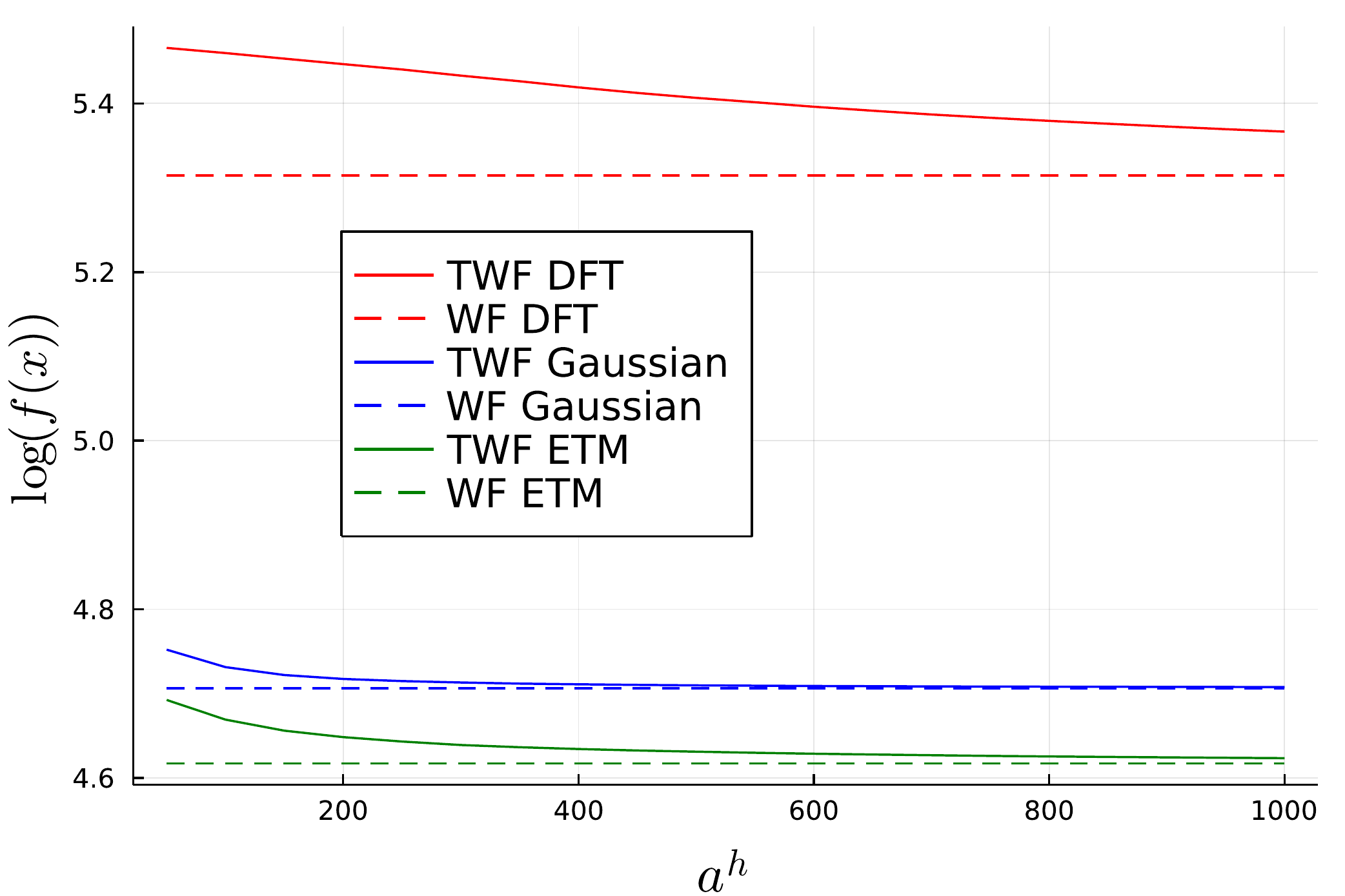}}
    \\
    \subfloat[Running time of WF vs. TWF.]{\includegraphics[width=0.9\linewidth]{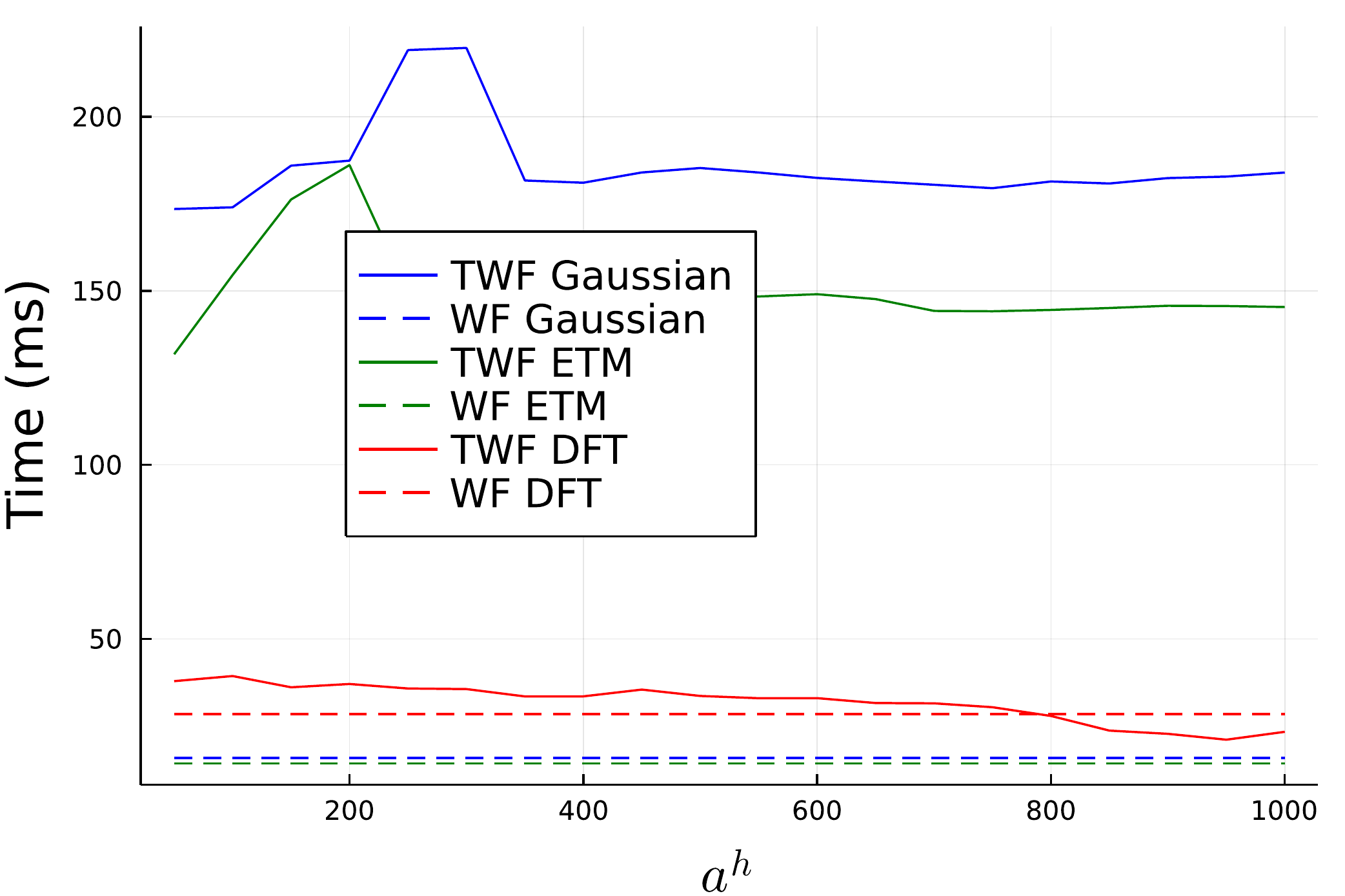}}
    \caption{Comparison of TWF and WF under random Gaussian, masked DFT and empirical transmission matrix settings. The logarithm of the Poisson 
    ML cost function was taken with base 10.
    All algorithms were ran 200 iterations.}
    \label{fig:trun-vs-notrun}
\end{figure}

\subsection{Derivation of an ADMM algorithm
for the Poisson phase retrieval problem}
\subsubsection{ADMM for Poisson ML problem}
With variable splitting $\vi = \baih \x$,
an augmented linearized 
Lagrangian of \eqref{e,cost} when $\bi = 0$ 
is given by
\begin{align}\label{admmb0,cost}
L_{\rho}(\v, \x; \bleta, \rho) & = 
\sumi \Big[(|\vi|^2 ) - \yi\log(|\vi|^2 )\Big]  \\
&+ \frac{\rho}{2} \sumi \Big[|\vi - \baih\x + \eta_i |^2 
- |\eta_i|^2 \Big], \nonumber
\end{align}
where $\bleta$ is the dual 
variable and $\rho > 0$ denotes the AL penalty parameter.

We followed the order of 
first updating $\v$, then $\x$ 
and finally $\bleta$. 
Because the updates for the
phase and magnitude of $\v$ are separable;
in particular,
the update for the phase of \v is
\begin{equation}\label{admmb0,signv}
    \mathrm{sign}(\vkk) = 
    \mathrm{sign}(\A \xk - \bleta_k).
\end{equation}
To update the magnitude  of \v,
similar to \cite{chang:18:tvb},
set $\ti=|\baih\xk - \eta_i|$;
then the 
update of a single
component $|\vi|$ is 
\comment{
determined
by the stationary condition:
\begin{equation}\label{admmb0,stat}
    0 = 2|\vi|-\frac{2\yi}{|\vi|}+\rho (|\vi|-\ti).
\end{equation}
\comment{ % i don't think we need this detail
Combining with the assumption 
that $|\vi|$ is positive
when $\bi = 0$, 
}
Equality \eqref{admmb0,stat} 
can be transformed
into a quadratic
}given by the following positive solution
of a quadratic polynomial:
\begin{equation}\label{admmb0,magv}
    |\vi| = \frac{\rho \ti+\sqrt{\rho^2\ti^2+8\yi(2+\rho)}}{2(2+\rho)}.
\end{equation}
Again, in the unregularized case,
the \x update is a simple least square (LS) problem
that can be optimized by CG 
or the following operation \eqref{admmb0,x} 
involving matrix inverse.
\begin{equation}\label{admmb0,x}
    \xkk = (\A'\A)^{-1} \A' (\vkk + \blmath{\eta}_k),
\end{equation}
Again,
if $\x \in \reals^N$,
then the \x update is
\[
\xkk =  (\real{\A'\A})^{-1} \realb{\A' (\vkk + \blmath{\eta}_k)}
.
\]

\subsubsection{Regularized ADMM}

For the regularized case
\eqref{e,Phi},
the \x update becomes
\begin{align}
\label{admm_x}
    \xkk &= \argmin{\x \in \FN}
    \frac{\rho}{2}\|\A \x - \vkk - \bleta_k\|_2^2  
    + \beta R(\x)
.
\end{align}
%or 
\comment{
\begin{align}\label{admm_x_bcd}
    \xkk, \zkk &= 
    \argmin{\x \in \FN, \bz \in 
    \complex^K}
    \frac{\rho}{2}\|\A \x - \vkk - \bleta_k\|_2^2  
    \nonumber \\
    &+ \beta \Big(\frac{1}{2} \|\T \x 
    - \bz \|_2^2 + \alpha \|\bz \|_1
    \Big),
\end{align}
\begin{equation}\label{admm_x_huber}
\xkk = \argmin{\x \in \FN}
    \frac{\rho}{2}\|\A \x - \vkk - \bleta_k\|_2^2 
    + \beta 1'h.(\T \x;\alpha)    
,\end{equation}
}
We solve this using CG or POGM,
depending on whether the regularizer $R$ is smooth or not.

The dual variable $\bleta$ ascent update \cite{boyd:11:doa} is simply
\begin{equation}\label{admmb0,eta}
    \bleta_{k+1} = \bleta_k + (\vkk - \A \xkk).
\end{equation}
For the case $\bi > 0$, everything is the 
same as the case $\bi=0$ except the update for 
$|\vi|$, for which one can verify the 
$|\vi|$ is instead a positive root
of the following cubic polynomial
\comment{
becomes
\begin{equation}\label{admmbn0,stat}
    0=2|\vi| -\frac{2\yi |\vi|}{|\vi|^2+\bi}
    + \rho (|\vi|-\ti),
\end{equation}
which reduces to the cubic
}
\begin{equation}\label{admmbn0,cubic}
    0 = (2+\rho)|\vi|^3 - \rho \ti |\vi|^2
    + (2\bi-2\yi+\rho \bi) |\vi| - \rho \bi \ti.
\end{equation}
Owing to the intermediate value theorem
and the non-negativity of $\rho$, 
$\bi$, $\ti$,
this cubic \eqref{admmbn0,cubic} 
can only possess 
one or three positive real roots.
If the cubic has one positive root, 
then the update of $|\vi|$ is simply 
to assign the single positive root.
If the cubic has three positive roots, 
we choose the root that minimizes 
the following Lagrangian term
based on
\eqref{lag,cost,bn0}:
%the most compared to the other two.
\begin{align}\label{lag,cost,bn0}
%L_{\rho}(\v, \x; \bleta, \rho) & = 
%\sumi \Big[
(|\vi|^2 +\bi) - \yi\log(|\vi|^2 +\bi)
%\Big]  
%\nonumber \\&
+ \frac{\rho}{2} 
%\sumi \Big[
(|\vi| - \ti )^2 
%- |\eta_i|^2
%\Big]
.\end{align}
A natural extension is to vary AL penalty parameter 
along with the variable update every few iterations. 
Reference \cite{boyd:11:doa} considered
the magnitude of primal residual 
$\br_{k+1} = \A\xkk - \vkk$
and dual residual $\bs_{k+1} = \rho \A' (\vkk - \vk)$,
as a principle to 
select penalty parameter to potentially improve
convergence for convex optimization problems. 
However, for non-convex problems, 
the penalty parameter $\rho$ is preferred 
to be sufficiently large to enable the convergence
of ADMM algorithm \cite{wang:19:gco}.
Thus, we used the following heuristic
to update $\rho$ every 10 iterations:
\begin{equation}
    \rho_{k+1} = 
    \left\{ 
    \begin{array}{cl}
    2 \rho_k, & \|r_k\| > 10 \|s_k\| \\
    \rho_k/2, & \|s_k\| > 100 \rho_k \|r_k\| \\
    \rho_k, & \text{otherwise}.
    \end{array}
    \right.
\end{equation}
Algorithm \ref{alg:admm} summarizes 
the ADMM algorithm described above.
\begin{algorithm}[ht!]
\SetAlgoLined
\SetInd{0.5em}{0.5em}
 \textbf{Input:} $\A, \y, \b, \x_0$ and 
 $n$ 
 (number of iterations)
 \\
 Initialize: \\
 $\v_0 \leftarrow \A \x_0$
 \\
 $\bleta_0 \leftarrow \v_0 - \A \x_0$
 \\
 \For{$k=0,...,n-1$}{
    Update $\text{sign}(\vk)$ by \eqref{admmb0,signv}
    \\
    \uIf{$\bi = 0$}{
    Update $|\vk|$ by \eqref{admmb0,magv}
    }
    \Else{
    Update $|\vk|$ by selecting root based on 
    \eqref{lag,cost,bn0}
    }
    \uIf{\textup{cost function is regularized}}{
        %\uIf{$\T$ \textup{is prox-friendly}}{
        Update $\xk$ by \eqref{admm_x} using CG or POGM
        %\eqref{admm_x_l1} using POGM
        %}
        %\Else{
        %Update $\xk$ by \eqref{admm_x_huber} using CG 
        %}
    }
    \Else{
    Update $\xk$ by \eqref{admmb0,x} or CG
    }
    Update $\bleta_k$ by \eqref{admmb0,eta}
 }
 \textbf{Output:} $\x_n$
 \caption{ADMM algorithm for the Poisson model}
 \label{alg:admm}
\end{algorithm}

\subsection{Computation Time Comparison Between CG and BS}

\fref{fig:NRMSE_vs_t_cgvsbs}
compares the convergence rates
of GS, MM and ADMM
that involve solving least squares problems.
In \fref{fig:NRMSE_vs_t_cgvsbs},
$M=3000$ and $N=100$
are both small enough
so that both backsubstitution (BS)
(backslash in Julia)
and CG are viable options
for solving the inner quadratic optimization problems.
In every case the CG version
of 250 iterations
decreased NRMSE faster than the BS version.

However, 
when \A is a DFT matrix,
including over-sampling and masks like in \eqref{e,masks},
one can verify that $\A'\A$
is a diagonal matrix
that is trivial to invert.
In this case
we used matrix inverse 
rather than CG 
to solve the least squares problems.
\label{subsec:supplots}
\begin{figure}[hbt!]
\begin{subfigure}[t]{\linewidth}
    \centering
    \includegraphics[width=\linewidth]{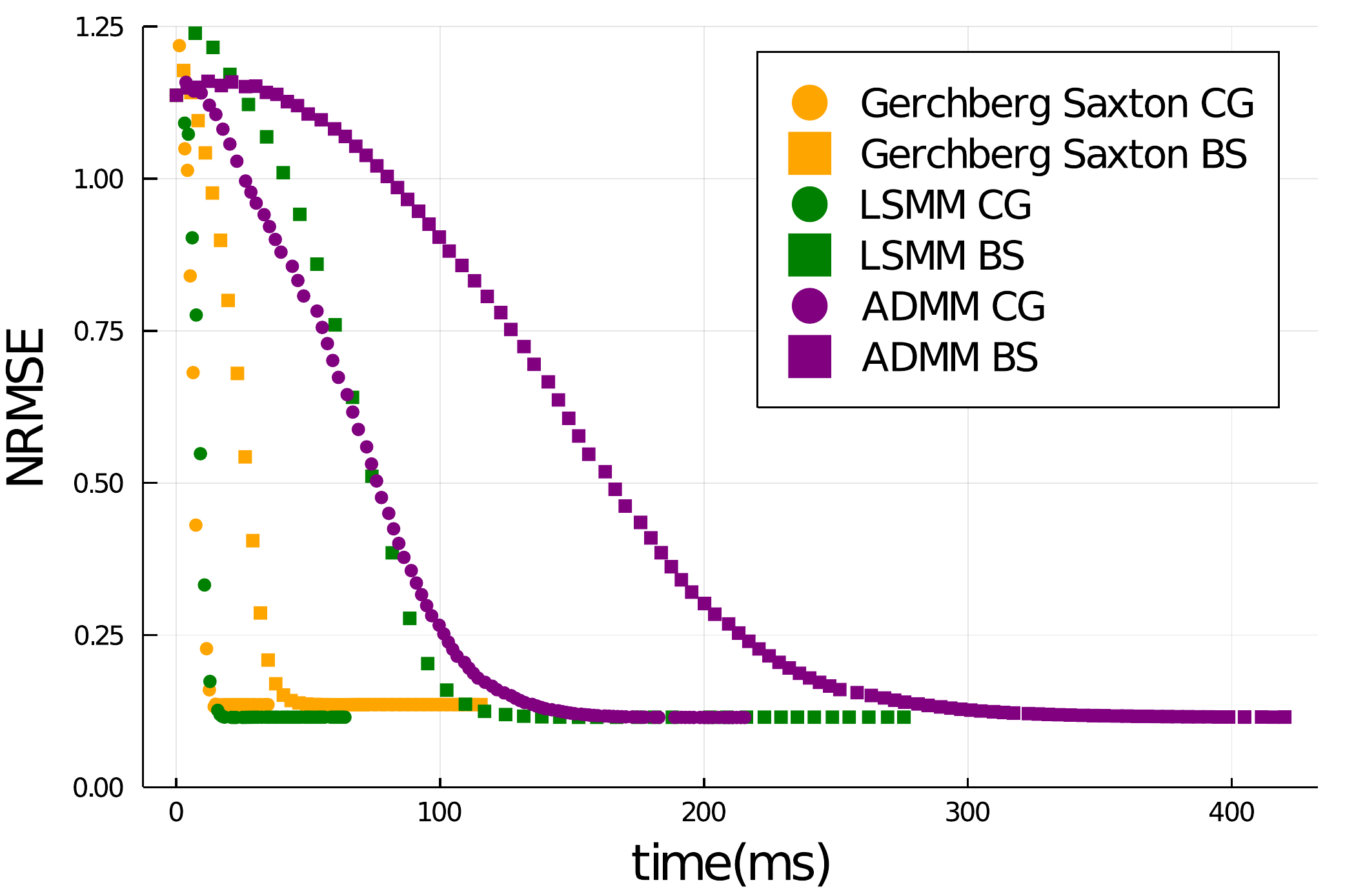}
    \caption{}
\end{subfigure}
\begin{subfigure}[t]{\linewidth}
    \centering
    \includegraphics[width=\linewidth]{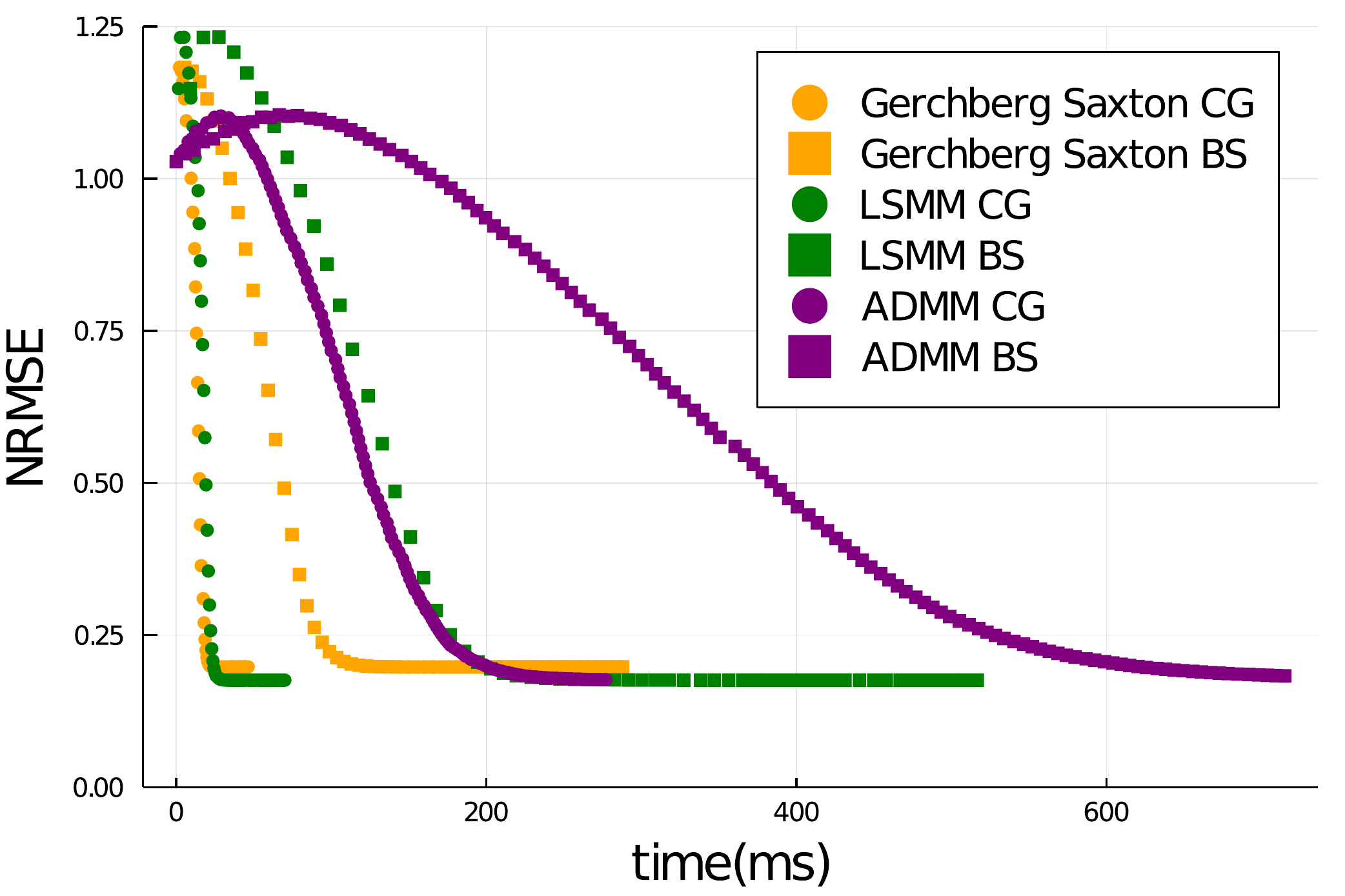}
    \caption{}
\end{subfigure}
    \caption{NRMSE vs. time (ms) using different
    update strategy. Here
    $\A$ is random Gaussian with 
    the average of $\baih \x$ equals to 2, 
    $M = 3000$
    and $\bi = 0.5$.
    Subfigure (a) 
    and (b) correspond to 
    experiments on 
    a real signal
    and 
    a complex signal,
    respectively.
    \comment{\red{rerun on server to remove "gaps".
    also how about "circles" for CG and "squares" for BS
    with matched colors so it is easier to compare?
    }}
    }
    \label{fig:NRMSE_vs_t_cgvsbs}
\end{figure}

\comment{
\subsection{MM for DFT Matrix Setting}
When  \A is a DFT matrix,
we conjectured
that
the dynamic range of the \yi values can be very wide,
\comment{
This is because of 
the Fourier series are all ones 
when $\tilde{n}$ is an integer 
multiplier of $\tilde{N}$, as illustrated in 
Fig.~\ref{fig:abs_y_fft}.
One issue evolved from this is the ill-posed 
curvature matrix.
}
leading to an ill-conditioned curvature matrix \W.
In particular,
the curvature changes  
dramatically across different coordinates,
and the MM algorithm based on a quadratic majorizer
using the maximum or improved curvature 
can converge very slowly.
Thus, 
for the DFT case
we imposed a threshold
on the elements of the curvature matrix \W
as follows:
\(
\tilde{\W} = \min(\W, \W_{\mu} + 3 \, \W_{\sigma})
,\)
where $\W_{\mu}$
and $\W_{\sigma}$ denote
the mean and the standard deviation
of the diagonal elements
of the original \W, respectively.
We used $\tilde{\W}$ 
to construct quadratic ``majorizers''
for the DFT case.
In addition, 
we found that 
it could be computationally expensive
to directly compute
the best Lipschitz constant
($\Lip = \|\A' \tilde{\W} \A\|_2$)
every time $\tilde{\W}$ changes,
so we used
$\Lip = \|\A'\A\|_2 \|\tilde{\W}\|_2$
in MM algorithms,
where one can verify 
\begin{equation}\label{ata,dft}
\|\A'\A\|_2 = \cdft^2 \tilde{N}^2 \max\Big\{\sum_{l=1}^L D_l\Big\}    
,\end{equation}
where $\cdft$ is the scaling factor applied to \A. % already part of A?

\subsection{Truncated vs. Non-truncated WF}
\fref{fig:trun_vs_notrun} compares truncated WF (TWF) vs. non-truncated WF under
Poisson noise model.
We found that, similar to results in \cite{bian:16:fpr},
the reconstruction error (NRMSE) for TWF is almost 
monotonically decreasing
as the truncate threshold parameter $a^h$
goes larger.
To achieve a comparable NRMSE with non-truncated WF,
one needs to set $a^h > 100$, however,
under this setting, 
less than 5\% indices of $\y$ are truncated
when calculating the gradient.
This implies that
almost all the measurements
are useful for the reconstruction
and therefore 
should not be truncated.
In addition, 
calculating the truncated indices
in each iteration
can be computationally inefficient,
especially when $M$ is large.
Considering all reasons described above,
we did not use gradient truncation (TWF) in this paper.
\begin{figure}[t!]
\centering
\subfloat[]{\includegraphics[width=\linewidth]{fig/fig,trun_vs_notrun/trun_vs_notrun_20210427.pdf}} \\
\subfloat[]{\includegraphics[width=\linewidth]{fig/fig,trun_vs_notrun/trun_vs_notrun_frac_kept_data_20210427.pdf}}
\caption{Comparison between truncated WF and non-truncated WF. 
$a^h$ was set from 10 to 500 with interval 10. System matrix $\A$ was
modeled as a random Gaussian. The average of $\baih \x$ is 2, with $\bi = 0.1$
and $M = 5000$.}
\label{fig:trun_vs_notrun}
\end{figure}
}

\subsection{Huber Function vs. Alternating Minimization}
For non-prox-friendly regularizers,
other than modifying by the Huber function,
an alternative can be 
introducing another variable 
and applying alternating minimization.
In particular,
The update of $\x$ in LSMM becomes
\begin{align}\label{quad_mm_bcd}
    \xkk, \zkk &= \argmin{\x \in \FN, 
    \z \in \complex^K}
    Q_k(\x,\z),
    \nonumber \\
    Q_k(\x,\z)
    &\defequ
    q(\x;\xk) + \beta \Big( \frac{1}{2}
    \|\T \x - \bz \|_2^2 + \alpha \|\bz\|_1 \Big),
    \nonumber
\end{align}
where one can alternatively update 
\x and \z.
The \x update
uses the closed-form solution that involves
matrix inverse or conjugate gradient.
%depending on the scale of  $N$.
The \z update is simply 
a soft-thresholding  operation.

Similarly,
the update of $\x$ in ADMM is
\begin{align}
    \xkk, \zkk &= 
    \argmin{\x \in \FN, \bz \in 
    \complex^K}
    \frac{\rho}{2}\|\A \x - \vkk - \bleta_k\|_2^2  
    \nonumber \\
    &+ \beta \Big(\frac{1}{2} \|\T \x 
    - \bz \|_2^2 + \alpha \|\bz \|_1
    \Big). \nonumber
\end{align}
We compared these two approaches 
(Huber function vs. alternating minimization)
and found that
using Huber function
was more efficient
than alternating minimization,
as evident in \fref{fig:huber_vs_altmin}.
\begin{figure}[ht!]
    \centering
    \includegraphics[width=\linewidth]{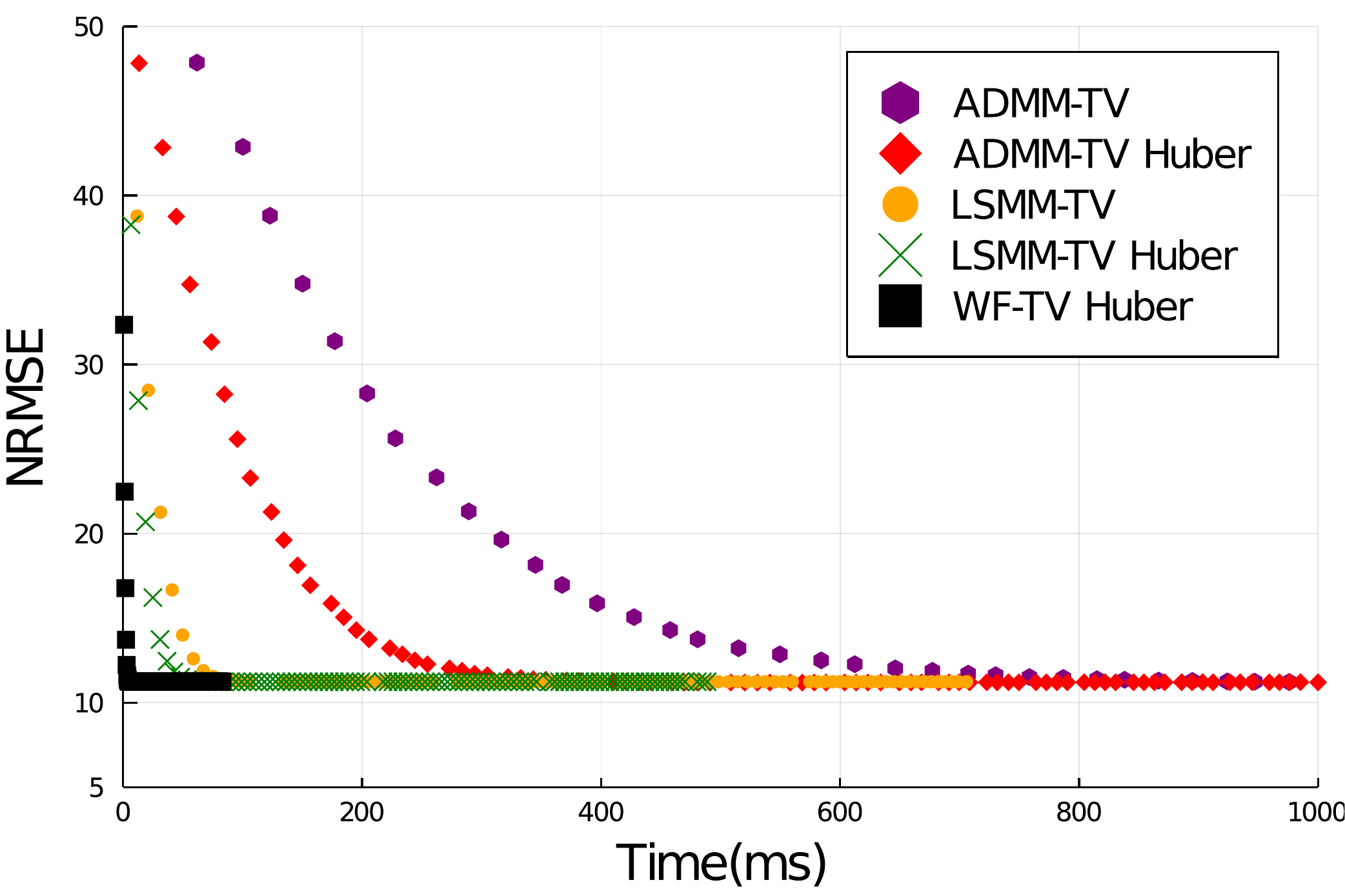}
    \caption{Speed comparison of WF, LSMM and ADMM
    using Huber function and alternating minimization 
    for regularization.
    LSMM-TV and ADMM-TV denote 
    the alternating minimization approach.
    }
    \label{fig:huber_vs_altmin}
\end{figure}

\comment{
\subsection{Experiments on an Empirical Transmission Matrix}
\label{subsec:exptm}
\begin{figure*}[ht!]
    \centering
    \includegraphics[width = \linewidth]{fig/A_etm_result.pdf}
    \caption{Reconstructed images and their corresponding 
    NRMSE compared to the true image (of size $16 \times 16$),
    using an empirical system matrix from \cite{metzler:17:cis}
    with $M=10000$. 
    \comment{ % not important to repeat
    $\x_{\text{true}}$ was designed to have 
    pixel values 
    ranging from 0 to 0.25 and $\bi$ was 
    set to 0.1.}
    }
    \label{fig:exp_A_etm}
\end{figure*}
\input{s,exp,tm}
}
% \clearpage
% \newpage

}
\end{document}